\documentclass{aa} 

\usepackage[svgnames]{xcolor}
\usepackage{graphicx}
\usepackage{caption}
\usepackage{subcaption}
\usepackage{cancel}
\usepackage{txfonts}
\usepackage{multirow}
\usepackage{nicefrac}
\usepackage[d]{esvect}
\usepackage{lipsum}
\usepackage{diagbox}
\usepackage{comment}
\usepackage[colorlinks=True,
            linkcolor=blue,
            citecolor= blue]{hyperref}

\begin{document} 

   \title{Angular momentum transport in a contracting stellar radiative zone embedded in a large scale magnetic field}

   \titlerunning{Angular momentum transport in a contracting stellar radiative zone embedded in a large scale magnetic field}

   \author{B. Gouhier,
           \inst{1}
           L. Jouve,
          \inst{1}
          \and
          F. Lignières
          \inst{1}
          }
          
   \authorrunning{B. Gouhier et al.}

   \institute{Institut de Recherche en Astrophysique et Planétologie (IRAP), Université de Toulouse,
              14 Avenue Edouard Belin, 31400 Toulouse, France\\
              \email{[bgouhier, ljouve, flignieres]@irap.omp.eu}
             }
   \date{Received 22 June 2021 / Accepted 7 January 2022}

\abstract
{Some contracting or expanding stars are thought to host a large-scale magnetic field in their radiative interior. By interacting with the contraction-induced flows, such fields may significantly alter the rotational history of the star. They thus constitute a promising way to address the problem of angular momentum transport during the rapid phases of stellar evolution.}
{In this work, we aim at studying the interplay between flows and magnetic fields in a contracting radiative zone.}
{We perform axisymmetric Boussinesq and anelastic numerical simulations in which a portion of radiative zone is modelled by a rotating spherical layer, stably stratified and embedded in a large-scale (either dipolar or quadrupolar) magnetic field. This layer is subject to a mass-conserving radial velocity field mimicking contraction. The quasi-steady flows are studied in strongly or weakly stably stratified regimes relevant for pre-main sequence stars and for the cores of subgiant and red giant stars. The parametric study consists in varying the amplitude of the contraction velocity and of the initial magnetic field. The other parameters are fixed with the guidance of the previous study of \cite{gouhier2020axisymmetric}.}  
{After an unsteady phase during which the toroidal field grows linearly and then back-reacts on the flow, a quasi-steady configuration is reached, characterised by the presence of two magnetically decoupled regions. In one of them, magnetic tension imposes solid-body rotation. In the other, called the dead zone, the main force balance in the angular momentum equation does not involve the Lorentz force and a differential rotation exists. In the strongly stably stratified regime, when the initial magnetic field is quadrupolar, a magnetorotational instability is found to develop in the dead zones. The large-scale structure is eventually destroyed and the differential rotation is able to build-up in the whole radiative zone. In the weakly stably stratified regime, the instability is not observed in our simulations but we argue that it may be present in stars.}
{We propose a scenario that may account for the post-main sequence evolution of solar-like stars, in which a quasi-solid rotation can be maintained by a large-scale magnetic field during a contraction timescale. Then, an axisymmetric instability would destroy this large-scale structure and enables the differential rotation to set in. Such a contraction driven instability could also be at the origin of the observed dichotomy between strongly and weakly magnetic intermediate-mass stars.}


   \maketitle

\section{Introduction}

Rotation is ubiquitous at every stage of stellar evolution. Yet, it is still often considered as a second order effect in stellar evolution models. A full description necessarily entails by a thorough study of the differential rotation and meridional circulation induced by the rotation, how they interact with other physical processes (e.g. magnetic field or internal gravity waves), and the ensuing potential instabilities. In his seminal work, \citet{zahn1992circulation} proposed a model for the transport of chemical elements and angular momentum (AM) (without magnetic field and internal gravity wave) that has been later implemented in state-of-the-art stellar evolution codes. One of the strong assumption is that the differential rotation in radiative zones is close to shellular (i.e. constant on a isobar) because of the anisotropic turbulence induced by shear instabilities in stably stratified conditions. This formalism has been successful at explaining a large number of stellar observations (see \cite{maeder2008physics} for a review). However a growing body of evidence shows that additional AM transport mechanisms are still needed. This is particularly true for contracting stars. During the pre-main sequence (PMS) or post-MS evolution, a spin-up of stellar cores is naturally produced. However, observations of surface rotation of stars in young clusters \citep{gallet2013improved} as well as asteroseismic studies of red giant stars \citep{eggenberger2012angular, ceillier2013understanding, marques2013seismic} tend to show that rotation rates are in fact strongly overestimated in models. In order to improve this formalism, the first thing to ask is what kind of flows are really expected in those contracting stars.

In a previous work \citep{gouhier2020axisymmetric}, we investigated the differential rotation and meridional circulation produced in a modelled contracting stellar radiative zone. This axisymmetric study included the effects of stable stratification and was conducted both under the Boussinesq and the anelastic approximations but the effects of a magnetic field were ignored. We showed that a radial differential rotation should be expected only in strongly stably stratified radiative zones such as the degenerate cores of subgiants. Indeed, any meridional circulation is inhibited by the strong buoyancy force and the characteristic amplitude of the differential rotation in the linear regime is found to be proportional to the ratio of the viscous to contraction timescales $\Delta \Omega / \Omega_0 \propto \tau_{\nu} / \tau_{\text{c}}$. In conditions relevant for the outside of the degenerate cores of subgiants and for PMS stars though, thermal diffusion weakens the stable stratification and allows a meridional circulation to exist, with a typical amplitude of the order of the contraction speed. The differential rotation profile then exhibits both a dependence in latitude and radius and its characteristic amplitude in the linear regime is found to be $\Delta \Omega / \Omega_0 \propto \tau_{\text{ED}} / \tau_{\text{c}}$, where $\tau_{\text{ED}}$ is the Eddington-Sweet timescale associated with the AM transport by the meridional flow. Both estimates, assuming a contraction timescale comparable to the Kelvin-Helmholtz timescale, tend to indicate that the amplitude of the differential rotation can be quite strong and that another process of AM transport should be invoked to reproduce the rotation rates of pre-MS and post-MS stars.

The presence of a large-scale magnetic field in such contracting radiative zones can drastically modify this picture. It is commonly agreed that radiative zones can host such fields (see \cite{braithwaite2017magnetic} for a review). Magnetic fields with surface intensities of a few hundred Gauss have indeed been detected in a fraction of intermediate-mass PMS stars \citep{alecian2013high}. These stars are most probably the progenitors of the MS chemically peculiar Ap/Bp  stars that host large-scale mostly dipolar fields with intensities ranging from $300$ G to $30$ kG \citep{donati2009magnetic}. Meanwhile, a distinct population of MS intermediate-mass stars, including the A-type star Vega and the Am-type stars Sirius, $\beta$ Ursae Majoris and $\theta$ Leonis, exhibits much weaker ($\sim$ 1 G) multi-polar magnetic fields \citep{lignieres2009first, petit2010rapid, petit2011detection, blazere2016detection}. This magnetic dichotomy could be explained if, during the PMS, contraction forces a differential rotation that destroys pre-existing large scale weak magnetic fields through magnetohydrodynamic (MHD) instabilities \citep{auriere2007weak, lignieres2013dichotomy, jouve2015three, jouve2020interplay}.

Unlike PMS stars, the radiative zone of post-MS stars is overlaid by a large convective envelope preventing any direct measurement of a magnetic field in the convectively stable regions. Recently, asteroseismology revealed a class of red giants exhibiting dipolar oscillation modes with a very weak visibility \citep{mosser2012characterization}. This phenomenon has been attributed to the presence of an internal magnetic field modifying the angular structure of the dipole waves thus leading to their trapping in the radiative core \citep{fuller2015asteroseismology}. This so-called greenhouse effect, supported by other authors \citep{stello2016suppression, stello2016prevalence, cantiello2016asteroseismic} is, however, seriously questioned because these modes preserve their mixed character at odds with Fuller's scenario \citep{mosser2017dipole}. In parallel, the possibility to detect magnetic fields through their effect on the oscillation frequencies is under investigation (see e.g., \cite{bugnet2021magnetic}). Awaiting observational evidence, theoretical work strongly supports the presence of magnetic fields in stars that possess a convective core (MS stars with M $\gtrsim$ 1.2 M$_{\sun}$). The numerical simulations of core-dynamos of MS A- and B-type stars \citep{brun2005simulations, augustson2016magnetic} indicate that magnetic fields with intensities ranging from $0.1 - 1.0$ MegaGauss can indeed be generated. Such a magnetic field could then relax into a large-scale stable configuration in the radiative interior of the post-MS stars where it would be buried for the rest of its evolution \citep{braithwaite2004fossil}. 

Large-scale magnetic fields are able to impose a quasi-solid rotation on very short timescales, even for very weak intensities \citep{ferraro1937non, mestel1987magnetic}. Besides, enforcing a quasi-solid rotation during $\sim 1$ Gyr after the end of the MS enabled \cite{spada2016angular} to reproduce the rotation rates of sub-giants, as measured by asteroseismology \citep{deheuvels2014seismic}. This result was also obtained by \cite{eggenberger2019asteroseismology} and an observational support was then provided by \cite{deheuvels2020seismic} who measured a near solid rotation in two young subgiants. Interestingly, the efficiency of the AM transport then seems to decrease up to the tip of the red giant branch (RGB) before increasing again \citep{spada2016angular, eggenberger2019asteroseismology, deheuvels2020seismic}. Those works suggest a scenario broken down into several key phases. First, a quasi-solid rotation is maintained  for some time through an efficient AM transport mechanism (possibly magnetic tension imposed by a large-scale field). Then, this mechanism would become inefficient and differential rotation  would build up again before another AM transport mechanism, such as turbulent transport induced by MHD instabilities \citep{spruit2002dynamo, rudiger2015angular, fuller2019slowing, jouve2020interplay} takes over. 

In this work, we intend to study the flows induced by a contracting radiative zone in the presence of a large-scale magnetic field, through axisymmetric MHD simulations. In particular, we focus on the structure of the steady-states differential rotation and on the ability of the magnetic field to transport AM. The paper is organised as follows: in Sect. \ref{governing_equations} we present the mathematical model, then in Sect. \ref{timescales_physical_processes}, the different timescales involved in our problem. The initial and boundary conditions as well as the numerical method are described in Sects. \ref{init_bound_cond} and \ref{numerical_model} respectively. In Sect. \ref{space_parameters} we provide the reader with the relevant timescales in the stellar context and the consequences for our numerical study. The results of the simulations in the viscous and Eddington-Sweet regimes are finally given in Sect. \ref{numerical_results}, and are then summarised in Sect. \ref{conclusion} where the astrophysical implications are also discussed. 

\section{Mathematical formulation}
\label{governing_equations}

In our previous work \citep{gouhier2020axisymmetric}, we investigated the differential rotation and meridional flows produced in a contracting stellar radiative zone. In this follow-up work, we add the effect of an initial large-scale magnetic field. To do so, we numerically solve the Boussinesq or anelastic magnetohydrodynamical (MHD) equations in a spherical shell filled with a stably-stratified fluid subject to a radial contraction and embedded in a magnetic field. In this section we present the governing equations that we numerically solve in the two aforementioned approximations.

In this study, the fluid contraction is modelled using a mass-conserving contraction velocity field defined by

\begin{equation}
\vv{V_f} = V_f \hspace*{0.02cm} (r) \hspace*{0.02cm} \vv{e}_r = - \displaystyle \frac{V_0 \hspace*{0.04cm} \rho_0 \hspace*{0.02cm} r{_0^2}}{\overline{\rho} r^2} \hspace*{0.02cm} \vv{e}_r
\label{Anelcontraction}.
\end{equation}

\noindent where $r$ is the radius and $\overline{\rho}$ the background density, $r_0$ and $\rho_0$ their respective values at the outer sphere and $V_0$ is the amplitude of the contraction velocity at the outer sphere. Using the Lantz-Braginsky-Roberts (LBR) approximation (\cite{lantz1992dynamical}, \cite{braginsky1995equations}), assuming a uniform kinematic viscosity $\nu$, thermal diffusion $\kappa$, magnetic diffusion $\eta$, and neglecting the centrifugal effects and local sources of heat, the dimensionless axisymmetric anelastic equations of a magnetised fluid \citep{jones2011anelastic} undergoing contraction read

\begin{equation}
\begin{array}{lll}
\vv{\nabla} \cdot \left \lbrack \tilde{\overline{\rho}} \left( \vv{\tilde{U}} + \vv{\tilde{V}}_f \right) \right \rbrack = 0 \quad \text{and} \quad \vv{\nabla} \cdot \vv{\tilde{B}} = 0
\label{anel_continuity_adim},
\end{array}
\end{equation}

\begin{equation}
\begin{array}{lll}
\hspace*{-0.15cm} R_o \left \lbrack \displaystyle \frac{\partial \vv{\tilde{U}}}{\partial t} + \left( \left( \vv{\tilde{U}} + \vv{\tilde{V}}_f \right) \cdot \vv{\nabla} \right) \left( \vv{\tilde{U}} + \vv{\tilde{V}}_f \right) \right \rbrack + 2 \vv{e}_z \times \left( \vv{\tilde{U}} + \vv{\tilde{V}}_f \right) = \\\\ \hspace*{-0.15cm} - \vv{\nabla} \left( \displaystyle \frac{\tilde{\Pi}^{'}}{\tilde{\overline{\rho}}} \right) +
 \displaystyle \frac{\tilde{S}^{'}}{\tilde{r}^2} \hspace*{0.05cm} \vv{e}_r +\displaystyle \frac{E}{\tilde{\overline{\rho}}} ~ \vv{\nabla} \cdot \vv{\vv{\tilde{\sigma}}} + 
 \left( \displaystyle \frac{L_u E}{P_m} \right)^2 \displaystyle \frac{R_o^{-1}}{\tilde{\overline{\rho}}}  \left \lbrack \left(\vv{\nabla} \times \vv{\tilde{B}} \right) \times \vv{\tilde{B}} \right \rbrack
 \label{anel_momentum_adim},
 \end{array}
\end{equation}

\begin{equation}
\begin{array}{lll}
\displaystyle \frac{\partial \vv{\tilde{B}}}{\partial t} = \vv{\nabla} \times \left \lbrack \left( \vv{\tilde{U}} + \vv{\tilde{V}}_f \right) \times \vv{\tilde{B}}  \right \rbrack + \displaystyle \frac{1}{R_m} \vv{\nabla}^2 \vv{\tilde{B}}
 \label{anel_induct_adim},
\end{array}
\end{equation}

\begin{equation}
\begin{array}{lll}
\hspace*{-0.15cm} \tilde{\overline{\rho}} \tilde{\overline{T}} \left \lbrack P_r R_o \left( \displaystyle \frac{\partial \tilde{S}^{'}}{\partial t} \right.  + \left( \left( \vv{\tilde{U}} + \vv{\tilde{V}}_f \right) \cdot \vv{\nabla} \right) \tilde{S}^{'} \right) + P_r \left( \displaystyle \frac{N_0}{\Omega_0} \right)^2 \left( \tilde{U}_r + \tilde{V}_f \right) \left. \displaystyle \frac{\text{d} \tilde{\overline{S}}}{\text{d} r} \right \rbrack \\\\ \hspace*{-0.15cm} = E ~ \vv{\nabla} \cdot \left( \tilde{\overline{\rho}} \tilde{\overline{T}} \vv{\nabla} \tilde{S}^{'} \right) + D_i Pe_c E^2 \left \lbrack \tilde{\mathcal{Q}}{_{\nu}} + P_m^{-1} \left(\displaystyle \frac{L_u E}{P_m R_o} \right)^2 \left( \vv{\nabla} \times \vv{\tilde{B}} \right)^2 \right \rbrack,
\end{array}
\label{anel_entropie_adim}
\end{equation}

\noindent where the diffusion of entropy is introduced in the energy equation instead of the diffusion of temperature (\cite{braginsky1995equations, clune1999computational}). The non-dimensional form (identified by the tilde variables) of these equations is obtained using the radius of the outer sphere $r_0$ as the reference lengthscale, the value of the contraction velocity field at the outer sphere $V_0$ as a characteristic velocity, and $\tau_{\text{c}} = r_0 / V_0$ as the reference timescale of the contraction. The frame rotates at $\Omega_0$, the rotation rate of the outer sphere and all the thermodynamics variables are expanded as a background value plus fluctuations, respectively denoted with an overbar and a prime. The background density and temperature field are non-dimensionalised respectively by the outer sphere density $\rho_0$ and temperature $T_0$, while the background gradients of temperature and entropy fields are adimensionalised using the temperature and entropy difference $\Delta \overline{T}$ and $\Delta \overline{S}$ between the two spheres. The pressure fluctuations are non-dimensionalised by $\rho_0 r_0 \Omega_0 V_0$ and the entropy fluctuations by $C_p \Omega_0 V_0/g_0$, where $C_p$ is the heat capacity and $g_0$ the gravity at the outer sphere. Finally we use the value of the surface poloidal field at the poles $B_0$ as the reference scale for the magnetic field.

In Eqs. \eqref{anel_continuity_adim}, \eqref{anel_momentum_adim}, \eqref{anel_induct_adim}, and \eqref{anel_entropie_adim}, $\vv{U}$ is the velocity field, $\tilde{\sigma}_{ij}$ is the dimensionless stress tensor, $\tilde{\mathcal{Q}}_{\nu}$ is the dimensionless viscous heating and the gravity profile is $\propto r^{-2}$. The reference state is non-adiabatic and a uniform positive entropy gradient is used to produce a stable stratification. It is related to the Brunt-Väisälä frequency defined by:

\begin{equation}
N_0 = \sqrt{\displaystyle \frac{g_0}{C_p} \displaystyle \frac{\Delta \overline{S}}{r_0}}
\label{Anel_Vaisala},
\end{equation}

\noindent and which controls the amplitude of this stable stratification. The magnitude of the deviation to the isentropic state is controlled by the parameter $\epsilon_s = \Delta \overline{S} / C_p$ chosen sufficiently small to ensure the validity of the anelastic approximation. With the dissipation number $D_i = g_0 r_0 / T_0 C_p$ it sets the background temperature and density profiles (see \cite{gouhier2020axisymmetric}).

These anelastic equations involve six independent dimensionless numbers, a Rossby number based on the amplitude of the contraction velocity $R_o = V_0 / \Omega_0 r_0$, the Ekman number $E = \nu / \Omega_0 r_0^2$, the Prandtl number $P_r = \nu / \kappa$, the ratio between the reference Brunt-Väisälä frequency and the rotation rate of the outer sphere $N_0/\Omega_0$, the magnetic Prandtl number $P_m = \nu / \eta$ and the Lundquist number $L_u = B_0 r_0 / \sqrt{\mu_0 \rho_0} \eta$ where $\mu_0$ is the vacuum permeability. From these dimensionless numbers, three additional parameters can then be defined: a contraction Reynolds number $Re_c = R_o/E$, a Péclet number $Pe_c = P_r \hspace*{0.05cm} Re_c$ and a magnetic Reynolds number $R_m = P_m \hspace*{0.05cm} Re_c$.

When the compressibility effects are neglected, except in the buoyancy term of the momentum equation, the Boussinesq approximation is recovered. In that case, using $\Omega_0 V_0 T_0 / g_0$ as the scale of the temperature deviations $\Theta'$, the dimensionless governing equations read

\begin{equation}
\vv{\nabla} \cdot \left( \vv{\tilde{U}} +  \vv{\tilde{V}}_f \right) = 0 \quad \text{and} \quad \vv{\nabla} \cdot \vv{\tilde{B}} = 0
\label{bouss_continuity_adim},
\end{equation}

\begin{equation}
\begin{array}{lll}
\hspace*{-0.15cm} R_o \left \lbrack \displaystyle \frac{\partial \vv{\tilde{U}}}{\partial t} + \left( \left( \vv{\tilde{U}} +  \vv{\tilde{V}}_f \right) \cdot \vv{\nabla} \right) \left( \vv{\tilde{U}} +  \vv{\tilde{V}}_f \right)\right \rbrack + 2 \vv{e}_z \times \left( \vv{\tilde{U}} +  \vv{\tilde{V}}_f \right) = \\\\ \hspace*{-0.15cm} - \vv{\nabla} \tilde{\Pi}^{'} + \tilde{\Theta}^{'} \hspace*{0.02cm} \vv{\tilde{r}} + E ~ \vv{\nabla}^2 \left( \vv{\tilde{U}} + \vv{\tilde{V}}_f \right) + \left( \displaystyle \frac{L_u E}{P_m} \right)^2 R_o^{-1} \left \lbrack \left(\vv{\nabla} \times \vv{\tilde{B}} \right) \times \vv{\tilde{B}} \right \rbrack
\label{bouss_momentum_adim},
\end{array}
\end{equation}

\begin{equation}
\displaystyle \frac{\partial \vv{\tilde{B}}}{\partial t} = \vv{\nabla} \times \left \lbrack \left( \vv{\tilde{U}} + \vv{\tilde{V}}_f \right) \times \vv{\tilde{B}}  \right \rbrack + \displaystyle \frac{1}{R_m} \vv{\nabla}^2 \vv{\tilde{B}}
 \label{bouss_induct_adim},
\end{equation}

\begin{equation}
\begin{array}{lll}
\hspace*{-0.15cm} P_r R_o \left \lbrack \displaystyle \frac{\partial \tilde{\Theta}^{'}}{\partial t} + \left( \left( \vv{\tilde{U}} +  \vv{\tilde{V}}_f \right) \cdot \vv{\nabla} \right) \tilde{\Theta}^{'} \right \rbrack + P_r \left( \displaystyle \frac{N_0}{\Omega_0} \right)^2 \left( \tilde{U}_r +  \tilde{V}_f \right) \displaystyle \frac{\text{d} \tilde{\overline{T}}}{\text{d} r} \\\\ = E ~ \vv{\nabla}^2 \tilde{\Theta}^{'}
\label{bouss_entropie_adim},
\end{array}
\end{equation}

\noindent where the gravity profile is now $\propto r$, the reference Brunt-Väisälä frequency is defined using the correspondence $\Delta \overline{S} / C_p = \Delta \overline{T} / r_0$ in Eq. \eqref{Anel_Vaisala} and the contraction velocity field Eq. \eqref{Anelcontraction} is simplified using $\overline{\rho} = \rho_0$. 

To conclude this section we note that when the anelastic approximation is used the parameter space is defined by eight dimensionless numbers: $\epsilon_s$, $D_i$, $Re_c$, $E$, $P_r$, $P_m$, $L_u$ and $N_0^2 / \Omega_0^2$. Instead, only the last six are necessary in the Boussinesq approximation. In this study, only $Re_c$, $L_u$ and the product $P_r \left( N_0/\Omega_0\right)^2$ will be varied.

\section{Timescales of physical processes}
\label{timescales_physical_processes}

In this section, we describe the various timescales involved in the transport of AM in our problem. We start by briefly recalling the relevant hydrodynamical timescales (see \cite{gouhier2020axisymmetric}) then we introduce two timescales associated with the presence of a magnetic field.

In this work, the dimensional form of the AM equation under the Boussinesq approximation reads

\begin{equation}
\begin{array}{lll}
\hspace*{-0.1cm}
\displaystyle \frac{\partial U_{\phi}}{\partial t} + \text{NL} + \underbrace{2 \Omega_0 U_s}_{\text{Coriolis term}} - \underbrace{\displaystyle \frac{1}{\mu_0 \rho_0} \left( \vv{B}_p \cdot \vv{\nabla} \right) B_{\phi}}_{\text{Lorentz force}} -\underbrace{\nu \hspace*{0.05cm} D^2 U_{\phi}}_{\text{Viscous term}} = \\\\
\underbrace{\displaystyle \frac{V_0 r{_0^2}}{r^3} \displaystyle \frac{\partial}{\partial r} \left(r U_{\phi} + r^2 \sin{\theta} \hspace*{0.05cm} \Omega_0 \right)}_{\text{Contraction}}
\label{angular_momentum_evolution},
\end{array}
\end{equation}

\noindent where $D^2$ is the azimuthal component of the vector Laplacian operator, $U_s = \cos{\theta} U_{\theta} + \sin{\theta} U_r$ is the velocity field perpendicular to the rotation axis and NL denotes the non-linear advection term. By balancing the partial time derivative with the contraction term in Eq. \eqref{angular_momentum_evolution} we recover the contraction timescale used to non-dimensionalise the governing equations in Sect. \ref{governing_equations}

\begin{equation}
\tau_{\text{c}} = \displaystyle \frac{r_0}{V_0}
\label{contraction_timescale},
\end{equation}

\noindent as well as its linear form 

\begin{equation}
\tau_{\text{c}}^{L} = \displaystyle \frac{r_0}{V_0} \displaystyle \frac{\Delta \Omega_0}{\Omega_0}
\label{linear_contraction_timescale},
\end{equation}

\noindent when $\Delta \Omega / \Omega_0 \ll 1$. In the anelastic approximation the contraction term in Eq. \eqref{angular_momentum_evolution} is multiplied by $\rho_0 / \overline{\rho}$ and the resulting timescale is then weighted by the background density profile

\begin{equation}
\tau_{\text{c}}^{A} = \left( \displaystyle \int_{r_i/r_0}^{1} \displaystyle \frac{\overline{\rho}}{\rho_0} \hspace*{0.05cm} \text{d} \left( r/r_0\right) \right) \tau_{\text{c}} \hspace*{0.2cm} \text{or} \hspace*{0.2cm} \tau_{\text{c}}^{L^A} = \left( \displaystyle \int_{r_i/r_0}^{1} \displaystyle \frac{\overline{\rho}}{\rho_0} \hspace*{0.05cm} \text{d} \left( r/r_0\right) \right) \tau_{\text{c}}^{L}
\label{anel_contraction_timescales},
\end{equation}

\noindent where $\tau_{\text{c}}^{L^A}$ denotes the linear version. The AM transport by contraction can be balanced either by the viscous processes on a viscous timescale $\tau_{\nu} = r_0^2 / \nu$, or by a meridional circulation of Eddington-Sweet type, in which case it redistributes the AM on the following timescale

\begin{equation}
\tau_{\text{ED}} = \displaystyle \frac{r_0^2}{\kappa} \left(\displaystyle \frac{N_0}{\Omega_0} \right)^2
\label{edd_timescale}.
\end{equation}

Ekman layers tend to develop at the spherical boundaries to accommodate the interior flow to the boundary conditions. In unstratified flows, they drive a global circulation. In stars, the stable stratification efficiently opposes this global flow although it can still exist in numerical simulations because the Ekman numbers can not reach stellar values. It then transports the AM on a spin-up timescale defined by 

\begin{equation}
\tau_{\text{E}} = \sqrt{\displaystyle \frac{r_0^2}{\Omega_0 \nu}}
\label{Ekman_timescale}.
\end{equation}

The relative importance of these AM redistribution processes is given by the ratio of the above timescales, namely:

\begin{equation}
\displaystyle \frac{\tau_{\text{ED}}}{\tau_{\nu}} = P_r \left( \displaystyle \frac{N_0}{\Omega_0}\right)^2 \text{;} \quad \displaystyle \frac{\tau_{\text{E}}}{\tau_{\nu}} = \sqrt{E} ~ \text{;} \quad \displaystyle \frac{\tau_{\text{E}}}{\tau_{\text{ED}}} = \displaystyle \frac{\sqrt{E}}{ P_r \left( \displaystyle \frac{N_0}{\Omega_0}\right)^2}
\label{ratio_comparison}.
\end{equation}

\noindent Two main dimensionless numbers thus appear, the Ekman number $E$ and the $P_r \left( N_0/\Omega_0\right)^2$ parameter. As first noticed by \cite{garaud2002rotationally}, the latter is of prime interest since it controls the flow dynamics. Thus, at fixed $P_r$ and $N_0$, the author shows that depending on the rotation rate value this parameters defines two rotation regimes (slow or fast). More in line with our work, \cite{garaud2008penetration1, garaud2008dynamics, garaud2009penetration2, wood2012transport, acevedo2013dynamics} have shown that it also controls the efficiency of the burrowing of the meridional circulation in radiative layers adjacent to convective regions. 
In particular, when $P_r \left( N_0/\Omega_0\right)^2 \gg 1$ this circulation is suppressed by the stable stratification, a situation similar to the one that we encountered in the viscous regime described in \cite{gouhier2020axisymmetric}. In our case, $E$ and $P_r \left( N_0/\Omega_0\right)^2$ allow us to distinguish three regimes of interest:

\begin{equation}
\begin{array}{lll}
P_r \left( \displaystyle \frac{N_0}{\Omega_0}\right)^2 \ll \sqrt{E} \ll 1 ~ \text{;} \quad \sqrt{E} \ll P_r \left( \displaystyle \frac{N_0}{\Omega_0}\right)^2 \ll 1 ~ \text{;} \quad \\\\ \hspace*{2.05cm}
\sqrt{E} \ll 1 \ll P_r \left( \displaystyle \frac{N_0}{\Omega_0}\right)^2 
.\end{array}
\label{Interest_Regime}
\end{equation}
 
\noindent As discussed in \cite{gouhier2020axisymmetric}, the two last regimes, namely the Eddington-Sweet regime $ \tau_{\text{E}} \ll \tau_{\text{ED}} \ll \tau_{\nu}$ and the viscous regime $ \tau_{\text{E}} \ll \tau_{\nu} \ll \tau_{\text{ED}}$, are the most relevant for stars. We thus focus on them for the magnetic study.

The magnetic field introduces two new timescales: the magnetic diffusion timescale 

\begin{equation}
\tau_{\eta} = \displaystyle \frac{r_0^2}{\eta}
\label{magnetic_diffusion_timescale},
\end{equation}

\noindent and the Alfvén timescale 

\begin{equation}
\tau_{\text{A}_\text{p}} = \displaystyle \frac{r_0 \sqrt{\mu_0 \rho_0}}{B_0}
\label{Alfven_timescale}.
\end{equation}

\noindent When the density contrast is taken into account (anelastic approximation), a new Alfvén timescale can be defined as

\begin{equation}
\tau_{\text{A}_\text{p}}^{A} = \left( \displaystyle \int_{r_i/r_0}^{1} \sqrt{\displaystyle \frac{\overline{\rho}}{\rho_0}} \hspace*{0.05cm} \text{d} \left( r/r_0\right) \right) \hspace*{0.05cm} \tau_{\text{A}_\text{p}}
\label{Anelastic_Alfven_timescale}.
\end{equation}

\section{Initial and boundary conditions}
\label{init_bound_cond}

\begin{figure}[!h]
\begin{center}
\includegraphics[width=4.4cm]{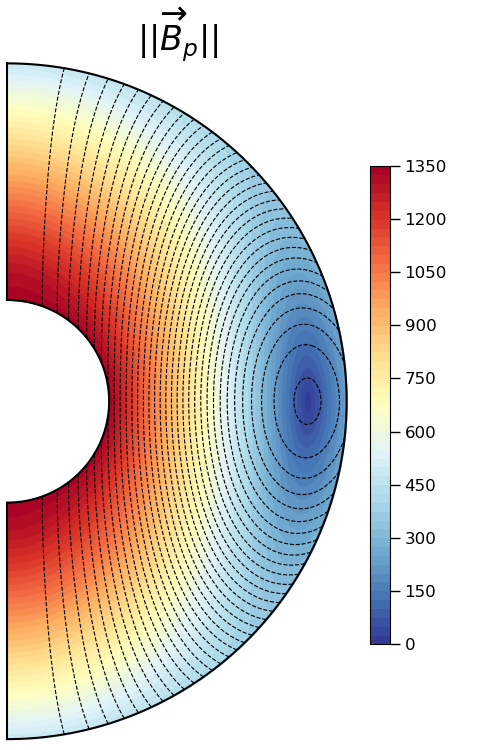}
\includegraphics[width=4.4cm]{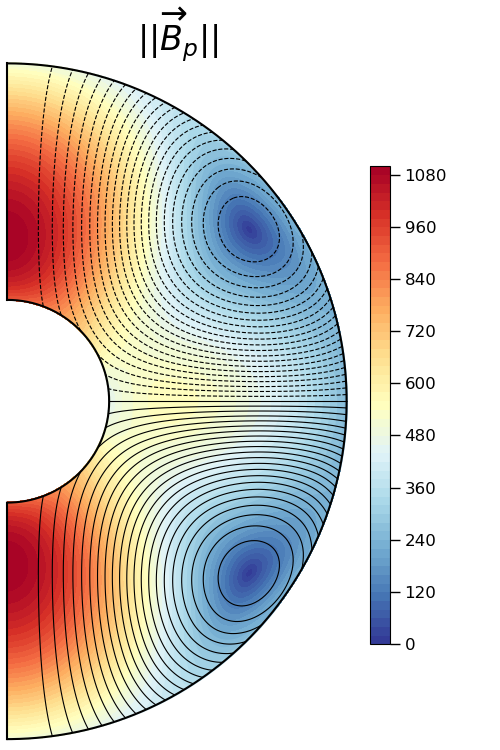}
\caption{Meridional cuts of the norm of the initial magnetic field (colour) for the dipole Eq.\eqref{dipole} (left) and the quadrupole Eq.\eqref{quadrupole} (right). The black lines show the poloidal field lines.}
\label{Initial_Pol_Mag}
\end{center}
\end{figure}

Initially, we impose a dipolar or a quadrupolar poloidal magnetic field. In both cases, the radial distribution of the magnetic field is such that the azimuthal current density does not depend on $r$, $\left( \partial j_{\phi} / \partial r =0 \right)$, to avoid possible numerical instabilities resulting from strong current sheets at the boundaries. For the dipole topology (left panel in Fig. \ref{Initial_Pol_Mag}) the initial field reads

\begin{equation}
\begin{array}{lll}
\vv{B}\left(r,\theta,t=0\right) = \displaystyle \frac{3 r B_0}{r_0 \left( 1 - \left(r_i/r_0\right)^4 \right)}  \cos{\theta} \left( 1 + \displaystyle \frac{r_i^4}{3 r^4} - \displaystyle \frac{4r_0}{3r}  \right) \vv{e}_r \\\\
- \displaystyle \frac{3 r B_0}{2 r_0 \left( 1 - \left(r_i/r_0\right)^4 \right)} \sin{\theta} \left( 3 - \displaystyle \frac{r_i^4}{3r^4} - \displaystyle \frac{8 r_0}{3r} \right) \vv{e}_{\theta}
.\end{array}
\label{dipole}
\end{equation}

\noindent For the quadrupole topology (right panel in Fig. \ref{Initial_Pol_Mag}) it takes the following form

\begin{equation}
\begin{array}{lll}
\vv{B}\left(r,\theta,t=0\right) = \displaystyle \frac{r B_0}{2 r_0 \left( 1 - \left(r_i/r_0\right)^5 \right) } \cdot \left( 2 \cos^2 \theta - \sin^2 \theta \right) \\\\ \left( \displaystyle \frac{r_i^5}{r^5} - 1 - 5 \ln \left( \displaystyle \frac{r_0}{r} \right) \right) \vv{e}_r
 - \displaystyle \frac{r B_0}{r_0 \left( 1 - \left(r_i/r_0\right)^5 \right) } \sin{\theta} \cos{\theta} \\\\ \left( 1 - \displaystyle \frac{r_i^5}{r^5} - \displaystyle \frac{15}{2} \ln \left( \displaystyle \frac{r_0}{r} \right) \right) \vv{e}_{\theta}
.\end{array}
\label{quadrupole}
\end{equation}

\noindent Figure \ref{Initial_Pol_Mag} shows that some field lines connect to the inner and outer boundaries while others are only connected to the outer boundary or even loop-back on themselves inside the domain. In both configurations, the norm of the magnetic field at the poles and at the outer sphere is $B_0$.

Insulated boundary conditions are imposed at the inner and outer spheres. For our axisymmetric setup, these conditions translate into 

\begin{equation}
\vv{B_p} = \vv{\nabla} \Phi \quad \text{and} \quad B_{\phi} = 0 \quad \text{at} \quad r = r_i, r_0
\label{insulated_bc},
\end{equation}

\noindent where $\Phi$ is a potential field.

The rotation rate is chosen to be initially uniform $\Omega(r,\theta,t=0) = \Omega_0$ in the Boussinesq approximation. In the anelastic case, the initial profile is

\begin{equation}
\Omega(r,\theta,t=0) = \overline{\rho}(r) ~ \Omega_0 \hspace*{0.05cm} \exp{\left( \displaystyle \frac{-\left(r - r_0 \right)}{\sigma}\right)}
\label{rot_profile_anel},
\end{equation}

\noindent where $\sigma$ controls the amplitude of the differential rotation. 

For all simulations we impose stress-free conditions at the inner sphere for the latitudinal and azimuthal velocity fields, and impermeability condition for the radial velocity field:

\begin{equation}
U_r = \displaystyle \frac{\partial}{\partial r} \left( \displaystyle \frac{U_{\phi}}{r} \right) = \displaystyle \frac{\partial}{\partial r} \left( \displaystyle \frac{U_{\theta}}{r} \right) = 0 \quad \text{at} \quad r = r_i
.\end{equation}

\noindent At the outer sphere we impose an impermeability condition on the radial velocity field and no-slip conditions on the latitudinal and azimuthal velocity fields:

\begin{equation}
U_r = U_{\theta} = U_{\phi} =0 
,\end{equation}

\noindent the rotation of the outer sphere being thus fixed to $\Omega_0$. The boundary layers induced by these conditions in the absence of magnetic field were analysed in \cite{gouhier2020axisymmetric}.

Finally in the Boussinesq approximation the temperature is prescribed at the inner and outer spheres and the initial  temperature field is the purely radial solution of the conduction equation. In the anelastic case, the entropy is also fixed at the boundaries. The initial stably stratified background density and temperature profiles are displayed in \cite{gouhier2020axisymmetric}.

\section{Numerical method}
\label{numerical_model}

The numerical study is carried out using the fully documented, publicly available code MagIC (\url{https://github.com/magic-sph/magic}) to solve the set of axisymmetric magneto-hydrodynamical equations in a spherical shell under the anelastic approximation \citep{gastine2012effects} (Eqs. \eqref{anel_momentum_adim}, \eqref{anel_induct_adim} and \eqref{anel_entropie_adim}) or under the Boussinesq approximation \citep{wicht2002inner} (Eqs. \eqref{bouss_momentum_adim}, \eqref{bouss_induct_adim} and \eqref{bouss_entropie_adim}). The solenoidal condition of Eqs. \eqref{anel_continuity_adim} and \eqref{bouss_continuity_adim} is ensured by a poloidal--toroidal decomposition for the mass flux and the magnetic field. Then the different fields are expanded on the basis of the spherical harmonics for the horizontal direction, and on the set of the Chebyshev polynomials for the radial direction. In particular, the Chebyshev discretisation guarantees a better resolution near the boundaries. The extent of the spherical shell can be reduced to a two-dimensional domain such as $\mathcal{D} = \left \lbrace r_i = 0.3 \leq r \leq r_0 = 1.0 \hspace*{0.03cm} ; \hspace*{0.1cm} 0 \leq \theta \leq \pi \right \rbrace$.

In the viscous regime, most of the simulations are performed with $N_r \times N_{\theta} = 127 \times 256$ while for the most resolved cases $N_r \times N_{\theta} = 193\times 512$. In the Eddington-Sweet regime, a higher resolution is needed and for most of the simulations $N_r \times N_{\theta} = 193 \times 512$ where, $N_r$ can be increased when the boundary layers need to be carefully resolved. 

\section{Space of parameters}
\label{space_parameters}

In this section, we estimate the relevant timescales in stars in order to constrain the regime of parameters of our numerical study.

\subsection{The stellar context}
\label{stellar_context}

The magnetic fields considered in this paper are large-scale fossil fields. They can be remnants of the proto-stellar phase, or the product of a core-dynamo buried in the radiative zone. At large scales, the ohmic diffusion timescale is around one Gyr \citep{braithwaite2017magnetic}, i.e. longer than the MS lifetime of the intermediate mass-stars. Dynamic processes such as the Aflvén waves propagation occur over much shorter timescales, and we will always have

\begin{equation}
\tau_{\text{A}_\text{p}} \ll \tau_{\eta}
\label{first_ordering}.
\end{equation}

\noindent Typical magnetic Prandtl numbers in stellar plasmas are $\sim 5 \times 10^{-3} - 10^{-2}$ \citep{rudiger2016instability} so that $\tau_{\eta} \ll \tau_{\nu}$. There is an exception though in the core of subgiants where the electrons are partially or fully degenerate. In that case, the Prandtl and magnetic Prandtl numbers increase because the thermal and magnetic diffusivities as well as the kinematic viscosity are dominated by the electron conduction \citep{garaud2015excitation}. The magnetic Prandtl number then ranges from $0.1$ to $10$ \citep{cantiello2011magnetic, garaud2015excitation, rudiger2015angular} and consequently $\tau_{\eta} \leq \tau_{\nu}$ or $\tau_{\nu} \leq \tau_{\eta}$.

In \cite{gouhier2020axisymmetric} we already found that contracting stars (PMS or subgiant stars) always lie in the regime

\begin{equation}
\tau_{\text{c}} \ll \tau_{\nu} , \tau_{\text{ED}}.
\label{second_ordering}
\end{equation}

\noindent In addition, we showed that the Eddington-Sweet regime is relevant for PMS stars and outside the degenerate core of subgiants. We thus have $P_r \left(N_0/\Omega_0\right)^2 \ll P_m \ll 1$ or $\tau_{\text{ED}} \ll \tau_{\eta} \ll \tau_{\nu}$ in these cases. By contrast, the degenerate cores of subgiants experience a viscous regime with higher $P_m$, such that $1 \sim P_m \ll P_r \left(N_0/\Omega_0\right)^2$ i.e. $\tau_{\nu} \sim \tau_{\eta} \ll \tau_{\text{ED}}$.
%

We can now wonder how the Alfvén time compares to the contraction time in stars. We lack precise information about the magnetic field intensities within stars but we can get some insight from the spectropolarimetric data of Herbig stars or from the asteroseismology of red giant stars combined to numerical simulations. On the one hand high-resolution spectropolarimetric surveys show that a small fraction of HAeBes hosts large-scale dipolar fields stronger than a hundred Gauss (\cite{wade2005discovery, alecian2013high}). For a typical Herbig star of $3$ \(\textup{M}_\odot\) and $3$ \(\textup{R}_\odot\) hosting a magnetic field of $300$ G, the Alfvén poloidal timescale, computed using the mean density of the star, would be of the order of a few tens of years. For these PMS stars the mass is between $2$ \(\textup{M}_\odot\) and $5$ \(\textup{M}_\odot\) and the Kelvin-Helmholtz timescale $\tau_{\text{KH}}$ typically ranges from $23$ to $1.2$ Myr \citep{maeder2008physics}. Then, assuming $\tau_{\text{c}} \approx \tau_{\text{KH}}$, implies that $\tau_{\text{A}_\text{p}} \ll \tau_{\text{c}}$.

On the other hand, the recent discovery of depressed dipole oscillation modes in red giants \citep{mosser2012characterization} has been assigned to a greenhouse effect resulting from a strong magnetic field $\sim 1$ MG trapping the gravity waves in the radiative core \citep{fuller2015asteroseismology}. Although this scenario is controversial (see e.g., \cite{mosser2017dipole}), the three-dimensional MHD simulations of convective core dynamos of \cite{brun2005simulations, augustson2016magnetic} also point towards magnetic field intensities of the order of $10^5-10^6$ G. Such intensities again lead to an Alfvén timescale much smaller than the contraction timescale. Even for a $1$ G magnetic field in a subgiant such as KIC $5955122$ which is a $1.1$ \(\textup{M}_\odot\) star of $2$ \(\textup{R}_\odot\) with a radiative interior extending to $0.74$ \(\textup{R}_\star\), we get an Alfvén timescale of $8 \cdot 10^3$ years. According to \cite{deheuvels2020seismic}, the instantaneous contraction time defined as $\Omega_{\text{core}} / \left(\text{d} \Omega_{\text{core}} / \text{d} t\right)$, where $\Omega_{\text{core}}$ is the mean rotation rate of the core, varies between $100$ Myr and $3$ Gyr in the subgiant phase. Its average value during this phase is $\sim 1$ Gyr, thus much higher than the Alfvén timescale corresponding to a $1$ G field.

\subsection{The numerical study}
\label{the_numerical_study}

We intend to perform numerical simulations in the timescale regimes thought to exist in stars. However the regime $\tau_{\text{c}} \ll \tau_{\nu}$, $\tau_{\text{ED}}$ is strongly non-linear and too challenging numerically \citep{gouhier2020axisymmetric}. The ratio $\tau_{\nu}/\tau_{\text{c}} = Re_c$ will instead vary in the range $0.1-5$ which will allow us to study the viscous regime in the linear and non-linear regimes, 
while the Eddington-Sweet regime will be studied in the linear and weakly non-linear regimes as $\tau_{\text{ED}}/\tau_{\text{c}} = P_r \left(N_0/\Omega_0\right)^2 Re_c$ varies between $10^{-3}$ and $0.5$.

The fact that $\tau_{\nu}/\tau_{\text{c}} \sim 1$ also constrains the magnetic Prandtl number. Indeed, to avoid a significant dissipation of the initial poloidal field during the simulation, the diffusion time $\tau_{\eta}$ must exceed the timescale  $\tau_{\text{c}} \sim \tau_{\nu}$ for the establishment of the stationary flow  which implies that $P_m$ has to be larger than one. Such magnetic Prandtl numbers are expected in the degenerate cores of subgiants but are not realistic in the radiative envelope of subgiants or PMS stars. As in \cite{charbonneau1993angular}, to prevent the diffusion of the poloidal field, an alternative option would have been to fix it. This is however not suited for the present problem where at large-scale, the field topology is modified by the contraction. 

Finally, simulations are run for $\tau_{\Omega} / \tau_{\text{c}} = Re_c E$, $\tau_{\Omega} / \tau_{\nu} = E$ and $\tau_{\Omega} / \tau_{\text{ED}} = E / P_r \left(N_0/\Omega_0\right)^2$ far from typical stellar values since realistic 
Ekman numbers are numerically unreachable. However, as shown in \cite{gouhier2020axisymmetric}, the flow dynamics do not critically depend on these ratios and we thus expect the model associated to our numerical simulations to remain valid for stars. Indeed, the first important parameter is $\tau_{\text{ED}} /\tau_{\nu} = P_r \left(N_0/\Omega_0\right)^2$ which determines if we are in the Eddington-Sweet or viscous regime. Realistic values can be used for this parameter. Another important ratio is $\tau_{\nu} / \tau_{\text{c}} = Re_c$ which governs the level of differential rotation. Realistic values for this last parameter are more difficult to reach numerically but we come back on the implications of this discrepancy in Sect. \ref{conclusion}.
%

To conclude, all the simulations performed in the viscous regime fulfil the following conditions:

\begin{equation}
\tau_{\Omega} \ll \tau_{\text{A}_\text{p}} \ll \tau_{\text{c}} \leq \tau_{\nu} \ll \tau_{\eta} \ll \tau_{\text{ED}}
\label{timescales_visc},
\end{equation}

\noindent or in terms of dimensionless numbers

\begin{equation}
E \ll \left( \displaystyle \frac{L_u}{P_m} \right)^{-1} \ll Re_c^{-1} \leq 1 \ll P_m \ll P_r \left( \displaystyle \frac{N_0}{\Omega_0} \right)^2
\label{dimensionless_numbers_visc}.
\end{equation}

\noindent For the Eddington-Sweet regime, we shall have

\begin{equation}
\tau_{\Omega} \ll \tau_{\text{A}_\text{p}} \ll \tau_{\text{ED}} < \tau_{\text{c}} \ll \tau_{\nu} \ll \tau_{\eta}
\label{timescales_edd},
\end{equation}

\noindent or equivalently

\begin{equation}
E \ll \left( \displaystyle \frac{L_u}{P_m} \right)^{-1} \ll P_r \left( \displaystyle \frac{N_0}{\Omega_0} \right)^2 <  Re_c^{-1}  \ll 1 \ll P_m
\label{dimensionless_numbers_edd}.
\end{equation}

\section{Numerical results}
\label{numerical_results}

\begin{table*}[!h]
\begin{center}
\begin{tabular}{c|c|c|c|c|c|c|c}
\hline
\hline
 \multirow{3}{*}{\textbf{Case}}& \multirow{3}{*}{\textbf{Topology}} & \multirow{3}{*}{\textbf{Contraction in induction equation}} &\multirow{3}{*}{$\boldsymbol{E}$} & \multirow{3}{*}{$\boldsymbol{P_r \left(\displaystyle \frac{N_0}{\Omega_0}\right)^2} $}  &  \multirow{3}{*}{$\boldsymbol{Re_c} $} & \multirow{3}{*}{$ \boldsymbol{L_u} $} & \multirow{3}{*}{$ \boldsymbol{\displaystyle \frac{\rho_i}{\rho_0} } $} \\ & & & & & & & \\ & & & & & & & \\
\hline
\hline
D$1$ & Dipole & No & $10^{-4}$ & $10^4$ & $10^{-1}$ & $10^4$ & $1$\\
D$2$ & Dipole & No & $10^{-4}$ & $10^4$ & $10^{-1}$ & $5\cdot10^4$ & $1$\\
D$3$ & Dipole & No & $10^{-4}$ & $10^4$ & $1$ & $5\cdot10^3$ & $1$\\
D$4$ & Dipole & No & $10^{-4}$ & $10^4$ & $1$ & $10^4$ & $1$\\
D$5$ & Dipole & No & $10^{-4}$ & $10^4$ & $1$ & $5\cdot10^4$ & $1$\\
D$6$ & Dipole & No & $10^{-4}$ & $10^4$ & $5$ & $10^4$ & $1$\\
D$7$ & Dipole & No & $10^{-4}$ & $10^4$ & $5$ & $5\cdot10^4$ &$1$\\
D$8$ & Dipole & Yes & $10^{-4}$ & $10^4$ & $10^{-1}$ & $10^4$ &$1$\\
D$9$ & Dipole & Yes & $10^{-4}$ & $10^4$ & $5\cdot 10^{-1}$ & $10^3$ &$1$\\
D$10$ & Dipole & Yes & $10^{-4}$ & $10^4$ & $5\cdot10^{-1}$ & $5\cdot10^3$ &$1$\\
D$11$ & Dipole & Yes & $10^{-4}$ & $10^4$ & $5\cdot10^{-1}$ & $10^4$ &$1$\\
D$12$ & Dipole & Yes & $10^{-4}$ & $10^4$ & $1$ & $5\cdot10^3$ &$1$\\
D$13$ & Dipole & Yes & $10^{-4}$ & $10^4$ & $1$ & $10^4$ &$1$\\
D$14$ & Dipole & Yes & $10^{-4}$ & $10^4$ & $2$ & $10^4$ &$1$\\
D$15$ & Dipole & Yes & $10^{-4}$ & $10^4$ & $1$ & $5\cdot10^3$ &$20.85$\\
D$16$ & Dipole & Yes & $10^{-4}$ & $10^4$ & $1$ & $10^4$ &$20.85$\\
D$17$ & Dipole & Yes & $10^{-4}$ & $10^4$ & $1$ & $5\cdot10^4$ &$20.85$\\
D$18$ & Dipole & Yes & $10^{-4}$ & $10^4$ & $1$ & $10^5$ &$20.85$\\
D$19$ & Dipole & Yes & $10^{-4}$ & $10^4$ & $5$ & $10^4$ &$20.85$\\
D$20$ & Dipole & Yes & $10^{-4}$ & $10^4$ & $5$ & $5\cdot10^4$ &$20.85$\\
D$21$ & Dipole & Yes & $10^{-5}$ & $10^{-1}$ & $10^{-1}$ & $5\cdot10^4$ &$20.85$\\
D$22$ & Dipole & Yes & $10^{-5}$ & $10^{-1}$ & $10^{-1}$ & $10^5$ &$20.85$\\
D$23$ & Dipole & Yes & $10^{-5}$ & $10^{-1}$ & $5\cdot10^{-1}$ & $5\cdot10^4$ &$20.85$\\
D$24$ & Dipole & Yes & $10^{-5}$ & $10^{-1}$ & $5\cdot10^{-1}$ & $10^5$  &$20.85$\\
D$25$ & Dipole & Yes & $10^{-5}$ & $10^{-1}$ & $1$ & $5\cdot10^3$  &$20.85$\\
D$26$ & Dipole & Yes & $10^{-5}$ & $10^{-1}$ & $1$ & $10^4$  &$20.85$\\
D$27$ & Dipole & Yes & $10^{-5}$ & $10^{-1}$ & $1$ & $5\cdot10^4$ &$20.85$\\
D$28$ & Dipole & Yes & $10^{-5}$ & $10^{-1}$ & $1$ & $10^5$ &$20.85$\\
D$29$ & Dipole & Yes & $10^{-5}$ & $10^{-1}$ & $2$ & $5\cdot10^4$  &$20.85$\\
D$30$ & Dipole & Yes & $10^{-5}$ & $10^{-1}$ & $2$ & $10^5$  &$20.85$\\
D$31$ & Dipole & Yes & $10^{-5}$ & $10^{-1}$ & $5$ & $10^5$ &$20.85$\\
D$32$ & Dipole & Yes & $10^{-5}$ & $10^{-2}$ & $10^{-1}$ & $5\cdot10^4$ &$20.85$\\
D$33$ & Dipole & Yes & $10^{-5}$ & $10^{-2}$ & $10^{-1}$ & $10^5$  &$20.85$\\
D$34$ & Dipole & Yes & $10^{-5}$ & $10^{-2}$ & $5\cdot10^{-1}$ & $5\cdot10^4$ &$20.85$\\
D$35$ & Dipole & Yes & $10^{-5}$ & $10^{-2}$ & $5\cdot10^{-1}$ & $10^5$ &$20.85$\\
D$36$ & Dipole & Yes & $10^{-5}$ & $10^{-2}$ & $1$ & $5\cdot10^4$ &$20.85$\\
D$37$ & Dipole & Yes & $10^{-5}$ & $10^{-2}$ & $1$ & $10^5$ &$20.85$\\
D$38$ & Dipole & Yes & $10^{-5}$ & $10^{-2}$ & $2$ & $5\cdot10^4$ &$20.85$\\
D$39$ & Dipole & Yes & $10^{-5}$ & $10^{-2}$ & $2$ & $10^5$ &$20.85$\\
\hline
\hline
Q$1$ & Quadrupole & No & $10^{-4}$ & $10^4$ & $1$ & $10^4$  &$1$\\
Q$2$ & Quadrupole & No & $10^{-4}$ & $10^4$ & $1$ & $5\cdot10^4$  &$1$\\
Q$3$ & Quadrupole & No & $10^{-4}$ & $10^4$ & $5$ & $5\cdot10^3$  &$1$\\
Q$4$ & Quadrupole & No & $10^{-4}$ & $10^4$ & $5$ & $10^4$  &$1$\\
Q$5$ & Quadrupole & No & $10^{-4}$ & $10^4$ & $5$ & $5\cdot 10^4$  &$1$\\
Q$6$ & Quadrupole & Yes & $10^{-4}$ & $10^4$ & $5 \cdot 10^{-1}$ & $10^3$  &$1$\\
Q$7$ & Quadrupole & Yes & $10^{-4}$ & $10^4$ & $5 \cdot 10^{-1}$ & $5\cdot10^3$  &$1$\\
Q$8$ & Quadrupole & Yes & $10^{-4}$ & $10^4$ & $5 \cdot 10^{-1}$ & $10^4$  &$1$\\
Q$9$ & Quadrupole & Yes & $10^{-4}$ & $10^4$ & $1$ & $5\cdot 10^3$  &$1$\\
Q$10$ & Quadrupole & Yes & $10^{-4}$ & $10^4$ & $1$ & $10^4$ &$1$\\
Q$11$ & Quadrupole & Yes & $10^{-4}$ & $10^4$ & $2$ & $10^4$ &$1$\\
Q$12$ & Quadrupole & Yes & $10^{-5}$ & $10^{-1}$ & $1$ & $5\cdot10^4$  &$20.85$\\
Q$13$ & Quadrupole & Yes & $10^{-5}$ & $10^{-1}$ & $1$ & $10^5$  &$20.85$\\
Q$14$ & Quadrupole & Yes & $10^{-5}$ & $10^{-1}$ & $2$ & $5\cdot10^4$  &$20.85$\\
Q$15$ & Quadrupole & Yes & $10^{-5}$ & $10^{-1}$ & $2$ & $10^5$  &$20.85$\\
Q$16$ & Quadrupole & Yes & $10^{-5}$ & $10^{-1}$ & $3$ & $5\cdot10^4$  &$20.85$\\
Q$17$ & Quadrupole & Yes & $10^{-5}$ & $10^{-1}$ & $5$ & $5\cdot10^4$ &$20.85$\\
\hline
\end{tabular}
\vspace*{0.2cm}
\caption{Parameters of the simulations performed at $P_m = 10^2$ in the viscous and Eddington-Sweet regimes.}
\label{parameters_viscous}
\end{center}
\end{table*}

We are mostly interested in the steady state differential rotation produced by the contracting flow in the stably stratified magnetised layer. We shall investigate separately the viscous regime $\sqrt{E} \ll 1 \ll P_r \left(N_0/\Omega_0\right)^2$ and the Eddington-Sweet regime $\sqrt{E} \ll P_r \left(N_0/\Omega_0\right)^2 \ll 1$. For the initial magnetic field we consider either a dipole or a quadrupole and we include or not the effect of the density stratification. For each configurations we vary the Lundquist and the contraction Reynolds numbers to study the effect of the amplitudes of the initial poloidal field and of the contraction. To help understand physically our numerical results we also performed simulations where the contraction term is artificially removed from the induction equation, thus preventing the advection of the magnetic field by the contraction velocity field. All the relevant simulations and their associated parameters are listed in Table \ref{parameters_viscous}. 
 
\subsection{Unsteady evolution}
\label{unsteady_evolution}

Before we focus on the steady states, we start with a brief description of the unsteady phase. 

\begin{figure*}[!h]
\begin{center}
\includegraphics[width=3.4cm]{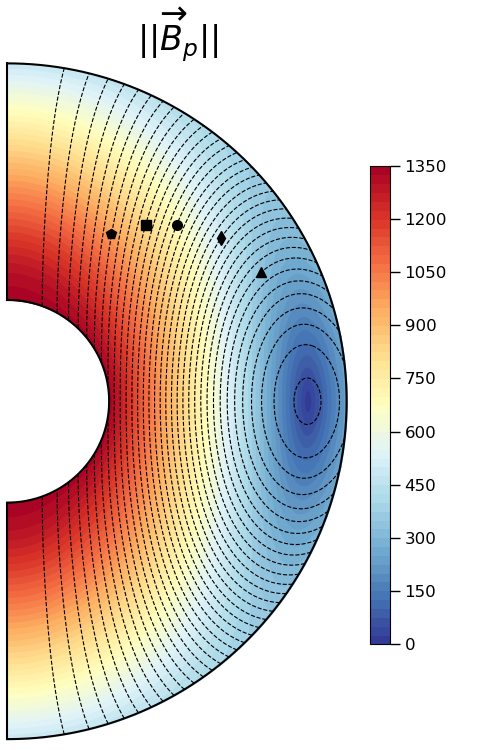}
\includegraphics[width=7cm]{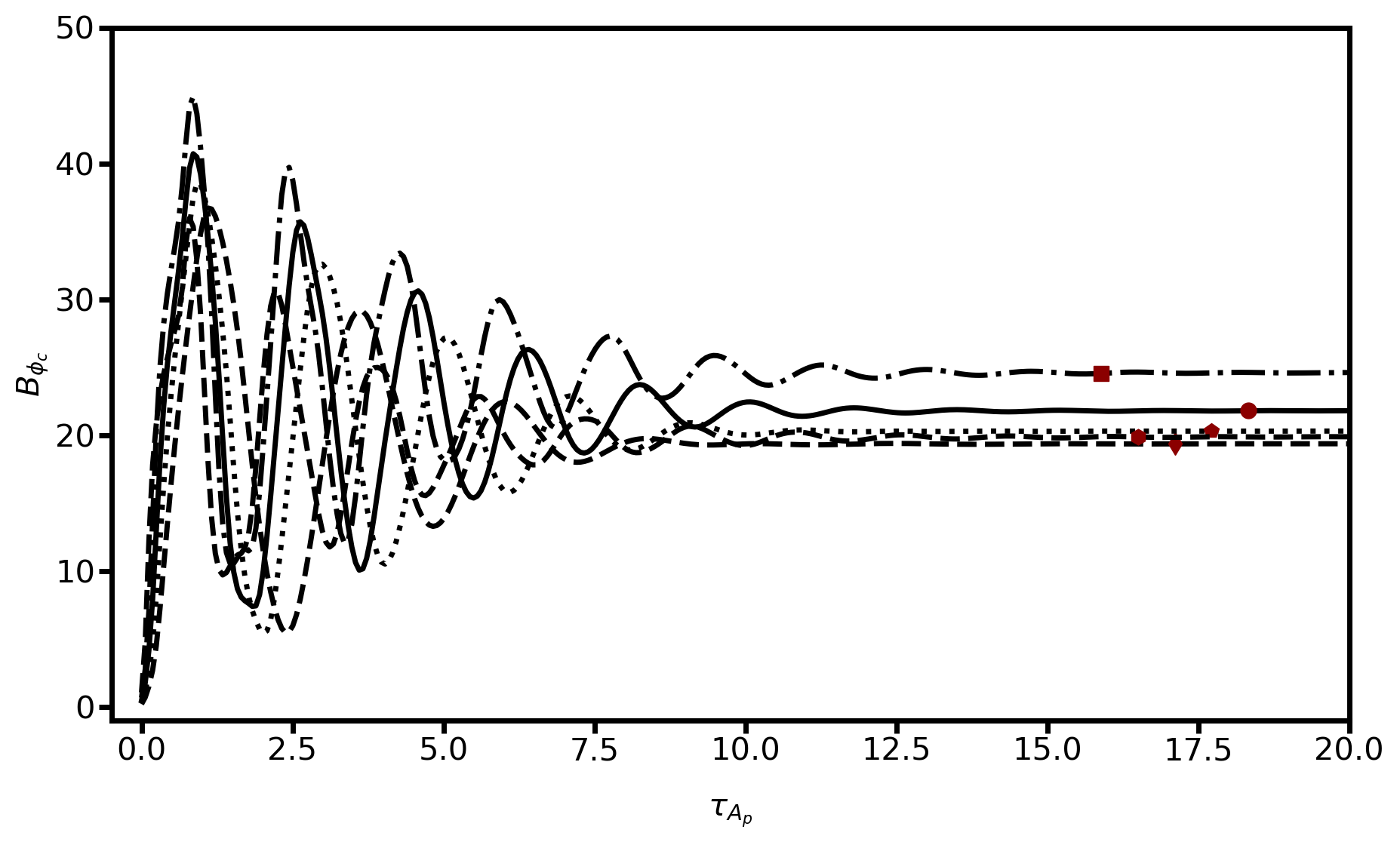}
\includegraphics[width=7cm]{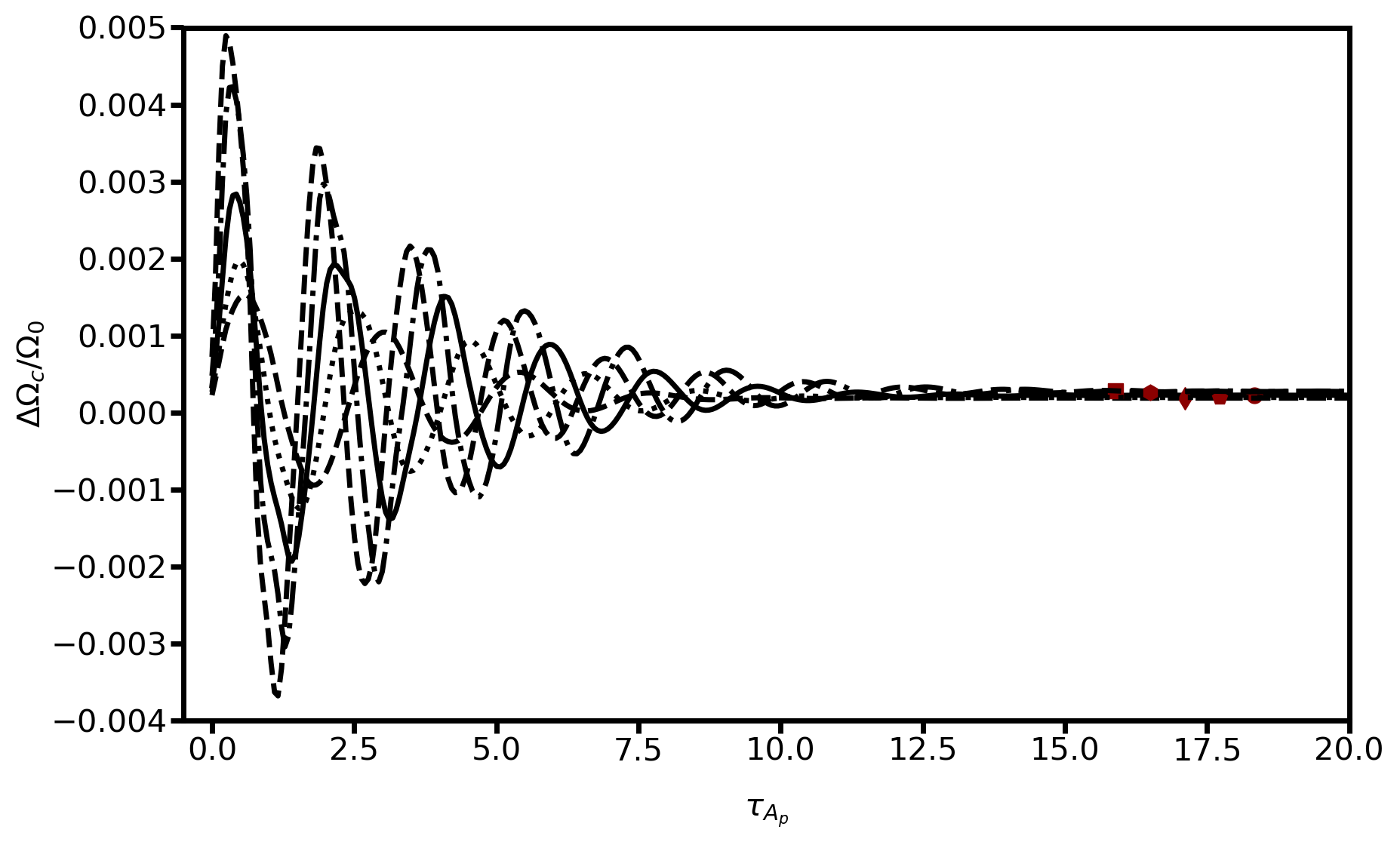}
\caption{Left panel: meridional cut of the norm of the poloidal magnetic field. The black dots show the position of $5$ control points located on different field lines. In the other two panels the temporal evolution of these points is followed during $20 \hspace*{0.05cm} \tau_{\text{A}_\text{p}}$, both for the toroidal field $B_{{\phi}_c}$ (middle panel) and the normalised differential rotation $\Delta \Omega_c / \Omega_0$ (right panel). The parameters are $E = 10^{-4}$, $P_r \left(N_0/\Omega_0\right)^2 = 10^4$, $Re_c = 1$, $L_u = 5\cdot10^4$ and $P_m =10^2$ (run D$5$ of Table \ref{parameters_viscous}).}
\label{phase_mixing}
\end{center}
\end{figure*}

Figure \ref{phase_mixing} illustrates for a particular run the typical evolution of the toroidal field $B_{{\phi}_c}$ (middle panel) and of the normalised differential rotation $\delta \Omega_c = \left( \Omega_c - \Omega_0\right) / \Omega_0$ (right panel) at some fixed locations in the domain (black points in the left panel). The run corresponds to a dipolar field in a viscous regime without advection of the field lines (run D$5$ in Table \ref{parameters_viscous}). We observe that as soon as the contraction produces a differential rotation, the initially zero toroidal field linearly grows by the $\Omega$-effect. After $\sim 1 \hspace*{0.05cm} \tau_{\text{A}_{\text{p}}}$, it saturates because the amplification of the toroidal field produces a Lorentz force that back-reacts on the differential rotation thus counteracting further $\Omega$-effect. This leads to the propagation of Alfvén waves along the poloidal field lines illustrated by the oscillations of the control points in Fig. \ref{phase_mixing}. As these waves oscillate independently from each other, they quickly get out of phase. This builds gradients of the toroidal magnetic field and of the differential rotation on sufficiently small scales so that they can be efficiently dissipated by the diffusion processes: this is the so-called phase mixing mechanism (see \cite{heyvaerts1983coronal} or \cite{cally1991phase}). In the absence of a process forcing the differential rotation this would lead to a uniformly rotating steady state. 

\subsection{Steady state in the viscous regime with a dipolar field}
\label{viscous_dipole}

In this section we investigate, in the viscous regime and for an initial dipolar field, the steady states obtained after the Alfvén waves have dissipated and the contraction has imposed some level of differential rotation. These states are actually quasi-steady because the initial poloidal field continues to slowly decrease through magnetic diffusion. Three parameters are fixed $P_r \left(N_0/\Omega_0\right)^2 = 10^{4}$, $E = 10^{-4}$ and $P_m =10^2$, while the contraction Reynolds number $Re_c$ varies between $10^{-1}$ to $5$ and Lundquist number between $10^3$ and $5\cdot10^4$ (see details in Table \ref{parameters_viscous}). The anelastic simulations have been performed with a density contrast $\rho_i / \rho_0 = 20.85$.

\begin{figure*}[!h]
\begin{center}
\includegraphics[width=4.4cm]{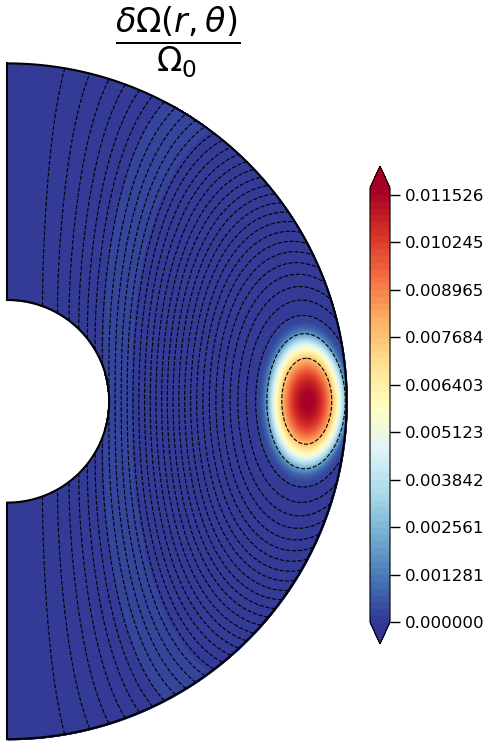}
\includegraphics[width=4.4cm]{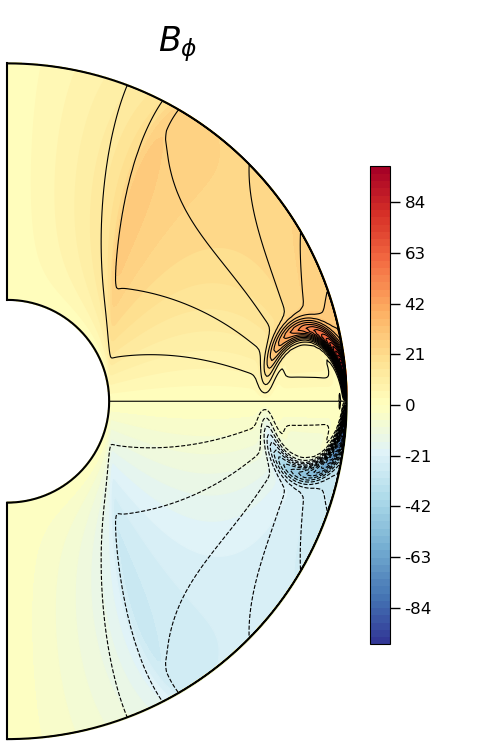}
\includegraphics[width=4.4cm]{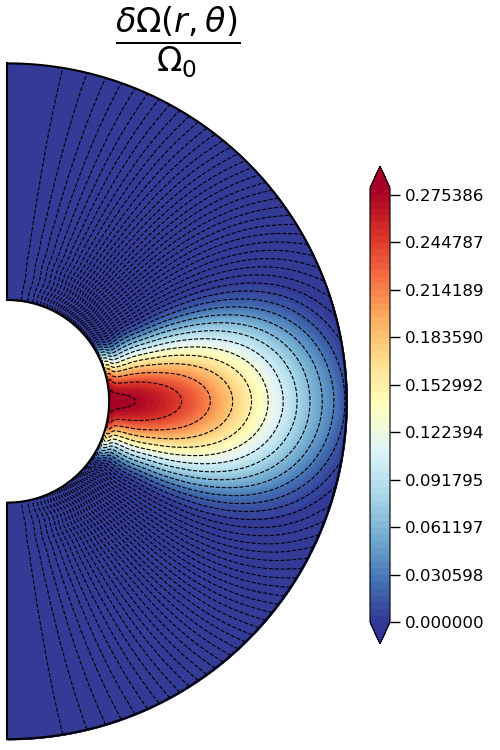}
\includegraphics[width=4.4cm]{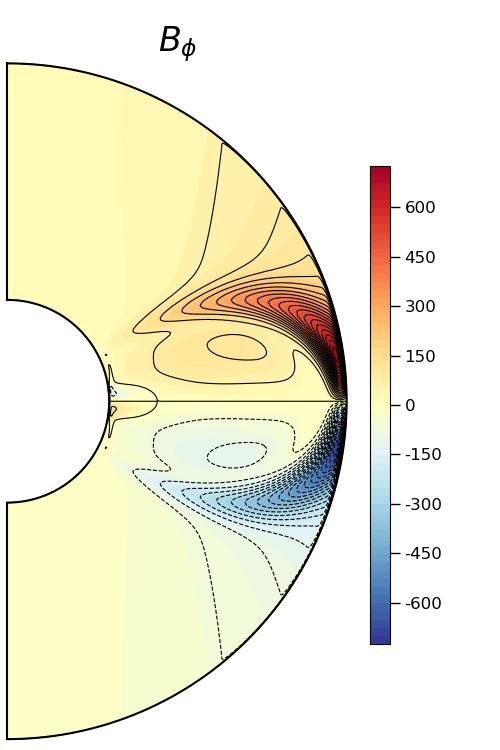}
\caption{Meridional cuts of the rotation rate normalised to the value at the outer sphere (first and third panels) and toroidal field (second and fourth panels), in the quasi-steady state. In black lines are also represented the poloidal field lines (first and third panels) and the streamlines associated with the electrical-current function defined in Appendix \ref{elec_current_fonc} (second and fourth panels). The dotted (solid) lines then correspond to an anticlockwise (clockwise) current circulation. In the first two panels the contraction does not advect the poloidal field lines and the quasi-steady state is achieved after $\sim 0.04 \hspace*{0.05cm} \tau_{\text{c}}$ (i.e. $\sim 20 \hspace*{0.05cm} \tau_{\text{A}_\text{p}}$, see Fig. \ref{phase_mixing}). In the last two panels such an advection is allowed and the quasi-steady configuration is reached after $\sim 1 \hspace*{0.05cm} \tau_{\text{c}}$ (i.e. after $\sim 100 \hspace*{0.05cm} \tau_{\text{A}_\text{p}}$ for this simulation). For these two cases the Lundquist numbers are respectively $L_u = 5 \cdot 10^4$ and  $L_u = 10^4$ (runs D$5$ and D$13$). The other parameters are identical, namely $P_r \left(N_0/\Omega_0\right)=10^4$, $E = 10^{-4}$, $Re_c = 1$ and $P_m = 10^2$.}
\label{Numerical_Results_Viscous}
\end{center}
\end{figure*}

Representative results are displayed in Fig. \ref{Numerical_Results_Viscous}, for a case without advection of the field lines at $Re_c = 1$ and $L_u = 5 \cdot 10^4$ (first two panels, run D$5$ of Table \ref{parameters_viscous}), and for a case with advection at $Re_c = 1$ and $L_u = 10^4$ (last two panels, run D$13$ of Table \ref{parameters_viscous}). From this figure we clearly distinguish two magnetically decoupled regions with different levels of differential rotation. In the first region the poloidal field lines are connected to the outer sphere and the flow is in quasi-solid rotation. In the second region, either the poloidal field lines loop-back on themselves inside the spherical shell (case without advection of the field lines) or they loop-back on the inner sphere (case  with advection of the field lines). As in \cite{charbonneau1993angular}  
we shall refer to these particular regions as the "dead zone" (DZ). The level of differential rotation is always significant in the DZs. With contraction of the field lines, the maximum differential rotation $\text{max} \left(\delta \Omega(r,\theta) / \Omega_0 \right)= \text{max}\left(\left(\Omega(r,\theta) - \Omega_0\right) / \Omega_0\right)$ is $30 \%$, while it is reduced to $\sim 1 \%$ in the other case. The snapshots of the toroidal magnetic field (second and fourth panels in Fig. \ref{Numerical_Results_Viscous}), show the presence of two anti-symmetric lobes of toroidal field in both hemispheres due to the $\Omega$-effect acting on the dipolar poloidal field. Interestingly the DZs are surrounded by a region of strong toroidal field. This corresponds to a Shercliff boundary layer \citep{shercliff1956flow} that develops between two magnetic regions that are forced to rotate at different rates. Indeed in our simulations, the first poloidal line delimiting the DZ and the neighbouring lines connected to the outer sphere are forced to rotate differentially and a layer involving strong toroidal fields accommodates this jump in rotation rate. We verified that, as expected \citep{roberts1967singularities}, the thickness of this Shercliff boundary layer scales with the inverse of the square root of the Hartmann number defined by $H_a = L_u / \sqrt{P_m}$. When we prevent the advection of the field lines (first panel in Fig. \ref{Numerical_Results_Viscous}), a second Shercliff layer is visible at the separation between the region where the field lines are connected to both the outer and inner spheres, and the region where they are connected to the outer sphere only.

\begin{figure*}[!h]
\begin{center}
\includegraphics[width=3.2cm]{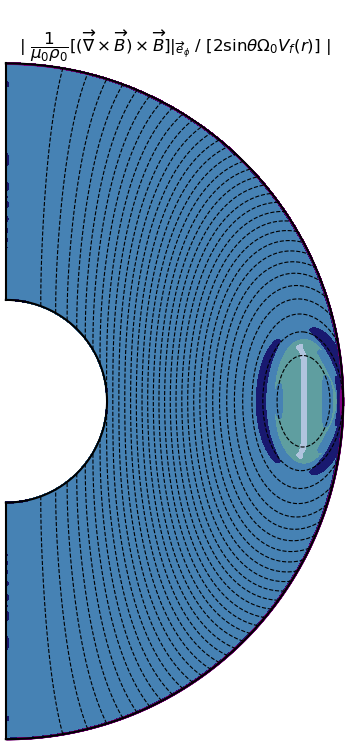}
\hspace*{0.7cm}
\includegraphics[width=3.2cm]{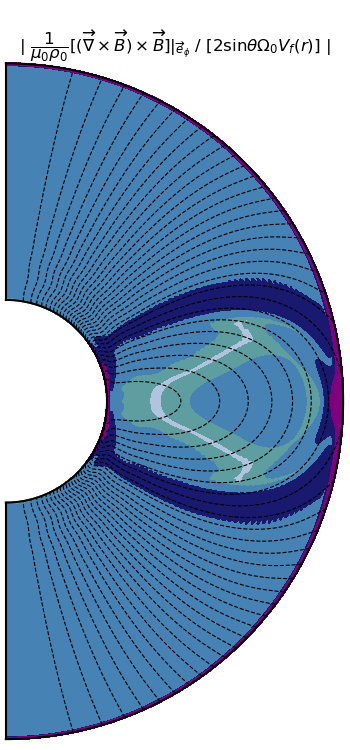}
\hspace*{0.7cm}
\includegraphics[width=3.2cm]{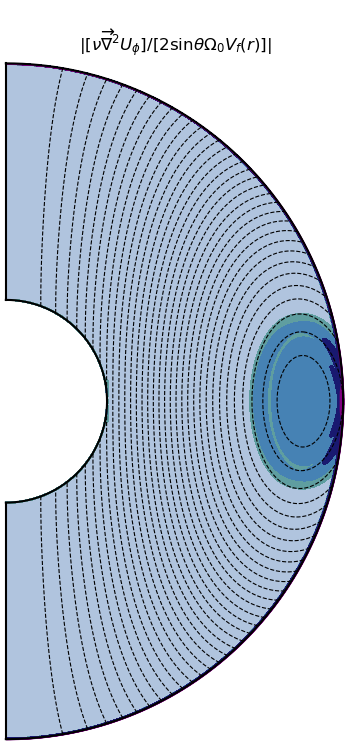}
\hspace*{0.7cm}
\includegraphics[width=4.6cm]{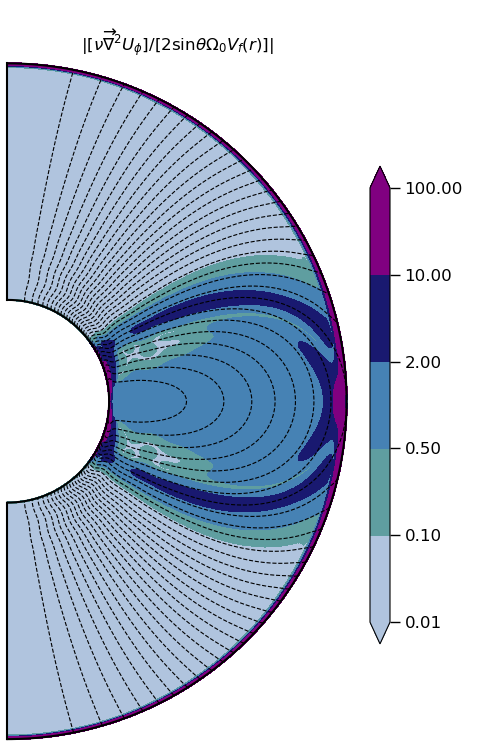}
\caption{$2$D maps comparing the relative importance of different azimuthally projected quantities: the Lorentz force with the contraction (first two panels) and the viscous term with the contraction (last two panels). The runs where the contraction term has been removed from the induction equation are presented in the first and third panels. In the other two, the effect of the contraction on the magnetic field is taken into account. Parameters are the same to those of Fig. \ref{Numerical_Results_Viscous}.}
\label{2D_map_lo_visc_cont}
\end{center}
\end{figure*}

To give a more detailed description of the dynamics outside and inside the DZ, we compare in Fig. \ref{2D_map_lo_visc_cont} the relative amplitudes of the different terms of the AM balance Eq. \eqref{angular_momentum_evolution}. From the first two panels of this figure, we see that outside the DZ, the quasi-steady configuration is characterised by a balance between the contraction term and the Lorentz force, that is:

\begin{equation}
-2 \sin{\theta} \hspace*{0.05cm} \Omega_0  \displaystyle \frac{V_0 r_0^2}{r^2} = \displaystyle \frac{1}{\mu_0 \rho_0} \left \lbrack \displaystyle \frac{B_{\theta}}{r \sin{\theta}} \displaystyle \frac{\partial}{\partial \theta} \left( \sin{\theta} B_{\phi} \right) + \displaystyle \frac{B_r}{r} \displaystyle \frac{\partial}{\partial r} \left( r B_{\phi} \right) \right \rbrack
\label{static_balance_eq}.
\end{equation}

\noindent On the contrary, the last two panels of Fig. \ref{2D_map_lo_visc_cont} show that inside the DZ the viscous term balances the contraction term, thus leading to

\begin{equation}
\begin{array}{lll}
\nu \left \lbrack r \displaystyle \frac{\partial^2}{\partial r^2} \left( \displaystyle \frac{\delta \Omega}{\Omega_0} \right) + 4 \displaystyle \frac{\partial }{\partial r} \left( \displaystyle \frac{\delta \Omega}{\Omega_0} \right) + \displaystyle \frac{1}{r} \displaystyle \frac{\partial^2 
}{\partial \theta^2} \left( \displaystyle \frac{\delta \Omega}{\Omega_0} \right) + \displaystyle \frac{3 \cot{\theta}}{r} \displaystyle \frac{\partial}{\partial \theta} \left( \displaystyle \frac{\delta \Omega}{\Omega_0} \right) \right \rbrack \\\\ = \displaystyle \frac{-2 V_0 r_0^2}{r^2}
,\end{array}
\label{differential_equation}
\end{equation}

\noindent where only the linear part of the contraction term has been retained, an approximation only valid if $\delta \Omega /\Omega_0 \ll 1$. In the following two sub-sections the differential rotation resulting from these two different balances is analysed.

\subsubsection{Region outside the dead zone}
\label{outside_dead_zone}

Although at first glance the flow seems to be in solid rotation outside the DZ, there is a rotation rate jump across a boundary layer at the outer sphere as well as a residual differential rotation along the poloidal field lines.

\begin{figure}[!h]
\begin{center}
    \centering
    \begin{subfigure}[b]{0.235\textwidth}
        \includegraphics[width=\textwidth]{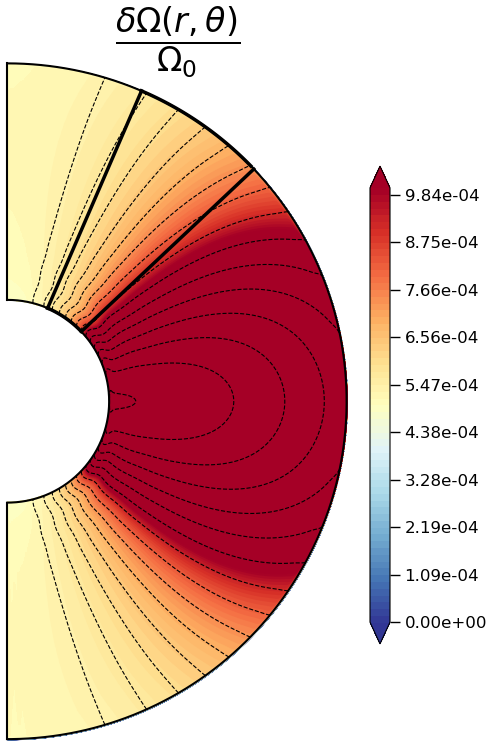}
        \caption{}
        \label{fig1}
    \end{subfigure}
    \hspace*{0.1cm}
    \hspace*{0.4cm}
    \begin{subfigure}[b]{0.435\textwidth}
        \includegraphics[width=\textwidth,height=5.4cm]{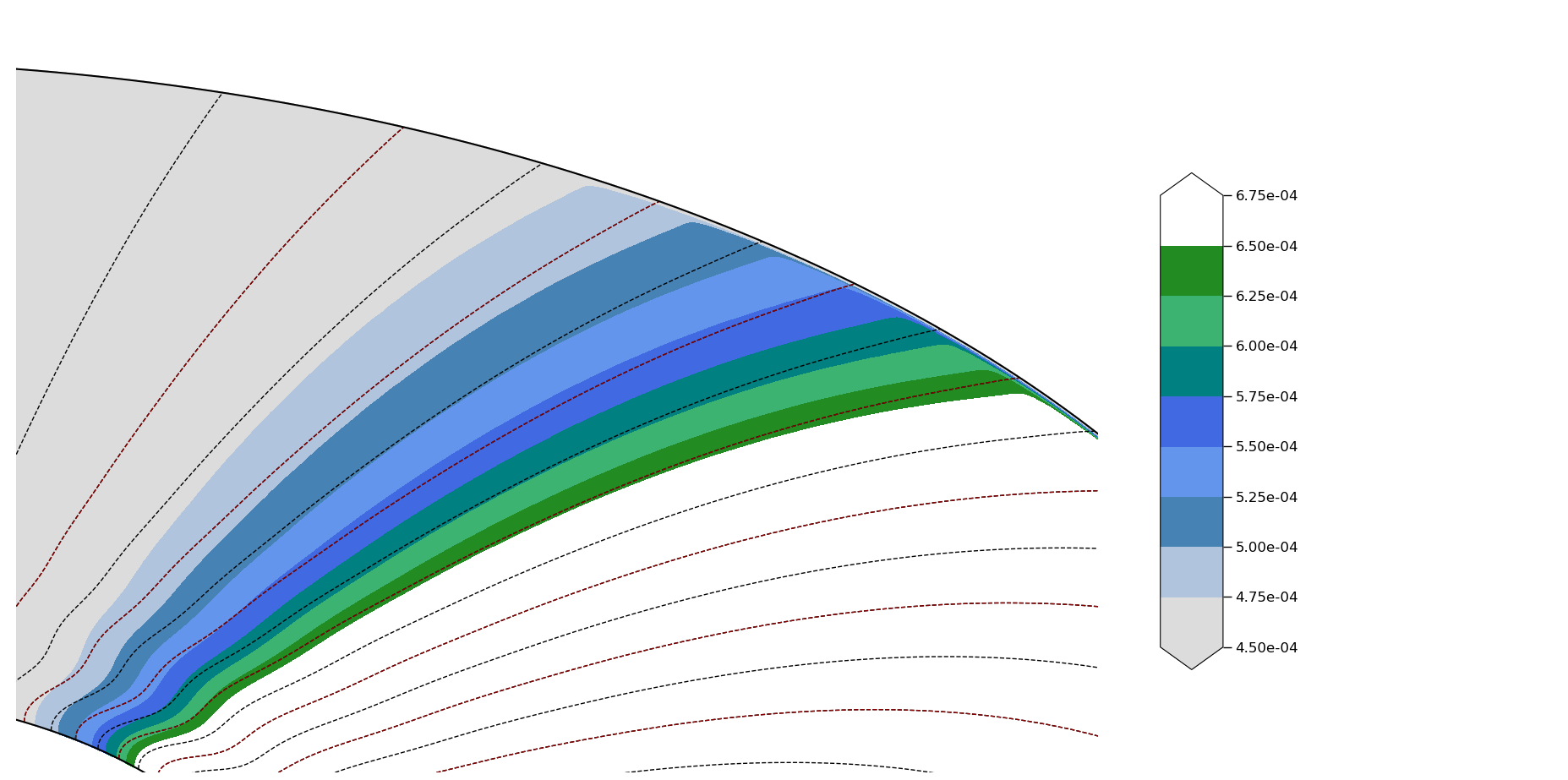}
        \caption{}
        \label{fig2}
    \end{subfigure}
\caption{Top panel (\ref{fig1}): meridional cut of the normalised rotation rate. This is the third panel of Fig. \ref{Numerical_Results_Viscous} presented on a smaller rotation rate scale. As a result, the DZ is saturated in colour. Bottom panel (\ref{fig2}): enlargement of the black-delimited zone displayed in the top panel. In each panels the poloidal field lines are also represented (black lines). For the sake of clarity, every other field line is plotted in red in the bottom panel. Parameters are the same as in Figs. \ref{Numerical_Results_Viscous} and \ref{2D_map_lo_visc_cont}.}
\label{forced_cuts}
\end{center}
\end{figure}

By rescaling the colour range of Fig. \ref{Numerical_Results_Viscous}, the top panel of Fig. \ref{forced_cuts} indeed shows that outside the DZ, the differential rotation between the interior flow and the outer sphere $\left(\Omega(r,\theta) -\Omega_0)/ \Omega_0 \right)$ is $\sim 6\cdot 10^{-4}$. In order to understand that value we first estimate the toroidal field amplitude outside the DZ using Eq. \eqref{static_balance_eq}. Assuming $B_r \sim B_{\theta} \sim B_0$ and $r \sim r_0$, we get

\begin{equation}
\displaystyle \frac{B_{\phi}}{B_0} \approx \displaystyle \frac{r_0 \mu_0 \rho_0 V_0 \Omega_0}{B_0^2} = \left( \displaystyle \frac{P_m}{L_u} \right)^2 \displaystyle \frac{Re_c}{E}
\label{estimate_bphi}.
\end{equation}

\begin{figure}[!h]
\begin{center}
\hspace*{-0.5cm}
\includegraphics[width=8.5cm]{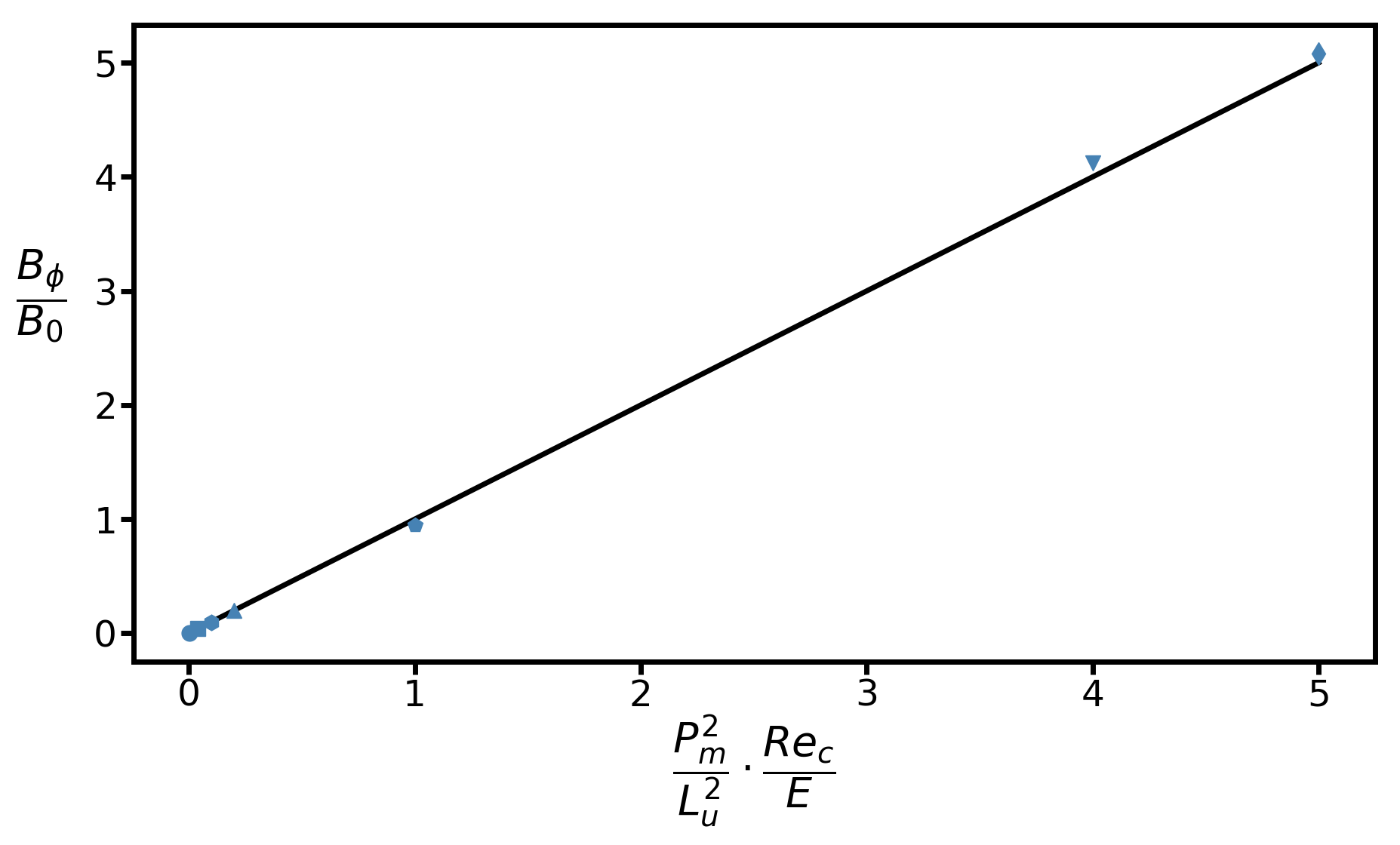}
\caption{Toroidal field $B_{\phi}$ normalised to the initial amplitude of the poloidal field $B_0$ as a function of $\left(P_m/L_u \right)^2 Re_c/E$ at the particular location $\theta = \pi/6$ and $r=0.65 \hspace*{0.05cm} r_0$. The different symbols correspond to the different runs D$1$ to D$7$ of Table \ref{parameters_viscous} (no contraction term in induction equation) namely, $Re_c = 10^{-1}$, $L_u = 10^4$ (circle); $Re_c = 10^{-1}$, $L_u = 5\cdot 10^4$ (square); $Re_c = 1$, $L_u = 5\cdot 10^3$ (hexagon); $Re_c = 1$, $L_u = 10^4$ (up triangle); $Re_c = 1$, $L_u = 5\cdot10^4$ (pentagon); $Re_c = 5$, $L_u = 10^4$ (down triangle) and $Re_c = 5$, $L_u = 5\cdot 10^4$ (diamond). The other parameters are fixed to $E=10^{-4}$, $P_r \left(N_0/\Omega_0\right)^2 = 10^4$ and $P_m = 10^2$.}
\label{scaling_law_bphi}
\end{center}
\end{figure}

\noindent This estimate is confirmed in Fig. \ref{scaling_law_bphi} where $B_{\phi}/B_0$ determined at a particular location, $\theta = \pi/6$, $r=0.65 \hspace*{0.05cm} r_0$, is compared with the right-hand side of Eq. \eqref{estimate_bphi} for the runs D$1$ to D$7$. 
 
\begin{figure}[!h]
\begin{center}
\hspace*{-0.5cm}
\includegraphics[width=8.5cm]{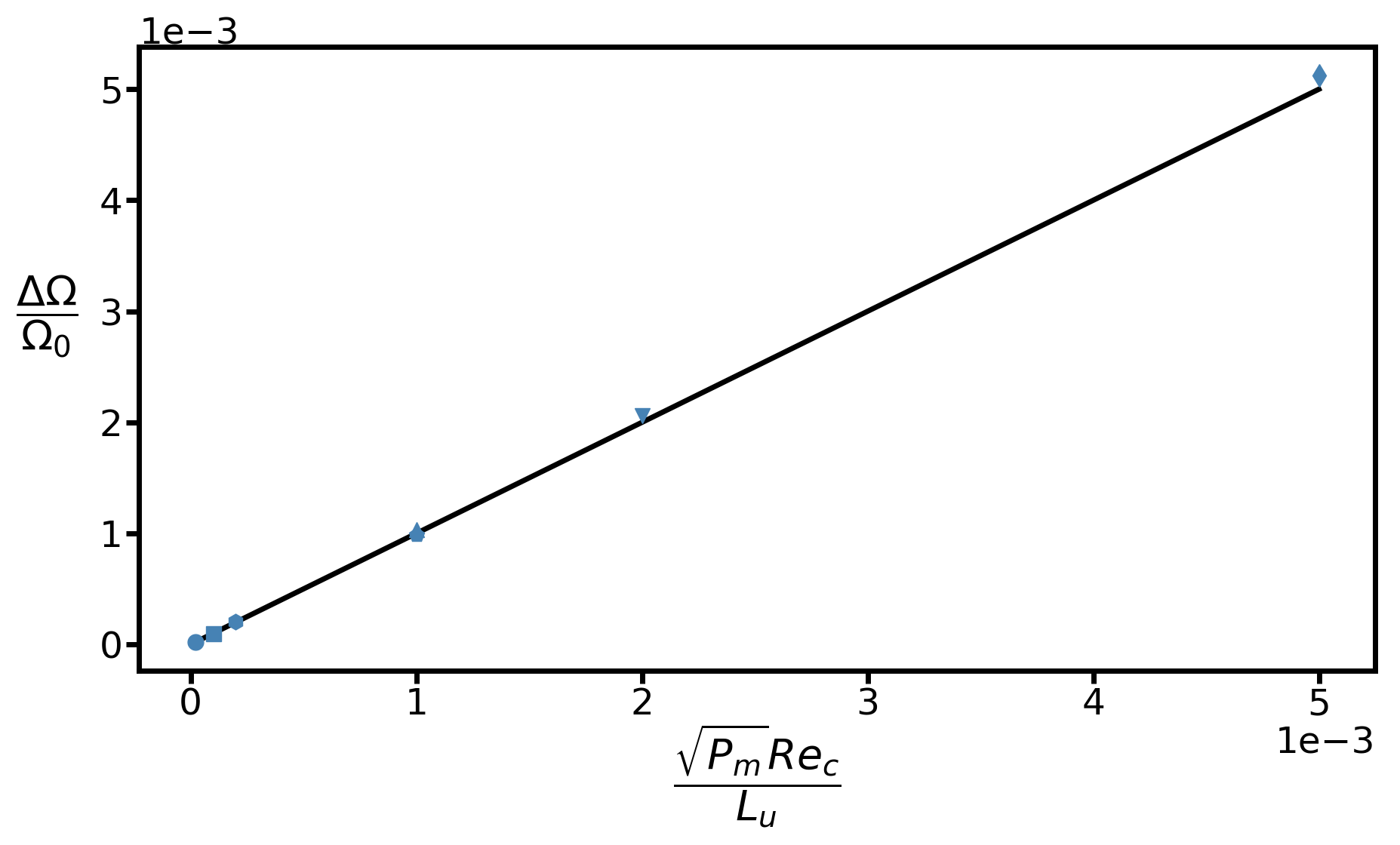}
\caption{Normalised differential rotation between a point outside the DZ and the outer sphere, $\Delta \Omega / \Omega_0 = \left(\Omega(r=0.65 \hspace*{0.05cm} r_0, \hspace*{0.05cm} \theta = \pi/12) - \Omega_0 \right)/ \Omega_0$, plotted as a function of $\sqrt{P_m} Re_c / L_u$ for the runs D$1$ to D$7$. The symbols are the same as in Fig. \ref{scaling_law_bphi}.} 
\label{scaling_law_rot_diff}
\end{center}
\end{figure}

This toroidal field is however unable to naturally match the vacuum condition imposed at the outer sphere ($B_{\phi} = 0$ at $r = r_0$). This is done across a $H_a^{-1}$ thickness magnetic boundary layer known as an Hartmann boundary layer. This layer is analysed in Appendix \ref{hartmann_equations} and the conclusion is that the $\mathcal{O} \left( P_m^2 Re_c / E L_u^2 \right)$ jump on $B_{\phi}/B_0$ induces a $\mathcal{O} \left( \sqrt{P_m} \hspace*{0.05cm} Re_c / L_u \right)$ jump on the differential rotation $\Delta \Omega / \Omega_0$ across the layer. This last scaling is confirmed in Fig. \ref{scaling_law_rot_diff} by computing $\Delta \Omega / \Omega_0$ at a particular location for the runs D$1$ to D$7$.

Figure \ref{fig2} shows a zoom on the zone delimited in black lines in Fig. \ref{fig1}. It illustrates the residual differential rotation that exists along each poloidal field lines. We note it as $\Delta \Omega_{\hspace*{0.05cm} \text{pol}} / \Omega_0$. Such a contrast of differential rotation is at odds with Ferraro's law of isorotation \citep{ferraro1937non} requiring a constant angular velocity along each poloidal field line. Ferraro's isorotation state is indeed found in our cases with no contraction in the induction equation. However, taking field advection into account, the $\Omega$-effect term can be balanced by the radial advection term in the induction equation, that is: 

\begin{equation}
V_f(r) \displaystyle \frac{\partial}{\partial r} \left( \displaystyle \frac{B_{\phi}}{r} \right) = \sin{\theta} \left(\vv{B}_p \cdot \vv{\nabla}\right) \delta \Omega
\label{modified_Ferraro}.
\end{equation}

\noindent Using the order of magnitude of the toroidal field derived in Eq. \eqref{estimate_bphi} together with Eq. \eqref{modified_Ferraro}, we end up with an estimate of this differential rotation

\begin{equation}
\displaystyle \frac{\Delta \Omega_{\hspace*{0.05cm} \text{pol}}}{\Omega_0} \approx \left( \displaystyle \frac{Re_c P_m}{L_u} \right)^2 = \left( \displaystyle \frac{\tau_{\text{A}_\text{p}}}{\tau_{\text{c}}} \right)^2
\label{estimate_diff_rot_ferraro}.
\end{equation}

\begin{figure}[!h]
\begin{center}
\hspace*{-0.5cm}
\includegraphics[width=8.5cm]{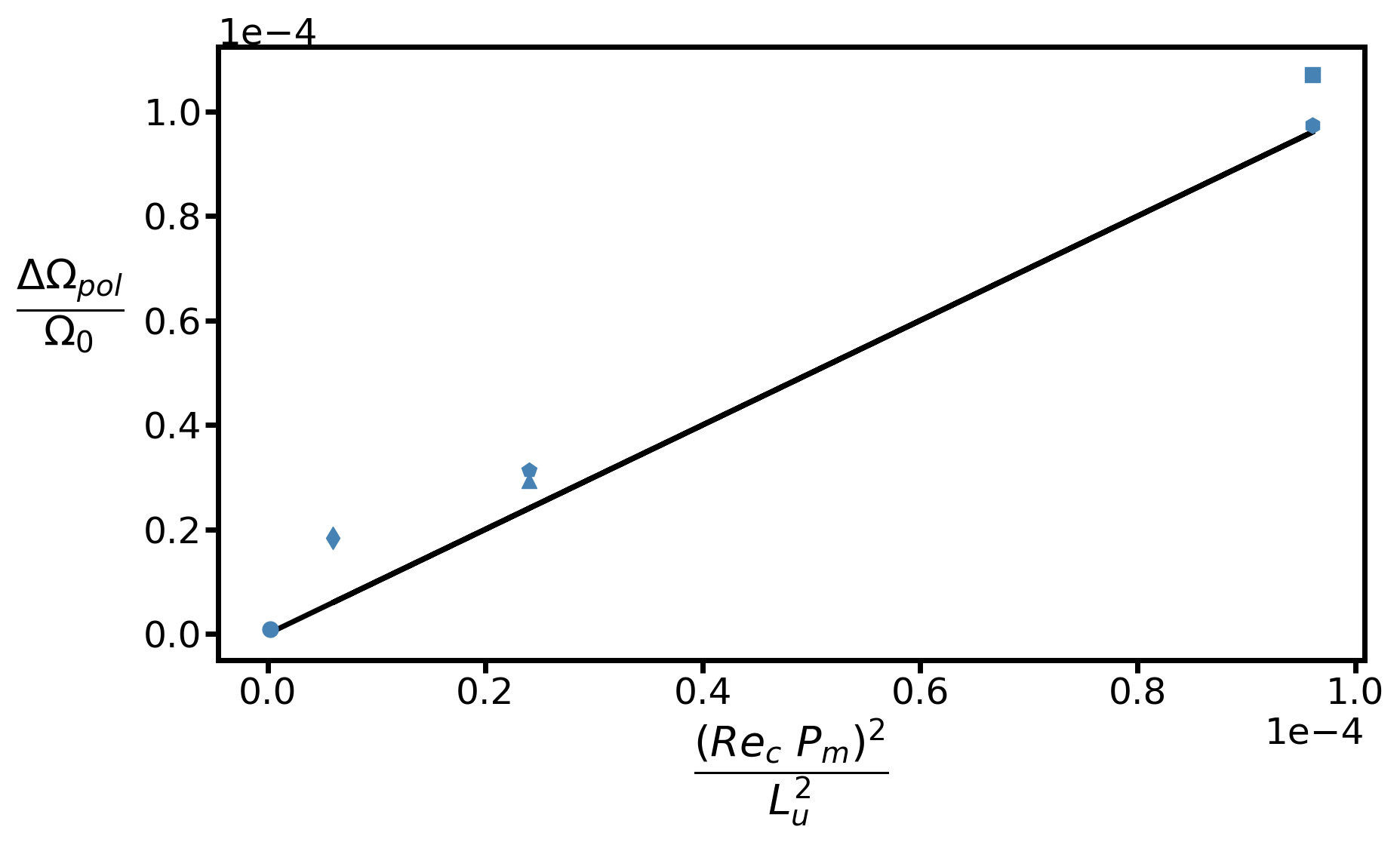}
\caption{Normalised differential rotation along a poloidal field line $\Delta \Omega_{\hspace*{0.05cm} \text{pol}} / \Omega_0$ as a function of $\left(Re_c P_m / L_u\right)^2$. This quantity is estimated between two sufficiently separated points along a field line for the runs D$8$ and, D$10$ to D$14$ of Table \ref{parameters_viscous}. The correspondences between parameters and symbols are as follows: Circle: $Re_c=10^{-1}$, $L_u = 10^4$; Triangle: $Re_c = 5\cdot10^{-1}$, $L_u = 5\cdot 10^3$; Diamond: $Re_c = 5\cdot10^{-1}$, $L_u = 10^4$; Square: $Re_c = 1$, $L_u = 5\cdot10^3$; Pentagon: $Re_c=1$, $L_u = 10^4$; Hexagon: $Re_c = 2$, $L_u = 10^4$. The other parameters are $E=10^{-4}$, $P_r \left(N_0/\Omega_0\right)^2 = 10^4$ and $P_m = 10^2$.}
\label{ferraro_violation}
\end{center}
\end{figure}

\noindent Figure \ref{ferraro_violation} shows the differential rotation taken between two ends of a poloidal field line as a function of the right-hand side of Eq. \eqref{estimate_diff_rot_ferraro}, for various $Re_c$ and $L_u$ respectively ranging from $10^{-1}$ to $2$ and from $5 \cdot 10^{3}$ to $10^{4}$. We can see that the level of differential rotation along a field line is indeed consistent with our estimate.

\begin{figure}[!h]
\begin{center}
\hspace*{-0.5cm}
\includegraphics[width=8.5cm]{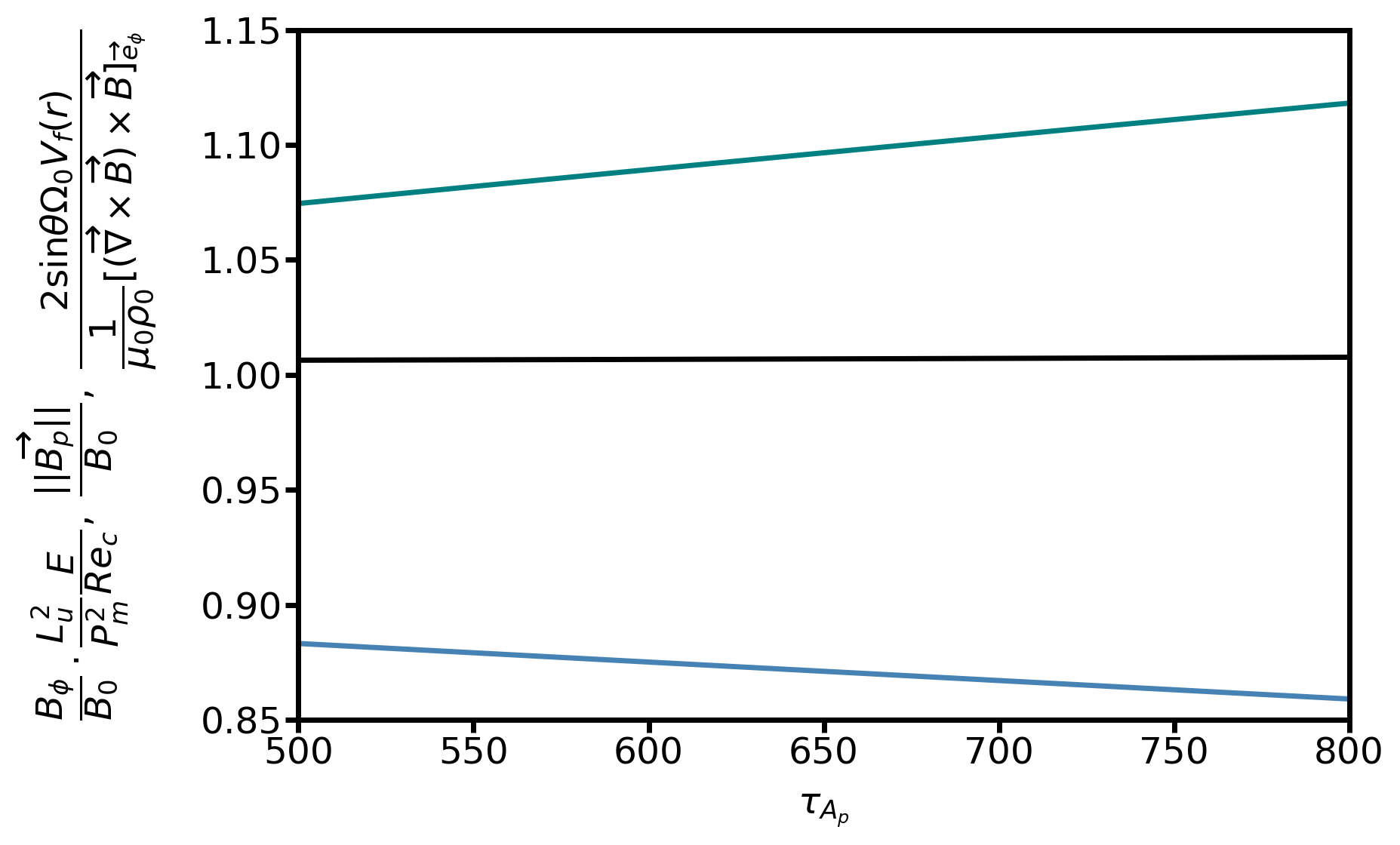}
\caption{Temporal evolution of different quantities followed over a period of $300 \hspace*{0.05cm} \tau_{\text{A}_\text{p}}$: the toroidal field normalised to the initial amplitude of the poloidal field and rescaled by $L_u^2 E / P_m^2 Re_c$ (green), the norm of the poloidal field normalised to the initial amplitude of the poloidal field (blue) and the ratio between the linear contraction term $\left(2 \sin{\theta} \hspace*{0.05cm} \Omega_0  V_f(r)\right)$ and the Lorentz force projected into the azimuthal direction (black). These quantities are evaluated at a particular location ($\theta = \pi/6$ and $r = 0.65 \hspace*{0.05cm} r_0$). For this simulation the contraction term has been removed from the induction equation but the results are exactly the same when it is included. The parameters are $Re_c = 1$, $L_u = 5 \cdot 10^4$, $P_m = 10^2$, $E = 10^{-4}$ and $P_r \left(N_0/\Omega_0\right)^2 = 10^4$ (run D$5$ of Table \ref{parameters_viscous}).}
\label{static_balance_evolve}
\end{center}
\end{figure}

Another consequence of the force balance Eq. \eqref{static_balance_eq} is that the toroidal field amplitude slowly increases over time. Figure \ref{static_balance_evolve} illustrates this by showing that in order to counteract the magnetic diffusion of the poloidal field (blue curve) and maintain this balance (black curve), the toroidal field amplitude is forced to increase (green curve). A similar situation was reported by \cite{charbonneau1993angular} where the magnetic stresses were in their case balanced by a wind torque.

\subsubsection{Dead zone}
\label{DZ_description}

We now turn to investigate the dynamics of the DZ. We already found that in this region the dominant balance is between the contraction and the viscous effects and is given by Eq. \eqref{differential_equation} in the linear regime $\Delta \Omega/\Omega \ll 1$. This implies that the order of magnitude of the differential rotation in the DZ, denoted $\Delta \Omega_{\hspace*{0.05cm} \text{DZ}}/\Omega_0$, is $\sim L_{\hspace*{0.05cm} \text{DZ}} V_f(r)/\nu$, where $L_{\hspace*{0.05cm} \text{DZ}}$ is a typical lengthscale of the DZ. While the magnetic field does not explicitly enter in this expression, it is of prime importance because its interaction with the contraction determines the shape and the location of the DZ. 

The differences observed in Fig. \ref{Numerical_Results_Viscous} concerning the level of differential rotation between the simulations with and without contraction of the field lines can thus be explained by the location and size of the DZ. Indeed, in the case without field line contraction, the DZ is smaller and located closer to the outer sphere, that is in a area where the contraction velocity is weaker, which leads to a smaller level of differential rotation. Another difference between these two cases is the time taken to reach the stationary solution. As this time is the viscous time based on the size of the DZ, it is not surprising that the stationary state is reached much more rapidly when the DZ occupies a small fraction of the spherical shell. Note that in the case with contraction of the field lines, the viscous time is of the same order as the contraction time $\tau_{\text{c}}$ because we are in the regime $\Delta \Omega /\Omega \sim 1$.

It is also possible to obtain a more precise determination of the DZ differential rotation, by deriving an approximate analytical solution of Eq. \eqref{differential_equation}. To this end, the DZ is first assimilated to a conical domain, $r \in \left \lbrack r_{i_{\hspace*{0.025cm} \text{DZ}}} \text{,} \hspace*{0.1cm} r_0 \right \rbrack, \theta \in [-\theta_{0} \text{,} \hspace*{0.1cm} \theta_{0}]$, outside of which the flow is in solid rotation. In the case where the domain connects to the inner sphere, we adopt a stress-free condition on the azimuthal velocity field at $r = r_i$. We then neglect the last term of the left-hand side in Eq. \eqref{differential_equation} as we found in our simulations that its contribution is small, particularly at low latitude. The homogeneous problem, which is separable in two radial and latitudinal eigenvalue sub-problems, allows us to construct a basis of orthogonal eigenfunctions that satisfy the boundary conditions on the conical domain. The details of the method are provided in Appendices \ref{resolution_no_induct} and \ref{resolution_induct_mod}

In the case where the contraction does not advect the field lines, the approximate analytical solution takes the following form

\begin{equation}
\begin{array}{lll}
\displaystyle \frac{\delta \Omega(r,\theta)}{\Omega_0} = Re_c \sum_{n=1}^{\infty} \sum_{k=1}^{\infty} A_{nk}
\left( \displaystyle \frac{r_0}{r} \right)^{3/2} \sin{\left(\left(\displaystyle \frac{n \pi}{\ln{\left(\displaystyle \frac{r_{i_{\hspace*{0.025cm} \text{DZ}}}}{r_0}\right)}}\right) \ln{\left(\displaystyle \frac{r}{r_0}\right)} \right)} \\\\ 
\hspace*{3.3cm} \cos{\left(\left( \displaystyle \frac{\left(2k - 1\right) \pi}{2 \theta_0} \right) \theta\right)} 
,\end{array}
\label{analytical_solution_dz_no_induct}
\end{equation}

\noindent wherein the expression of the coefficient $A_{nk}$ can be found in Appendix \ref{resolution_no_induct}.

\begin{figure}[!h]
\begin{center}
\hspace*{-0.5cm}
\includegraphics[width=8.5cm]{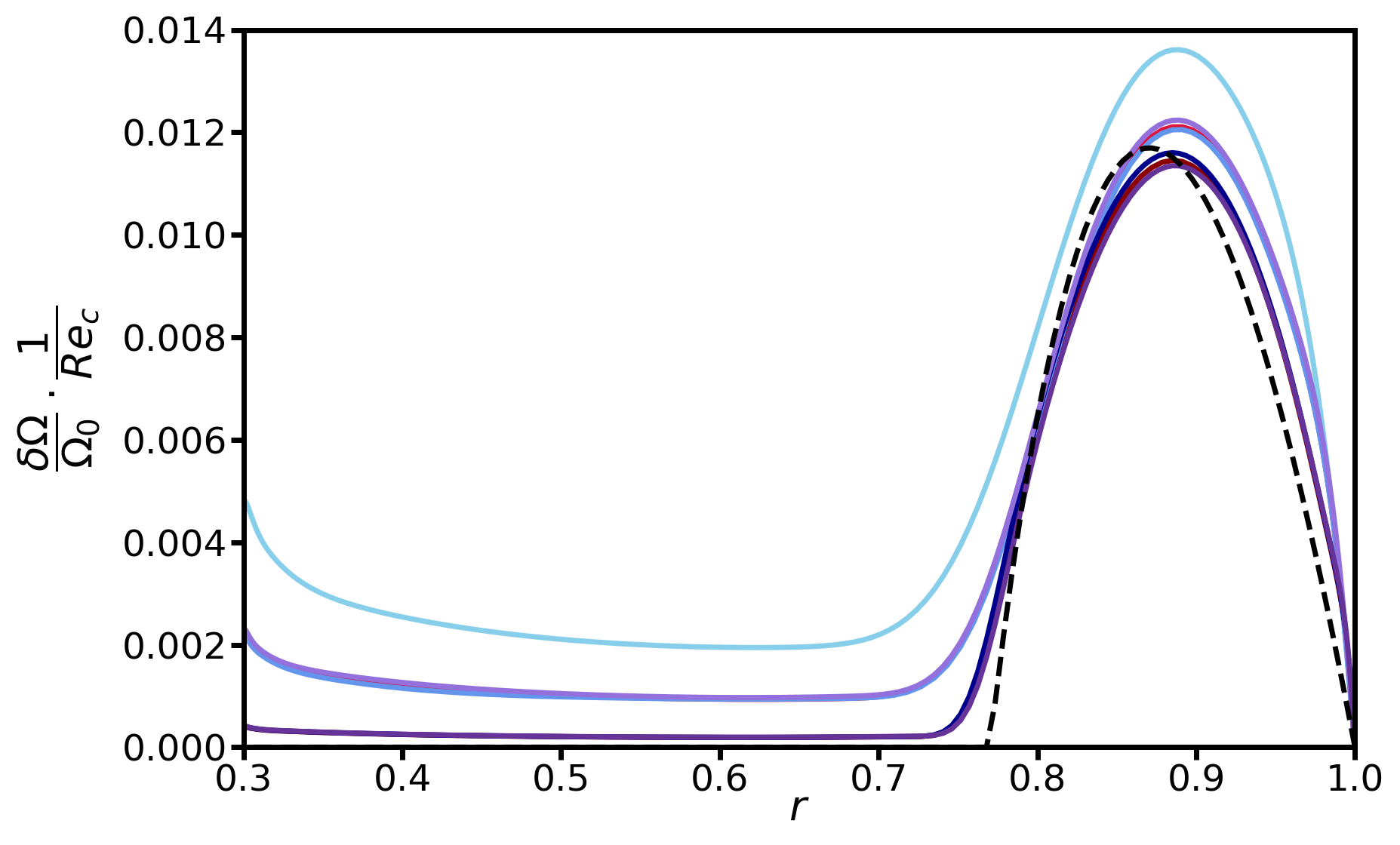}
\caption{Equatorial differential rotation as a function of radius when the contraction term is removed from the induction equation. Numerical solutions (runs D$1$ to D$7$) are plotted in colour at $Re_c = 10^{-1}$ (red), $Re_c = 1$ (blue) and $Re_c = 5$ (purple), for various Lundquist numbers ranging from $10^4$ to $5\cdot10^4$ (from the lightest to the darkest). For $Re_c = 1$ an additional simulation is presented at $L_u = 5 \cdot 10^3$. The curves are rescaled by $Re_c$ then compared to the analytical solution Eq. \eqref{analytical_solution_dz_no_induct} displayed in black dashed lines. The other parameters are $E =10^{-4}$, $P_r \left(N_0/\Omega_0 \right)^2 = 10^4$ and $P_m = 10^2$.}
\label{Analytical_Solution_No_Induct_Mod}
\end{center}
\end{figure}

\noindent This solution computed for a chosen conical domain is compared in Fig. \ref{Analytical_Solution_No_Induct_Mod} with the results of the numerical simulations performed at various $Re_c$ and $L_u$ respectively ranging from $10^{-1}$ to $5$ and from $5\cdot10^3$ to $5\cdot 10^4$. A good agreement is found between the numerical and analytical solutions. 
The differential rotation scales as $Re_c$ and this was expected because the condition $\delta \Omega/\Omega \ll 1$ for the linear approximation of the contraction term is fulfilled in the simulations. We also find that the differential rotation is almost independent of the initial amplitude of the magnetic field: an increase of $L_u$ of one order of magnitude causes only a small change on $\Delta \Omega_{\text{DZ}} / \Omega_0$. This is consistent with the fact that, in the regime considered here, the shape and location of the DZ do not depend on the magnetic field. We rather observe that a higher magnetic field affects the rotation rate outside the DZ, or equivalently the differential rotation outside the DZ. Again this is expected as the jump in rotation rate across the outer sphere Hartmann layer decreases with the field amplitude.

When the contraction term is introduced in the induction equation, the solution of Eq. \eqref{differential_equation} becomes

\begin{equation}
\begin{array}{lll}
\displaystyle \frac{\delta \Omega(r,\theta)}{\Omega_0} = Re_c \sum_{n=1}^{\infty} \sum_{k=1}^{\infty} 
A_{nk} \left( \displaystyle \frac{r_0}{r} \right)^{3/2} 
\sin{\left( \displaystyle \frac{1}{2} \sqrt{\left|9 + 4 \hspace*{0.05cm} \mu_n \right|} \ln{\left( \displaystyle \frac{r}{r_0}\right)} \right)} \\\\
\hspace*{3.3cm} \cos{\left(\left( \displaystyle \frac{\left(2k - 1\right) \pi}{2 \theta_0} \right) \theta\right)}
,\end{array}
\label{analytical_solution_dz_induct_mod}
\end{equation}

\noindent where the expressions of the coefficients $A_{nk}$ and $\mu_n$ are given in Appendix \ref{resolution_induct_mod}.

\begin{figure}[!h]
\begin{center}
\hspace*{-0.5cm}
\includegraphics[width=8.5cm]{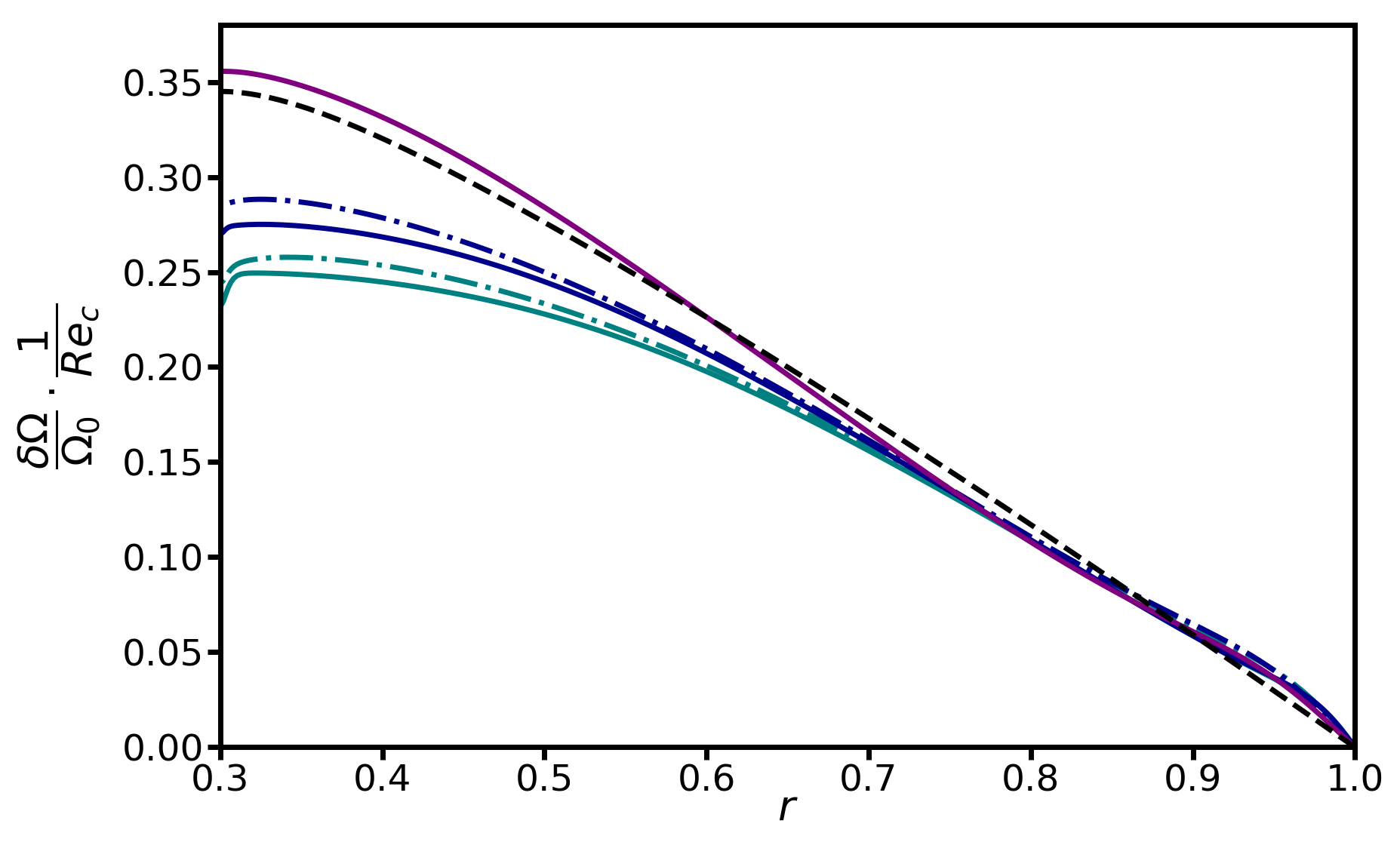}
\caption{Same as Fig. \ref{Analytical_Solution_No_Induct_Mod} but for cases where contraction acts on the field lines. The simulations presented here are the runs D$10$ to D$14$ of Table \ref{parameters_viscous} performed at $Re_c = 5\cdot10^{-1}$ (green), $Re_c = 1$ (blue) and $Re_c=2$ (purple). For all these cases $L_u = 10^{4}$ but for $Re_c = 5\cdot10^{-1}$ and $1$, additional simulations are shown at $L_u = 5\cdot10^{3}$ in dash-dot lines. The analytical solution Eq. \eqref{analytical_solution_dz_induct_mod} is displayed in black dashed lines.}
\label{Analytical_Solution_Induct_Mod}
\end{center}
\end{figure}

\noindent This solution computed for a chosen conical domain is compared in Fig. \ref{Analytical_Solution_Induct_Mod} with the numerical results obtained at various $Re_c$ and $L_u$. As previously, the initial amplitude of the magnetic field causes only small changes on the level of differential rotation. However, some discrepancies are visible. First of all, the numerical curves obtained at different $Re_c$ do not overlap, which indicates that for these higher levels of differential rotation ($\delta \Omega/\Omega_0 \sim 30\%$) the contraction term can no longer be linearised, a necessary condition to derive the analytical solution. A second limitation is the assumed conical shape of the DZ. This shape is indeed  more representative of the actual DZ for run D$14$ performed at $Re_c=2$, where we can observe that the analytical solution tends to reproduce the expected differential rotation, than for simulations D$10$ to D$13$ obtained at $Re_c = 0.5$ and $1$. 

\subsubsection{Effect of the density stratification}
\label{viscous_anel}

\begin{figure}[!h]
\begin{center}
\includegraphics[width=4.4cm]{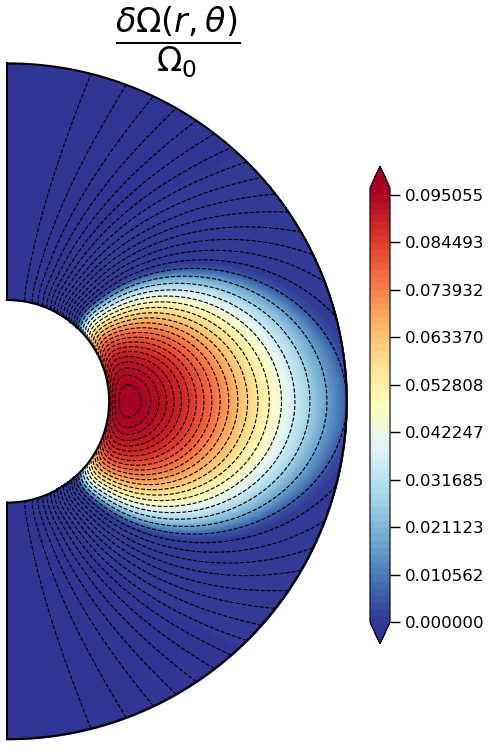}
\includegraphics[width=4.4cm]{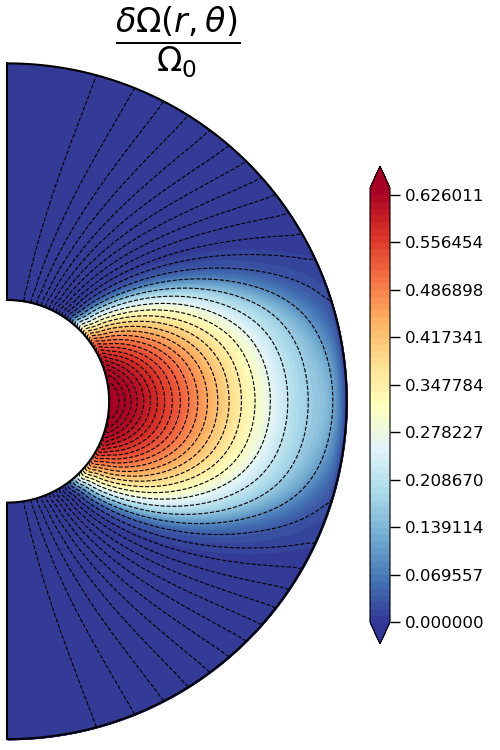}
\caption{Meridional cuts of the normalised rotation rate in two anelastic cases with $\rho_i/\rho_0 = 20.85$. The poloidal field lines are represented in black lines. For these simulations the contraction term is included in the induction equation. In the first panel $Re_c = 1$ and $ L_u = 10^5$ (run D$18$ of Table \ref{parameters_viscous}), and in the second one $Re_c = 5$ and $L_u = 5\cdot10^4$ (run D$20$ of Table \ref{parameters_viscous}). The other parameters are $P_r \left(N_0/\Omega_0\right)=10^4$, $E = 10^{-4}$ and $P_m = 10^2$.}
\label{Rot_Diff_Viscous_Anel}
\end{center}
\end{figure}

We have performed simulations in the anelastic approximation for a fixed density contrast $\rho_i/\rho_0 = 20.85$. Two typical results are presented in Fig. \ref{Rot_Diff_Viscous_Anel} at $Re_c = 1$, $L_u = 10^5$ and $Re_c=5$, $L_u = 5\cdot10^4$ (left and right panels respectively). For these runs where the contraction advects the field lines, we find again two separate regions, with a differential rotation still mostly located in the DZ. This zone is now more extended in latitude compared to the Boussinesq case and the main difference between the calculations at two different $Re_c$ lies in the level of differential rotation in the DZ. Indeed, when this level in the left panel is compared to its Boussinesq counterpart in the third panel of Fig. \ref{Numerical_Results_Viscous} we notice that it has been divided by a factor $\sim 3$. This is explained by the density stratification effect on the contraction velocity field. Indeed, as we already found in \citep{gouhier2020axisymmetric}, the normalised differential rotation $ \Delta \Omega_{\hspace*{0.05cm} \text{DZ}} / \Omega_0$ resulting from the balance between the viscous and contraction effects is weighted by the inverse of the background density profile:

\begin{equation}
\displaystyle \frac{\Delta \Omega_{\hspace*{0.05cm} \text{DZ}}^{\hspace*{0.05cm} \text{A}}}{\Omega_0} \approx \left( \displaystyle \int_{1}^{r_i/r_0} \displaystyle \frac{\rho_0}{\overline{\rho}} \hspace*{0.05cm} \text{d} \left( r / r_0 \right) \right) \displaystyle \frac{\Delta \Omega_{\hspace*{0.05cm} \text{DZ}}^{\hspace*{0.05cm} \text{B}}}{\Omega_0}
\label{rot_diff_estimate_anel_viscous},
\end{equation}

\noindent where the index $\text{A}$ stands for "anelastic" and $\text{B}$ for "Boussinesq". For $\rho_i/\rho_0=20.85$ and $r \in \left \lbrack r_i = 0.3 ; \hspace*{0.05cm} r_0 =1 \right \rbrack$, using the characteristic amplitude of the differential rotation given in the third panel of Fig. \ref{Numerical_Results_Viscous}, we obtain $\Delta \Omega_{\hspace*{0.05cm} \text{DZ}}^{\hspace*{0.05cm} \text{A}}/\Omega_0 = 0.095$, in agreement with the rotation rate shown in the left panel of Fig. \ref{Rot_Diff_Viscous_Anel}. A good estimate of the rotation rate displayed in the right panel is then readily obtained by multiplying this result by $Re_c$. 

\subsection{Quadrupolar field in the viscous regime}
\label{viscous_quadrupole}

We now consider an initial quadrupolar magnetic field as given by Eq. \eqref{quadrupole} and displayed in the right panel of Fig. \ref{Initial_Pol_Mag}. The outcome will be completely different as the differential rotation resulting from the contraction is subject to an instability that will be discussed in detail below.

\begin{figure*}[!h]
\begin{center}
\includegraphics[width=4.4cm]{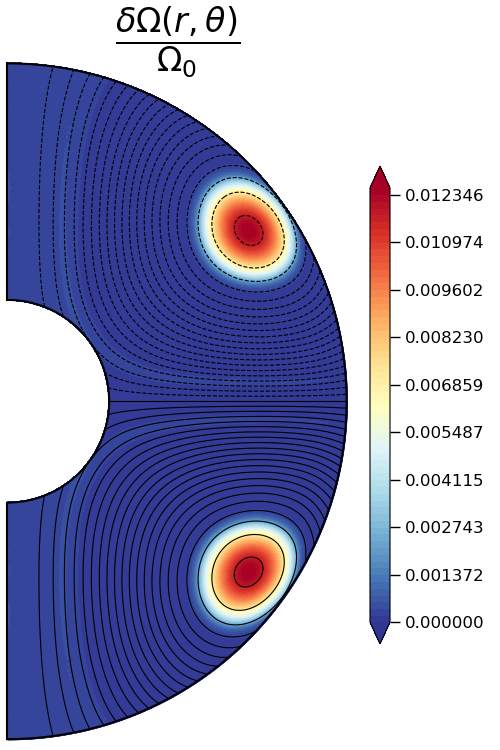}
\includegraphics[width=4.4cm]{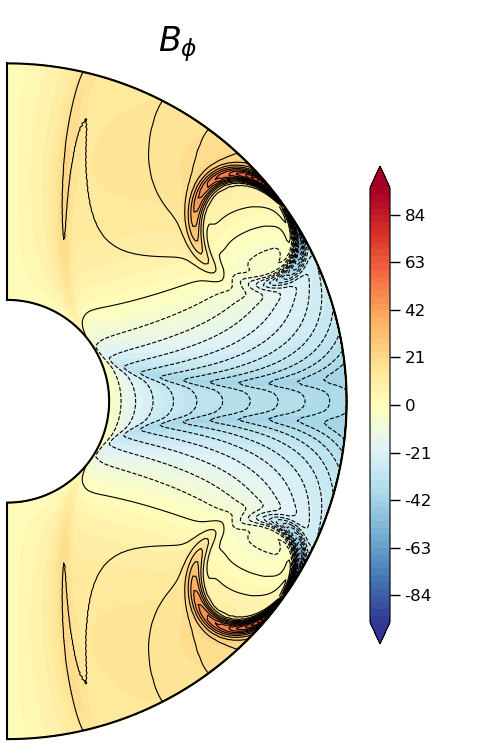}
\includegraphics[width=4.4cm]{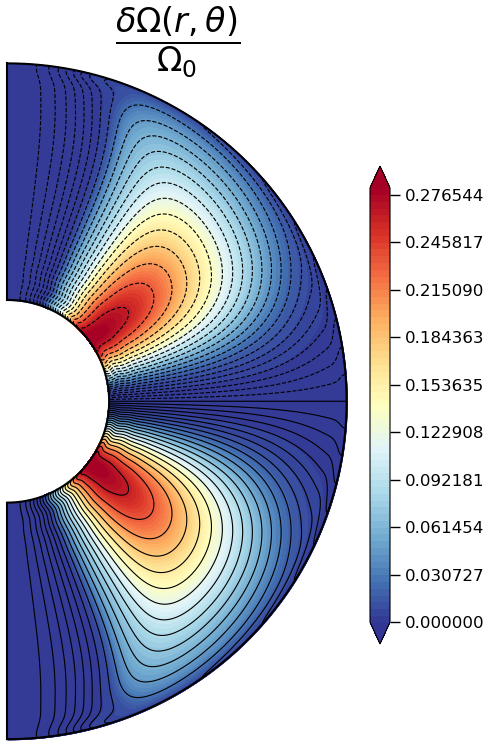}
\includegraphics[width=4.4cm]{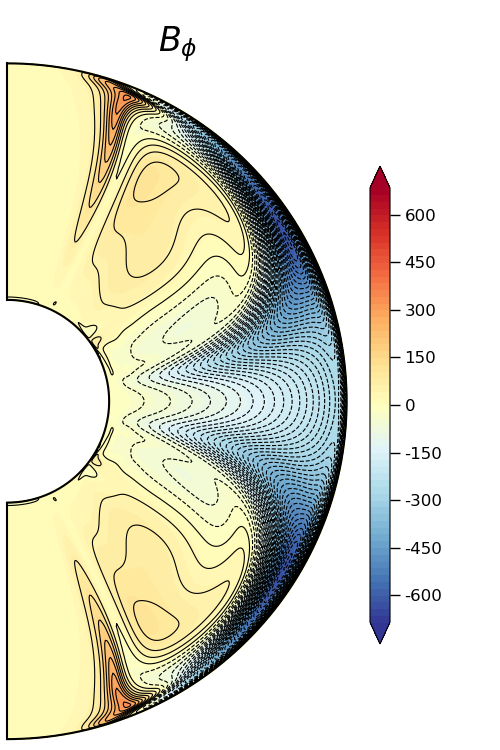}
\caption{Same as Fig. \ref{Numerical_Results_Viscous} but for the initially quadrupolar magnetic field. In the first two panels $L_u=5\cdot 10^4$ while $L_u=10^4$ in the last two. The other parameters are $Re_c=1$, $E=10^{-4}$, $P_r \left(N_0/\Omega_0\right)^2 = 10^4$ and $P_m = 10^{2}$ (runs Q$2$ and Q$10$ of Table \ref{parameters_viscous}).}
\label{Numerical_Results_Viscous_Quadrupole}
\end{center}
\end{figure*}

Figure \ref{Numerical_Results_Viscous_Quadrupole} displays typical states obtained after a contraction timescale. The two left panels show a case without contraction of the field lines while the two right panels correspond to simulations where contraction is included in the induction equation. At this stage, the similarities with the dipolar case are striking. First we observe the quasi-solid rotation region outside the DZs. When the advection of the field lines is prevented, these DZs are located near the outer sphere while they connect to the inner sphere when contraction is present. In addition, for a given set of parameters, when the dipolar and quadrupolar cases are compared with each other, the same levels of differential rotation are found. For the runs presented here, the maximum amplitude of the normalised differential rotation inside the DZs is again $\sim 30 \%$ when the contraction acts on the field lines and falls to $\sim 1 \%$ otherwise, similar to the cases presented in Fig. \ref{Numerical_Results_Viscous}. By observing the second and fourth panels of Fig. \ref{Numerical_Results_Viscous_Quadrupole} we note, again, the presence of magnetic boundary layers namely, the Shercliff layers separating the poloidal field lines constrained to rotate at different rates, and the Hartmann layer at the outer sphere. However compared to the dipolar case, after  $\sim 1 \hspace*{0.05cm} \tau_{\text{c}}$ the quadrupolar configuration is the seat of an axisymmetric instability. 

\begin{figure}[!h]
\begin{center}
\hspace*{-0.5cm}
\includegraphics[width=8.5cm]{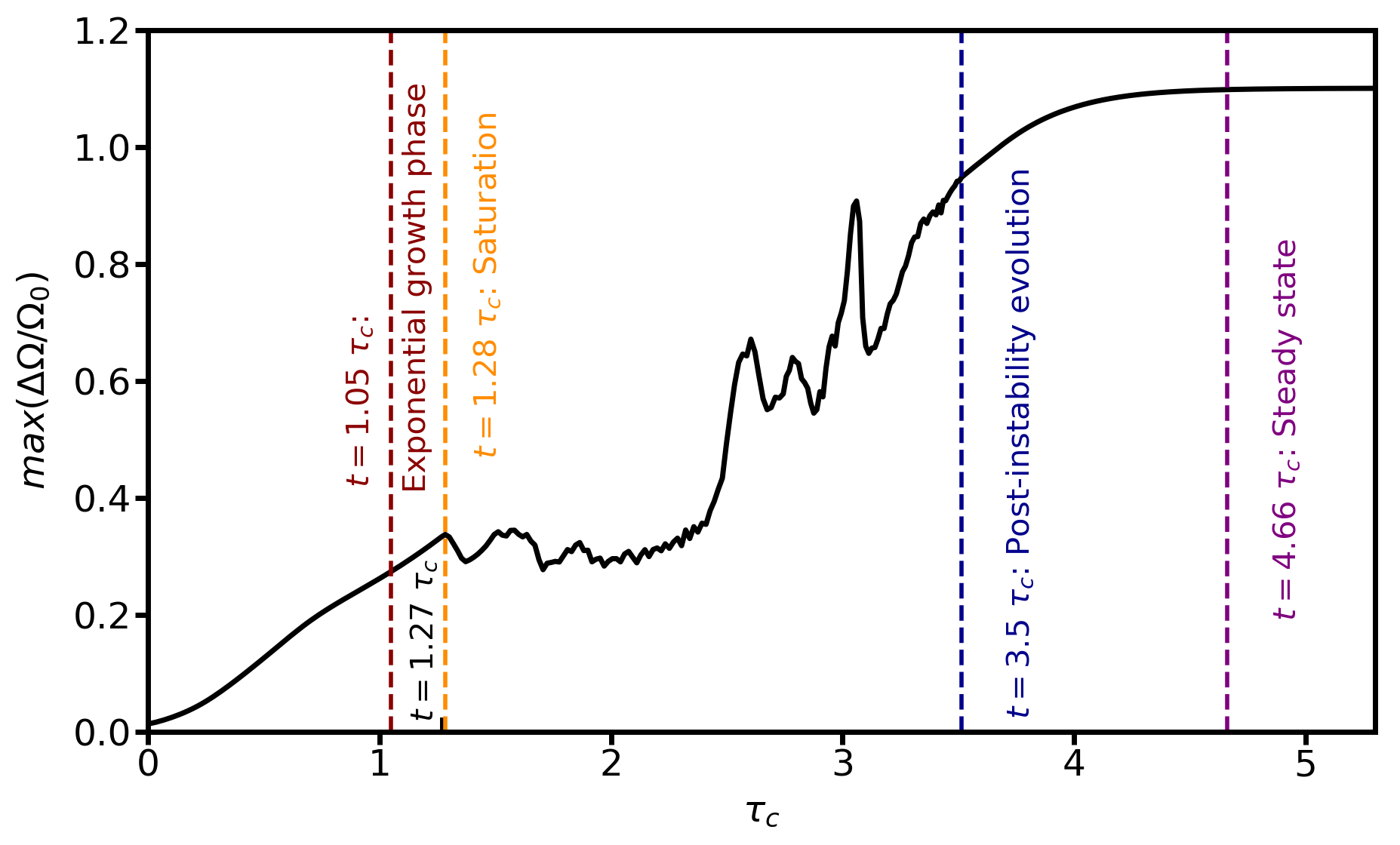}
\caption{Maximum value of the normalised differential rotation $\text{max} \left( \Delta \Omega / \Omega_0 \right)$ as a function of the contraction timescale as defined in Eq. \eqref{contraction_timescale}. The main evolution steps are highlighted by coloured dashed lines: the red line corresponds to the start of the exponential growth phase, orange to the time at which the instability saturates thus marking the beginning of the non-linear evolution. Then blue marks the start of the post-instability evolution and finally purple denotes the time at which the hydrodynamic steady state is reached. Parameters are $Re_c = 1$, $L_u = 10^4$, $E = 10^{-4}$, $P_r \left(N_0/\Omega_0\right)^2 = 10^{4}$ and $P_m = 10^2$ (run Q$10$ of Table \ref{parameters_viscous}, contraction term in induction equation).}
\label{max_rot_diff_evolution}
\end{center}
\end{figure}

The evolution of the maximum differential rotation is shown in Fig. \ref{max_rot_diff_evolution} where the different steps that will be described thereafter are highlighted. In particular, the differential rotation first builds up before an instability starts to kick in at $t \sim 1 \hspace*{0.05cm} \tau_{\text{c}}$ (red dashed lines). Between $\sim 1 \hspace*{0.05cm} \tau_{\text{c}}$ and $\sim 3.5 \hspace*{0.05cm} \tau_{\text{c}}$ (blue dashed lines) the instability grows, saturates and strongly modifies the flow and field as we shall see later. Finally, after $3.5 \hspace*{0.05cm} \tau_{\text{c}}$, the configuration evolves more smoothly until a final steady state is reached at $\sim 4.7 \hspace*{0.05cm} \tau_{\text{c}}$ (purple dashed lines).

\subsubsection{Description of the instability}
\label{description_instability}

\begin{figure*}[!h]
\begin{center}
\includegraphics[width=3.1cm]{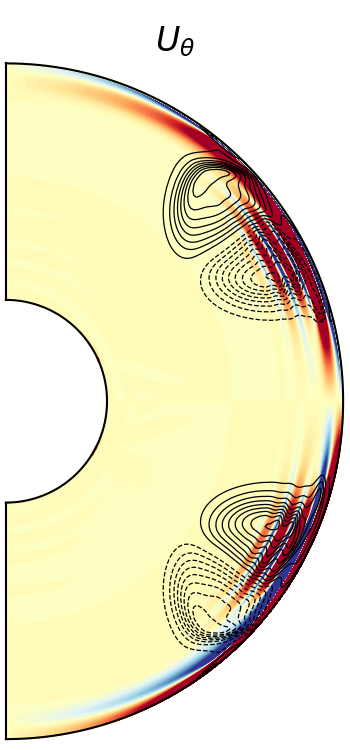}
\hspace*{0.35cm}
\includegraphics[width=4.4cm]{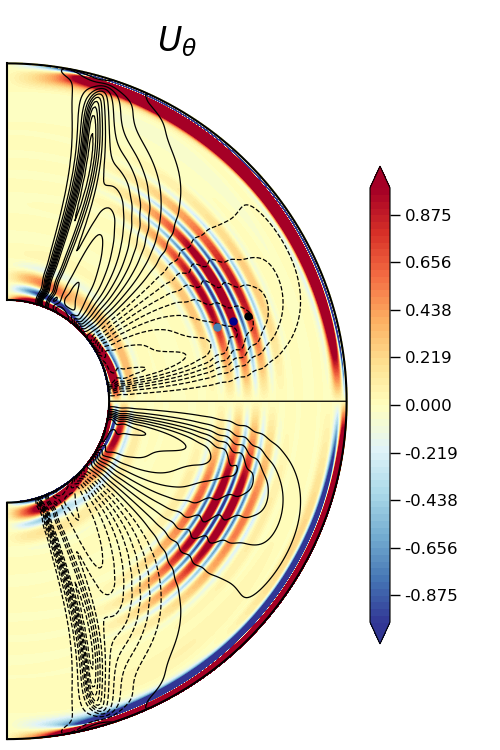}
\hspace*{-0.25cm}
\includegraphics[width=9.5cm]{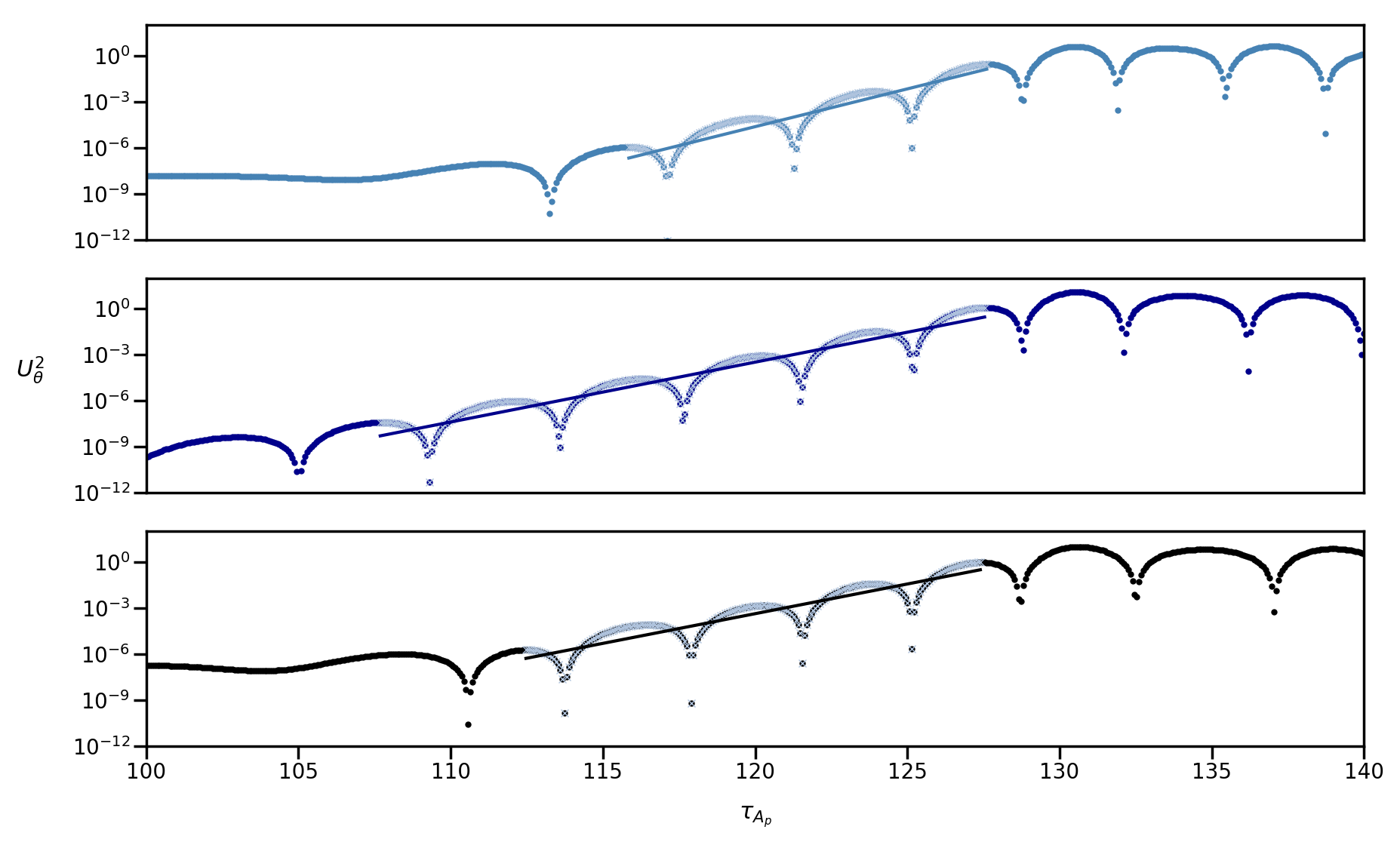}
\caption{First two panels: snapshots of the latitudinal velocity field $U_{\theta}$ taken during the developing of the instability, at $t = 14.1 \hspace*{0.05cm} \tau_{\text{A}_\text{p}}$ when the contraction does not act on the field lines (first panel) and at $t=127.4 \hspace*{0.05cm} \tau_{\text{A}_\text{p}}$ when it does (second panel). In these cuts are also represented the contours of the latitudinal shear $\partial \ln{\Omega} / \partial \theta$ in black, with dashed lines corresponding to a negative shear and full lines to a positive one. In the second panel, three control points (in light blue, blue and black) are chosen at the location where the instability grows. Third panel: temporal evolution of the square of the latitudinal velocity field at these selected points (with the same colour code), as a function of the Alfvén poloidal time. In each subplot a coloured straight line is also presented as the result of a linear regression used to deduce a growth rate associated with the instability. The parameters are those of the runs Q$4$ and Q$10$ of Table \ref{parameters_viscous} namely, $Re_c = 5$ (first panel) and $Re_c=1$ (two last panels), with $E=10^{-4}$, $P_r \left(N_0/\Omega_0\right)^2 = 10^4$, $P_m = 10^{2}$ and $L_u = 10^4$.}
\label{insta}
\end{center}
\end{figure*}

Figure \ref{insta} shows, in colour, the structure of the unstable modes when the contraction does not act on the field lines (first panel) and when it acts on them (second panel). In these meridional cuts are also represented, in black, the contours of the latitudinal shear $\partial \ln{\Omega} / \partial \theta$. Dashed (solid) lines represent negative (positive) values of this gradient. Contrary to the dipolar case, we now have a region of negative shear in the Northern hemisphere and positive shear in the South. As can be observed, this is precisely at these locations that the perturbations grow. We already note that this instability is not of the centrifugal type because the shear in these regions is not strong enough, that is $\left |\partial \ln{\Omega} / \partial \ln{s}\right | < 2$. The figure also shows that the unstable modes are characterised by small radial length scales and large horizontal length scales which implies that the meridional motions are predominantly horizontal. The last panel of Fig. \ref{insta} displays the local evolution of the square of the latitudinal velocity field at the fixed points marked on the second panel (in light blue, blue and black). We can see that the kinetic energy of the perturbations grows exponentially for about $\sim 20 \hspace*{0.05cm} \tau_{\text{A}_\text{p}}$ before saturation occurs. This time interval corresponds to the linear phase of the instability. The oscillating behaviour that adds to the exponential growth is due to the fact that the perturbations propagate.

\begin{table*}[!h]
\begin{center}
\begin{tabular}{c|c|c|c|c|c|c|c}
\hline
\hline
 \hspace*{-0.2cm} \multirow{3}{*}{\textbf{Case}}& \multirow{3}{*}{\textbf{Contraction in induction equation}} &\multirow{3}{*}{$\boldsymbol{Re_c}$} & \multirow{3}{*}{$\boldsymbol{L_u} $}  &  \multirow{3}{*}{$\boldsymbol{\displaystyle \frac{\sigma_1}{|<q>| \hspace*{0.05cm} \Omega}} $} & \multirow{3}{*}{$ \boldsymbol{\displaystyle \frac{\sigma_2}{|<q>| \hspace*{0.05cm} \Omega}} $} &  \multirow{3}{*}{$\boldsymbol{\displaystyle \frac{\sigma_3}{|<q>| \hspace*{0.05cm} \Omega}} $} & \multirow{3}{*}{$ \boldsymbol{ \overline{\displaystyle \frac{\sigma}{|<q>| \hspace*{0.05cm} \Omega}}} $} \\ & & & & & & \\ & & & & & & \\
\hline
\hline
Q$3$ &No& $5$ & $5\cdot 10^3$ & $2.5\cdot 10^{-2}$ & $2.4\cdot 10^{-2} $ & $1.9\cdot 10^{-2}$ & $2.2\cdot 10^{-2}$ \\
Q$4$ &No& $5$ & $10^4$ & $3.9\cdot 10^{-2}$ & $3.6\cdot 10^{-2} $ & $3.4\cdot 10^{-2}$ & $3.6\cdot 10^{-2}$ \\
Q$9$ &Yes& $1$ & $5\cdot 10^3$ & $1.1 \cdot 10^{-2}$ & $8.8  \cdot 10^{-3} $ & $8.1  \cdot 10^{-3}$ & $9.2 \cdot 10^{-3}$ \\
Q$10$ & Yes & $1$ & $10^4$ & $1.8 \cdot 10^{-2}$ & $1.5 \cdot 10^{-2}$ & $1.5 \cdot 10^{-2}$ & $ 1.6 \cdot 10^{-2}$ \\
Q$11$ & Yes & $2$ & $10^4$ & $1.8 \cdot 10^{-2}$ & $2.1 \cdot 10^{-2}$ & $1.5\cdot 10^{-2}$& $1.8 \cdot 10^{-2}$ \\
\hline
\end{tabular}
\vspace*{0.2cm}
\caption{Ratio between the growth rate $\sigma_{i=1, 2, 3}$ of the instability and the product of the absolute value of the surface-averaged shear parameter $|<q>|$ by the local rotation rate $\Omega$. The $\sigma_{i=1, 2, 3}$ are the growth rates associated to the different control points visible in the first panel of Fig. \ref{insta} for the cases Q$3$ and Q$4$, and in the second panel for the runs Q$9$ to Q$11$. The surface-averaged value of the shear parameter is obtained by taking a surface enclosing the location of the instability then by calculating a surface integral such as $<q> \hspace*{0.05cm} = 1/S \iint \left(\partial \ln{\Omega} / \partial \ln{\theta}\right) \text{d} S$. From these ratios an arithmetic mean is determined and denoted by $\overline{\left(\sigma / |<q>| \Omega \right)}$ in the present Table. The parameters $P_r \left(N_0/\Omega_0\right)^2$, $E$ and $P_m$ are respectively fixed to $10^4$, $10^{-4}$ and $10^2$ for each run.}
\label{growth_rates}
\end{center}
\end{table*}

Growth rates have been determined from these plots. Their values normalised by the product of the surface-averaged shear parameter $<q> \hspace*{0.05cm} = 1/S \iint \left( \partial \ln{\Omega} / \partial \theta \right) \text{d}S$ to the mean local rotation rate are listed in Table \ref{growth_rates}. We observe that when the contraction Reynolds number is multiplied by two, the shear rate and the growth rate are both doubled. This shows that the growth rate of the instability seems to be proportional to the shear rate. Moreover, for the simulation Q$5$ performed at higher $L_u = 5 \cdot 10^4$, the instability is not triggered, clearly indicating that a strong enough poloidal field has a stabilising effect. Finally, while runs Q$4$, Q$10$ and Q$11$ carried out for $L_u = 10^4$ are unstable, runs Q$1$ and Q$8$ performed for the same $L_u$, but at lower $Re_c$, are stable. This shows that, at a lower shear rate, a lower poloidal field is required to stabilise the flow. These findings, namely the requirement that the rotation decreases away from the rotation axis and the facts that the growth rates are proportional to the shear and that stabilisation occurs above a certain magnetic tension are all in agreement with an MRI-type instability \citep{balbus1991powerful}.

Our numerical results also exhibit more subtle effects that are not accounted for in the local WKB approach of the standard MRI (SMRI) \citep{balbus1991powerful, balbus1994stability, menou2004local}. First, the growth rates determined in Table \ref{growth_rates} are significantly lower than the maximum growth rate $\sigma_{\text{max}} = |q| \hspace*{0.05cm} \Omega/ 2$, where $q = \text{d} \ln{\Omega} / \text{d} \ln{s} $, predicted by these studies (e.g. \cite{balbus1994stability}). Second, a comparison between runs Q$9$ and Q$10$ shows that the growth rate increases with $L_u$ whereas the predicted $\sigma_{\text{max}}$ does not depend on the poloidal field. Finally, as noted earlier the perturbations propagate in the domain while the phase velocity of the modes is zero in the SMRI.

We now argue that these differences are due to the effect of the stable stratification and to the presence of a toroidal field. In \cite{balbus1994stability} the maximum growth rate is determined by assuming a zero latitudinal wavenumber. This avoids any stabilising effect of the stratification because the buoyancy force has no effect on purely horizontal motions. It follows that the most unstable radial scale is inversely proportional to the poloidal field amplitude and does not depend on the stable stratification. Even if the unstable motions found in our simulations are predominantly horizontal, assuming a zero latitudinal wavenumber is too extreme because our background flow is not uniform in latitude so that the perturbation must have and indeed has a finite wavelength in this direction. As a consequence, the stable stratification is expected to play a role in determining the most unstable radial lengthscale and the associated maximum growth rate. We indeed found that the radial wavenumbers of the unstable modes are always larger the theoretical value from \cite{balbus1994stability} and that it is very little dependent on the amplitude of the poloidal field. The effect of the stable stratification may thus potentially explain why the growth rates found in our simulations are significantly smaller than $\sigma_{\text{max}}$. We note that the thermal diffusion must be taken into account to consider the effect of the stable stratification. Indeed, we estimated that the thermal diffusion time scale $\tau_{\kappa}^{*} = \kappa^{-1} / \left(k_r^2 + k_{\theta}^2\right)$  associated with the observed unstable modes is about one order of magnitude smaller than the buoyancy time scale $\tau_{B}^{*} = \sqrt{k_r^2 + k_{\theta}^2} N^{-1} / k_{\theta}$, which implies that thermal diffusion will play an important role in determining the amplitude of the buoyancy force (e.g., \cite{lignieres1999shear}).

To interpret the increase of the growth rate with $L_u$ (runs Q$9$ and Q$10$), we first recall that in the context of the SMRI the toroidal field is assumed to be zero. This is not the case in our simulations where, according to Eq. \ref{static_balance_eq}, the amplitude of the stationary toroidal field decreases when the initial poloidal field (and thus $L_u$) increases. As reviewed in \cite{rudiger2018stability}, introducing a toroidal field leads to the so-called helical MRI (HMRI) \citep{hollerbach2005new}. Following the dispersion relation derived by \cite{kirillov2014local}, the toroidal field can be either stabilising or destabilising depending on the sign of $V_{A_{\phi}} / s =  B_{\phi} / \left( r \sin{\theta} \sqrt{\mu_0 \rho_0} \right)$. Moreover, the HMRI differs from the SMRI through the non-vanishing phase velocity of the unstable modes and a different phase shift between the perturbed fields. This phenomenon has been mentioned for the first time by \cite{knobloch1992stability} and then found experimentally by \cite{stefani2006experimental} 
and in numerical simulations \citep{petitdemange2013axisymmetric}.

\begin{table*}[!h]
\begin{center}
\begin{tabular}{c|c|c|c|c|c|c|c|c}
\hline
\hline
 \hspace*{-0.2cm} \multirow{3}{*}{\textbf{Case}}& \multirow{3}{*}{\textbf{Contraction in induction equation}}&\multirow{3}{*}{$\boldsymbol{Re_c}$}  & \multirow{3}{*}{$\boldsymbol{L_u} $}  &  \multirow{3}{*}{$\boldsymbol{\displaystyle \frac{V_{1 \hspace*{0.05cm} {\text{phase}}}}{V_{f_1}}} $} & \multirow{3}{*}{$ \boldsymbol{\displaystyle \frac{V_{2 \hspace*{0.05cm} {\text{phase}}}}{V_{f_2}}} $} &  \multirow{3}{*}{$\boldsymbol{\displaystyle \frac{V_{3 \hspace*{0.05cm} {\text{phase}}}}{V_{f_3}}} $} & \multirow{3}{*}{$ \boldsymbol{ \overline{\displaystyle \frac{V_{\text{phase}}}{V_f}}}$}& \multirow{3}{*}{$ \boldsymbol{ \overline{ \displaystyle \frac{V_{B_{\phi}}}{V_f} }}$}  \\ & & & & & & &\\ & & & & & & & \\
\hline
\hline
Q$3$ &No& $5$ & $5\cdot 10^3$ & $0.28$ & $0.33$ & $0.35$ & $0.32$ &$-0.68$\\
Q$4$ &No& $5$ & $10^4$ & $0.07$ & $0.09$ & $0.11$ & $0.09$ &$-0.91$\\
Q$9$ &Yes& $1$ & $5\cdot 10^3$ & $1.78$ & $2.12 $ & $2.49$ & $2.13$ &$1.13$\\
Q$10$ &Yes& $1$ & $10^4$ & $1.52$ & $1.77$ & $2.20$ & $1.83$ &$0.83$\\
Q$11$ &Yes& $2$ & $10^4$ & $2.00$ & $2.00$ & $2.34$& $2.11$ &$1.11$\\
\hline
\end{tabular}
\vspace*{0.2cm}
\caption{Ratio of the phase velocity $V_{ \hspace*{0.05cm} {\text{phase}}_{\hspace*{0.05cm} i=1, 2, 3} }$ and the local contraction velocity field $V_{f_{\hspace*{0.05cm} i=1, 2, 3}}$ estimated at a given control point as explained in Table \ref{growth_rates}. An arithmetic mean is obtained from these ratios and denoted by $\overline{V_{\text{phase}} / V_f}$. We also have listed the arithmetic mean of the ratio between the velocity field caused by the addition of the toroidal field and the contraction velocity field. The parameters $E$, $P_r \left(N_0/\Omega_0\right)^2$ and $P_m$ are respectively $10^{-4}$, $10^4$ and $10^2$.}
\label{phase_velocities}
\end{center}
\end{table*}

In our simulations the quantity $V_{A_{\phi}} / s$ changes sign in the region where the instability is triggered so that the WKB-type analysis of \cite{kirillov2014local} is not directly conclusive. Our results nevertheless would imply that the toroidal field has a stabilising effect since the growth rate increases when the intensity of $B_{\phi}$ decreases. As noted earlier, we did observe that the perturbations propagate in our simulations. Their phase velocity $V_{\hspace*{0.05cm} \text{phase}}$ normalised by the contraction velocity field are listed in Table \ref{phase_velocities}. As expected, comparing the runs  Q$10$ and Q$9$, the phase velocity is higher when the toroidal field is higher. 

\begin{figure}[!h]
\begin{center}
\hspace*{-0.25cm}
\includegraphics[width=8.5cm]{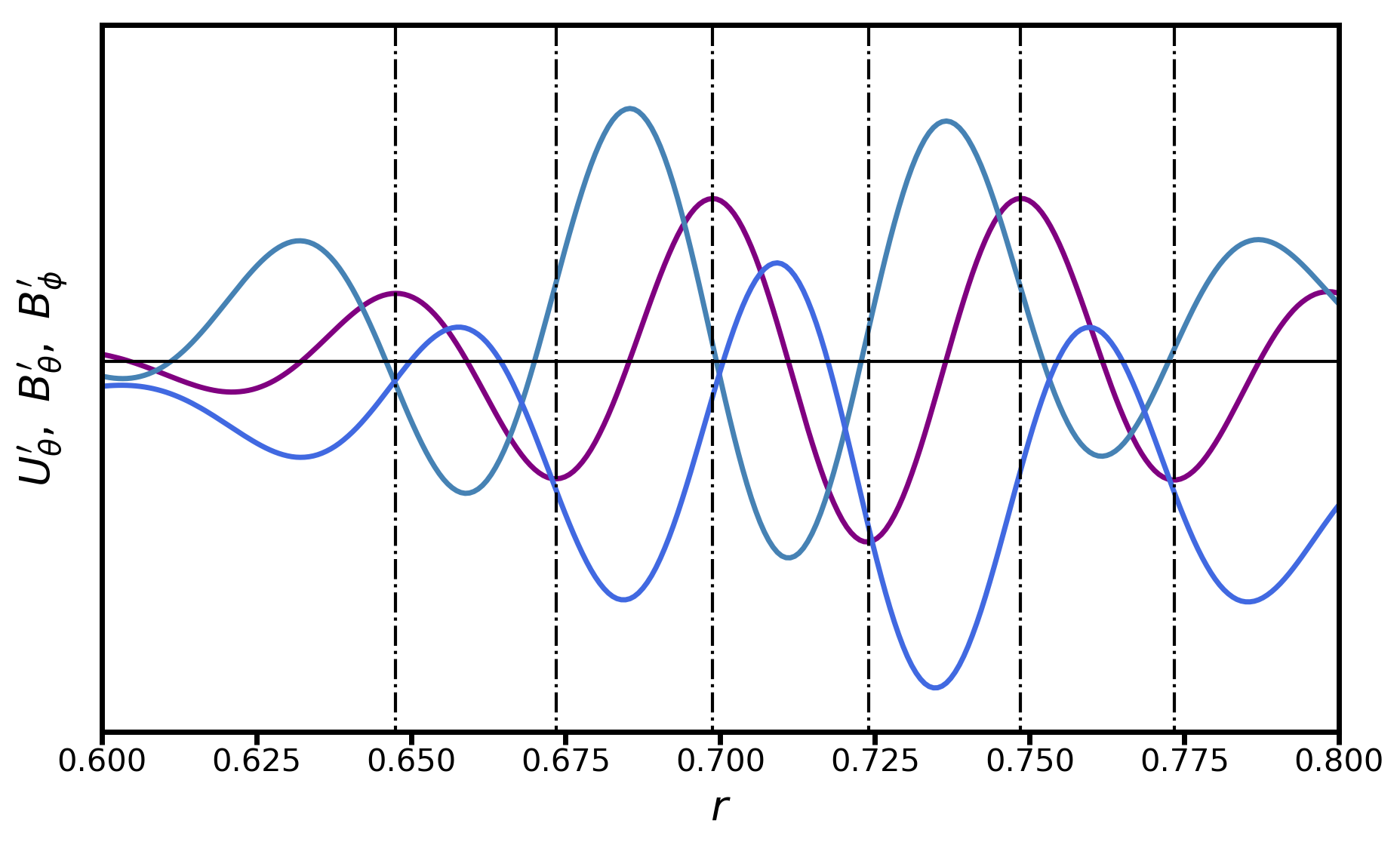}
\caption{Perturbed fields: latitudinal velocity field (purple), latitudinal (green) and azimuthal (blue) magnetic fields as a function of radius at $\theta \approx 2 \pi /5$ (i.e. at latitude $\pi/10$). The parameters are $Re_c = 1$, $L_u = 10^4$, $E=10^{-4}$, $P_r \left(N_0/\Omega_0\right)^2 = 10^4$ and $P_m = 10^2$ (run Q$10$ of Table \ref{parameters_viscous}).}
\label{phase_shift_perturbations}
\end{center}
\end{figure}

\noindent Moreover, contrary to a SMRI, we observe neither exact phase quadrature between the latitudinal velocity perturbation and the latitudinal and azimuthal magnetic perturbations, nor exact opposition phase between the perturbed latitudinal and toroidal magnetic fields \citep{petitdemange2013axisymmetric}. This is visible in Fig. \ref{phase_shift_perturbations} where the different perturbed quantities, the latitudinal velocity (purple) as well as the latitudinal (green) and azimuthal (blue) magnetic fields are plotted as a function of the radius.

We conclude that the instability triggered in the DZ of the quadrupole is of the MRI-type and that its full description would require a detailed modelling of the effects of the stable stratification and of the toroidal field. 

\subsubsection{Non-linear evolution}
\label{nonlinear_phase}

\begin{figure}[!h]
\begin{center}
\includegraphics[width=4.4cm]{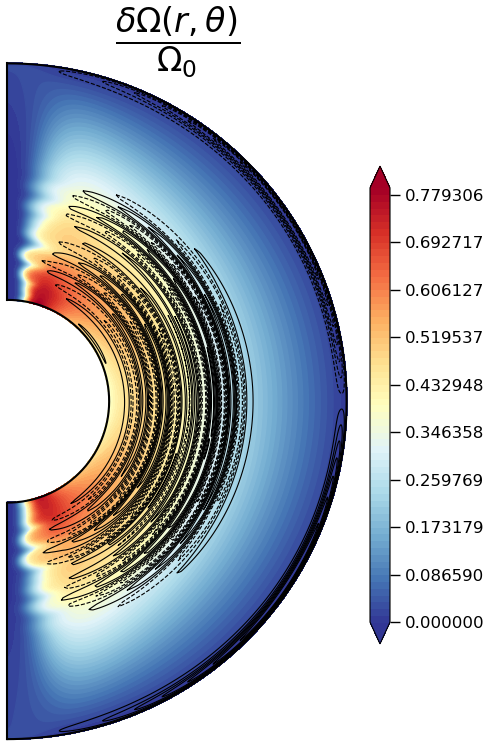}
\includegraphics[width=4.4cm]{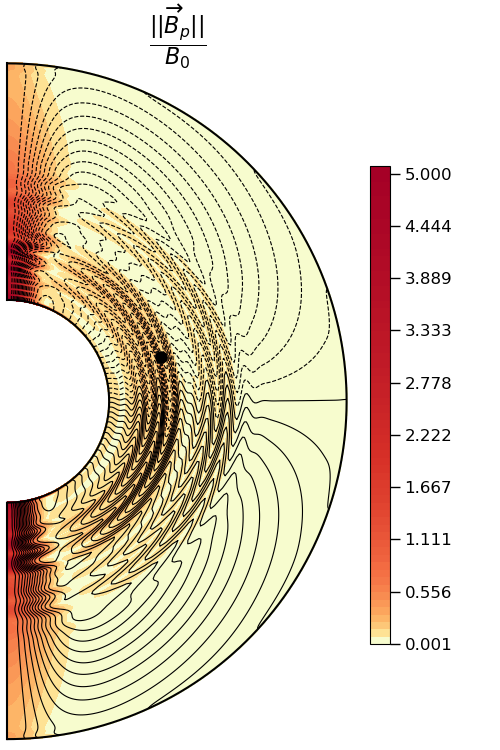}
\caption{Left panel: meridional cut of the normalised differential rotation $\delta \Omega / \Omega_0$ (colour) with the streamlines associated with the meridional circulation $\protect\vv{U} = U_r \protect\vv{e}_r + U_{\theta} \protect\vv{e}_{\theta}$ (black contours). The dashed (solid) lines correspond to an anticlockwise (clockwise) circulation. Right panel: meridional cut of the normalised norm of the poloidal magnetic field $\left \| \protect\vv{B}_p \right \| / B_0$ (colour) and its associated field lines (in black). The fixed point located at $r = 0.47 \hspace*{0.05cm} r_0$ and $\theta \approx 2 \pi / 5$ corresponds to the location where the norm of the poloidal field is plotted in Fig. \ref{bpol_evolve_unstable_vs_stable} for a stable and an unstable case. These snapshots have been taken at $t = 3 \hspace*{0.05cm} \tau_{\text{c}} $. Parameters are $Re_c = 1$, $L_u = 10^4$, $E = 10^{-4}$, $P_r \left( N_0/\Omega_0\right)^2 = 10^{4}$, $P_m = 10^{2}$ (run Q$10$ of Table \ref{parameters_viscous}).}
\label{local_to_global}
\end{center}
\end{figure}

After the exponential growth phase, the evolution becomes non-linear and the instability saturates. Figure \ref{local_to_global} displays the flow structure obtain at $\sim 3 \hspace*{0.05cm} \tau_{\text{c}}$ for the run Q$10$ ($Re_c = 1$, $L_u=10^4$), through the rotation rate (colour) and the meridional circulation (black) in the left panel, as well as the norm of the poloidal field (colour) and its associated field lines (black) in the right panel. We observe that the instability proceeds via a multi-cellular meridional circulation, radially confined by the stable stratification and latitudinally extended in both hemispheres. From Fig. \ref{local_to_global} we see that the poloidal field lines (right panel) are dragged around by this meridional circulation everywhere it is present (left panel), and then warped. The poloidal field thus behaves like a passive scalar advected and mixed by the multi-cellular circulation. This process creates small scales on which the magnetic diffusion can efficiently act to dissipate the poloidal field. 

\begin{figure}[!h]
\begin{center}
\hspace*{-0.5cm}
\includegraphics[width=8.5cm]{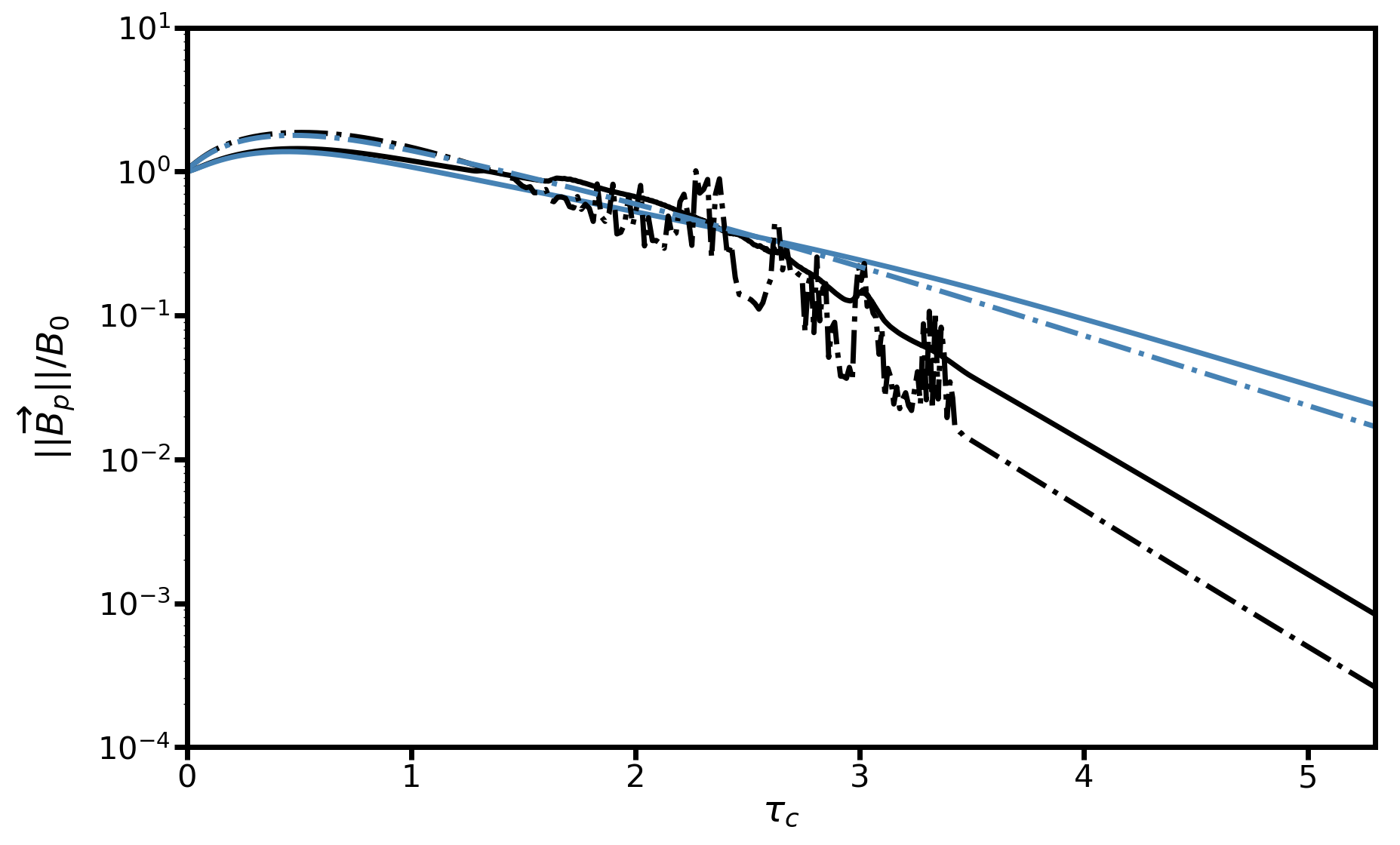}
\caption{Temporal evolution of the norm of the poloidal magnetic field normalised to $B_0$, as a function of the contraction timescale $\tau_{\text{c}}$. Plain curves represent a volume-averaged evolution and the dashed-dotted ones a local evolution at the fixed point displayed in black in Fig. \ref{local_to_global} ($r = 0.47 \hspace*{0.05cm} r_0$ and $\theta \approx 2 \pi / 5$). The stable and unstable configurations, $Re_c = 0.5$ and $1$, are respectively distinguished by their blue and black colours. The other parameters are $E = 10^{-4}$, $P_r \left( N_0/\Omega_0\right)^2 = 10^4$, $P_m = 10^2$ and $L_u = 10^4$ (runs Q$8$ and Q$10$ of Table \ref{parameters_viscous}).}
\label{bpol_evolve_unstable_vs_stable}
\end{center}
\end{figure}

In Fig. \ref{bpol_evolve_unstable_vs_stable} we compare the evolution of the norm of the poloidal field in an unstable case (in black) and in a stable case (in blue). The field norm is determined either through a volume average (solid line) or at a fixed point (displayed in black in Fig. \ref{local_to_global}) in the middle of the unstable region (dashed lines). The slope of these curves enables us to estimate a diffusion rate, and so a diffusion lengthscale, of the poloidal field. For the unstable configuration, this rate is determined during the saw-tooth evolution ranging from $\sim 2.25 \hspace*{0.05cm} \tau_{\text{c}}$ to $\sim 3.5 \hspace*{0.05cm} \tau_{\text{c}}$. By denoting $\omega_{\hspace*{0.025cm} \text{stab}} = \eta / L_{\hspace*{0.025cm} \text{stab}}^2$ and $\omega_{\hspace*{0.025cm} \text{unst}} = \eta / L_{\hspace*{0.025cm} \text{unst}}^2$ the diffusion rates of the stable and unstable runs respectively, we obtain $\omega_{\hspace*{0.025cm} \text{unst}}/\omega_{\hspace*{0.025cm} \text{stab}} \approx 42$, hence a diffusion of the poloidal field $42$ times faster in the unstable case. In other words, the motions driven by the instability induce an effective diffusion at a lengthscale $6.5$ smaller than the diffusion acting in the stable case ($L_{\hspace*{0.025cm} \text{stab}}/ L_{\hspace*{0.025cm} \text{unst}}= 6.5$).

Although we shall see below that the differential rotation increases as a result of the instability, the toroidal field does not grow through the $\Omega$-effect because the poloidal field is too weak in this phase. Then, the toroidal field experiences a diffusive-like decay, similar to the poloidal field.

\subsubsection{Post-instability description}

\begin{figure*}[!h]
\begin{center}
\includegraphics[width=4.4cm]{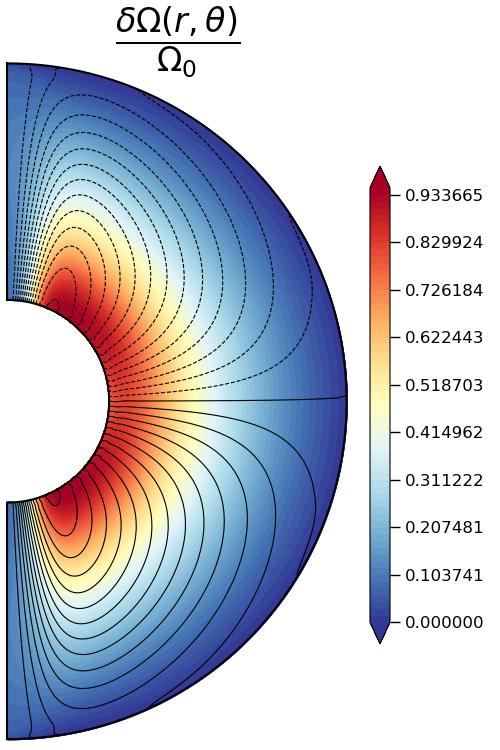}
\includegraphics[width=4.4cm]{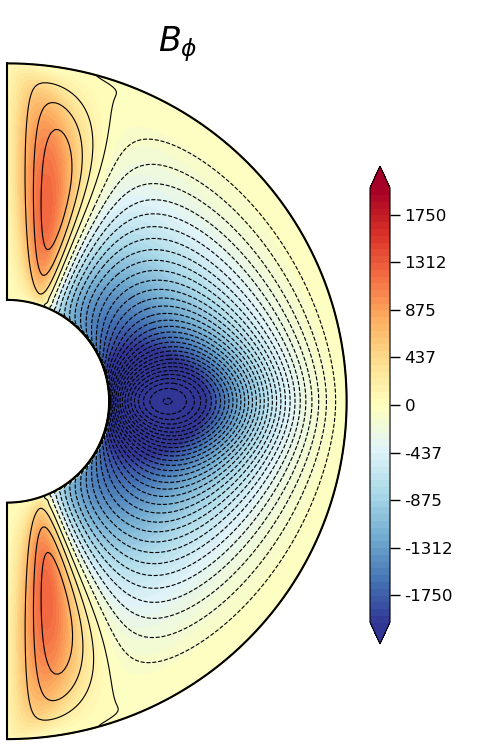}
\includegraphics[width=4.4cm]{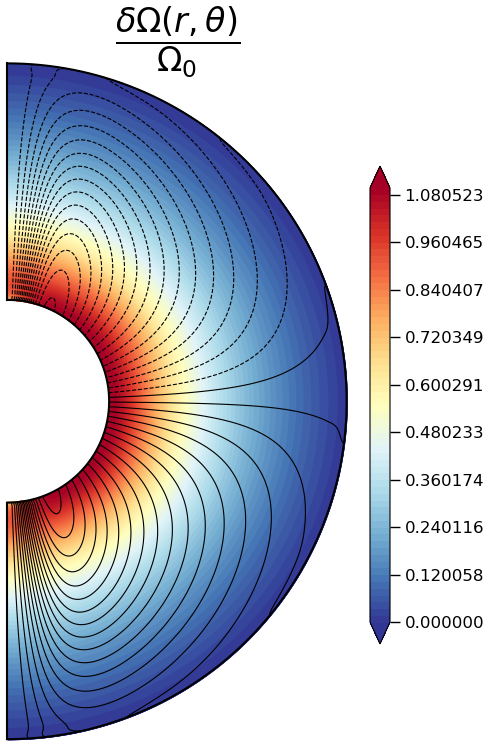}
\includegraphics[width=4.4cm]{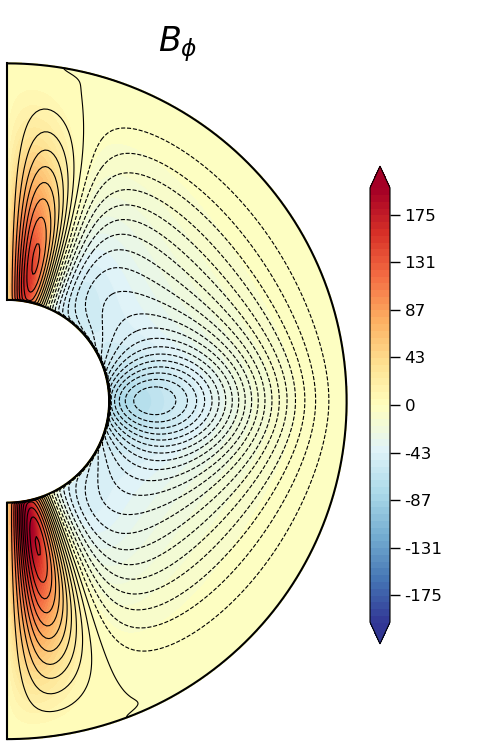}
\caption{Meridional cuts of the normalised differential rotation $\delta \Omega / \Omega_0$ and of the toroidal field $B_{\phi}$ obtained during the post-instability evolution at $\sim 3.5 \hspace*{0.05cm} \tau_{\text{c}}$ (first two panels), and when the hydrodynamic steady state is reached at $\sim 4.66 \hspace*{0.05cm} \tau_{\text{c}}$ (last two panels). Black contours represent either the poloidal field lines (first and third panels) or the streamlines associated with the electrical-current function (second and fourth panels). Parameters are $E= 10^{-4}$, $P_r \left( N_0/\Omega_0\right)^2 = 10^4$, $P_m = 10^2$, $Re_c = 1$ and $L_u = 10^4$ (run Q$10$ of Table \ref{parameters_viscous}).}
\label{post_insta}
\end{center}
\end{figure*}

By destroying the poloidal field, the instability allowed a reconfiguration of the flow structure. This is shown at $t = 3.5 \hspace*{0.05cm} \tau_{\text{c}}$ in the first two panels of Fig. \ref{post_insta} then at $t = 4.66 \hspace*{0.05cm} \tau_{\text{c}}$ in the last two. From the first panel, we observe that the maximum level of differential rotation is now three times higher than before the instability. The reason is as follows: from a DZ to another, the large-scale structure of the poloidal field has been destroyed by the instability. As a result, even if a significant level of toroidal field still exists in this region, as displayed in the second panel of Fig. \ref{post_insta}, the Lorentz force remains weak between the two DZs. The domain within which the contraction is balanced by the viscous effects thus becomes larger and, as expected, the differential rotation increases. By contrast, the poloidal field amplitude is still significant near the rotation axis and the Lorentz force imposes a very weak differential rotation $\mathcal{O} \left(\left( Re_c P_m / L_u \right)^2 \right)$ (see Sect. \ref{outside_dead_zone}) in this region.

Coming back to the numerical results presented in Fig. \ref{post_insta}. The first two panels show that the magnetic topology has also completely changed after the development of the instability. A comparison between the third panel of Fig. \ref{Numerical_Results_Viscous_Quadrupole} and the first panel of Fig. \ref{post_insta} shows that the field lines which looped back on themselves before the instability have now been moved towards the poles. In addition, the toroidal field is now very weak close to the outer sphere. An Hartmann layer is thus no longer needed to connect to the vacuum condition at the outer sphere. Likewise, the Shercliff layers have been removed with the dissipation of the poloidal field. Interestingly, we note that the new magnetic configuration, characterised by its positive lobe of toroidal field located near the rotation axis in both hemispheres, is from now on likely to be unstable to a non-axisymmetric instability of Tayler-type (see e.g., \cite{spruit1999differential}).
 
\begin{figure}[!h]
\begin{center}
\hspace*{-0.5cm}
\includegraphics[width=8.5cm]{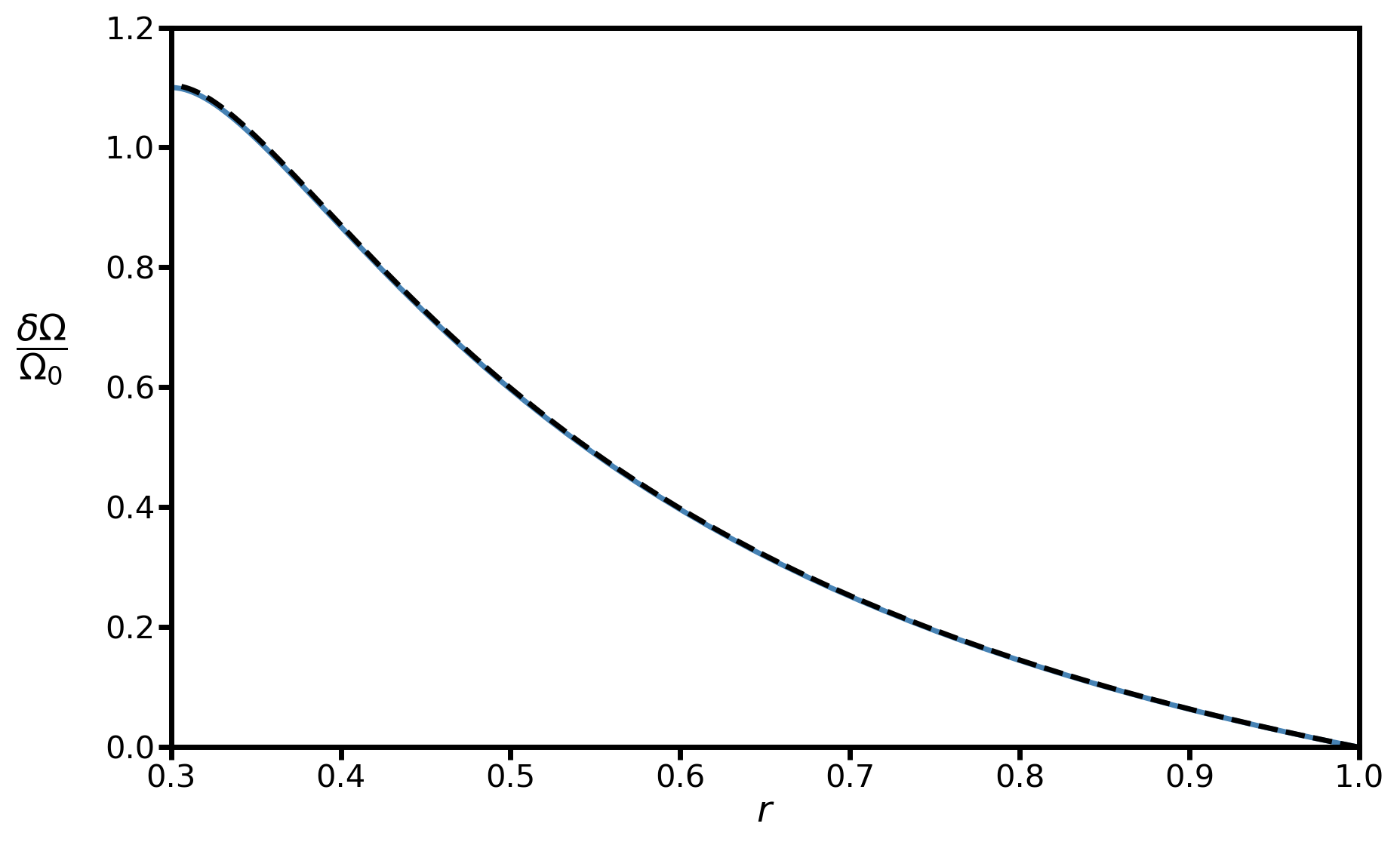}
\caption{Normalised differential rotation $\delta \Omega / \Omega_0$ as a function of radius. The analytical solution derived by solving the balance between the viscous and contraction terms in Eq. \eqref{angular_momentum_evolution}, is plotted in black dashed lines. It is compared to the numerical solution obtained at $4.66 \hspace*{0.05cm} \tau_{\text{c}}$ and plotted in blue solid line. The parameters used are the same as in Fig. \ref{post_insta}.}
\label{rot_diff_after_insta}
\end{center}
\end{figure}

After $\sim 4.66 \hspace*{0.05cm} \tau_{\text{c}}$, the third panel of Fig. \ref{post_insta} shows that the differential rotation is mostly radial and occupies the whole shell. As a consequence, its amplitude further increased. In Fig. \ref{rot_diff_after_insta} we plotted in black dashed line the analytical solution corresponding to the balance on the full sphere between the viscous and contraction terms of Eq. \eqref{angular_momentum_evolution}. This solution, derived in \cite{gouhier2020axisymmetric}, perfectly matches the numerical solution in blue, thus showing that the hydrodynamic steady state is recovered. In conclusion, the magnetic field now has a negligible effect on the flow dynamics as supported by the fourth panel of Fig. \ref{post_insta} where we can see that the amplitude of the toroidal field has been divided by more than $10$. 

\subsection{Steady state in the Eddington-Sweet regime}
\label{edd_dipole}

We now focus on the Eddington-Sweet regime ($\sqrt{E} \ll P_r \left(N_0/\Omega_0\right)^2 \ll 1$) considering a dipolar or a quadrupolar field as the pre-existing field. All simulations are performed in the anelastic approximation and include the contraction term in the induction equation. The contrast of density between the inner and the outer spheres is fixed to $20.85$, and the Ekman and magnetic Prandtl numbers are respectively equal to $10^{-5}$ and $10^2$. Our parametric study in this regime study consists in varying $Re_c$ from $10^{-1}$ to $5$ and $L_u$ from $5 \cdot 10^3$ to $10^5$ for $P_r \left(N_0/\Omega_0\right)^2 = 10^{-2}$ and $10^{-1}$ (see details in Table \ref{parameters_viscous}). 

We basically found three types of steady states: the first one is characterised, as in the viscous case, by two magnetically decoupled regions, one of them including a DZ where the contraction enforces a certain level of differential rotation. This state is the most relevant in a stellar context because it is obtained for the lowest values of $\tau_{\text{A}_{\text{p}}} / \tau_{\text{ED}}$ and $\tau_{\text{A}_{\text{p}}} / \tau_{\text{c}}$ and these ratios are a priori very small in magnetic contracting stars as discussed in Sect. \ref{stellar_context}. As we increase these ratios, we find another type of solution where the differential rotation and the meridional circulation are no longer confined within the DZ while the field topology is unchanged (for higher $\tau_{\text{A}_{\text{p}}} / \tau_{\text{ED}}$) and finally a state where the advection of the poloidal field destroys the dead zone and significantly reconfigures the magnetic field and the rotation profile (for higher $\tau_{\text{A}_{\text{p}}} / \tau_{\text{c}}$). In the following, these last two solutions will be described briefly as they are thought to be less relevant for our purpose, although physically interesting.

\subsubsection{Meridional circulation and differential rotation confined to the dead zone}
\label{relevant_Edd_regime}

\begin{figure*}[!h]
\begin{center}
\includegraphics[width=4.4cm]{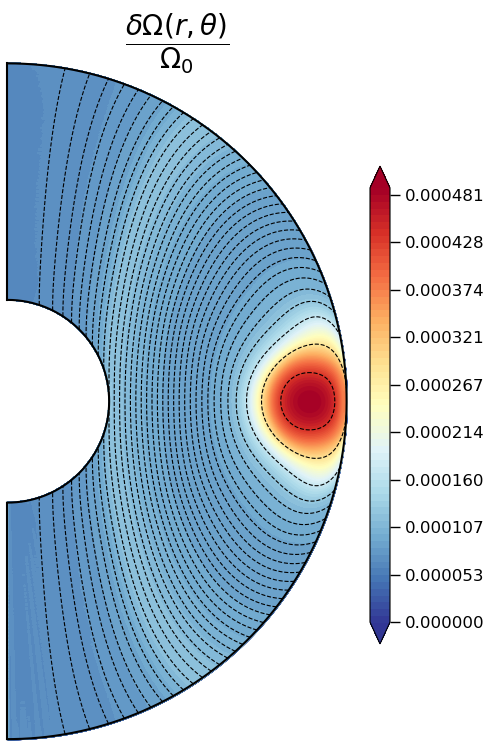}
\includegraphics[width=4.4cm]{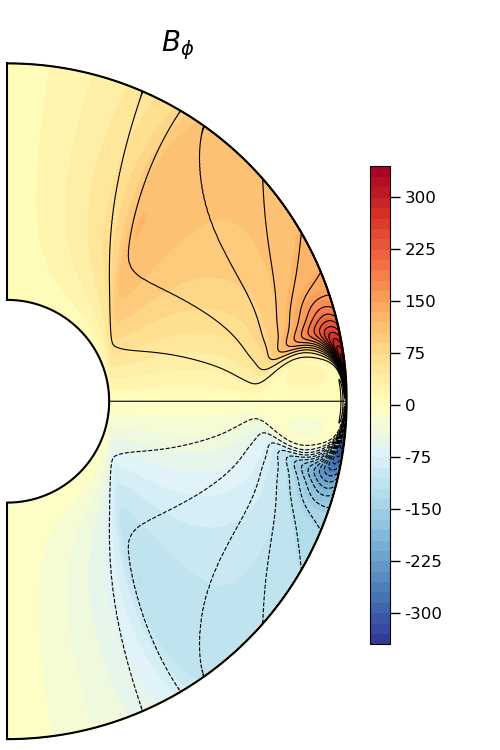}
\includegraphics[width=4.4cm]{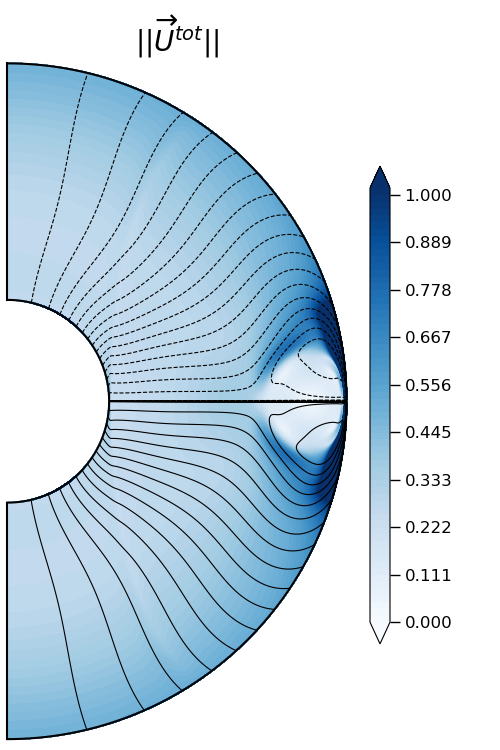}
\includegraphics[width=4.4cm]{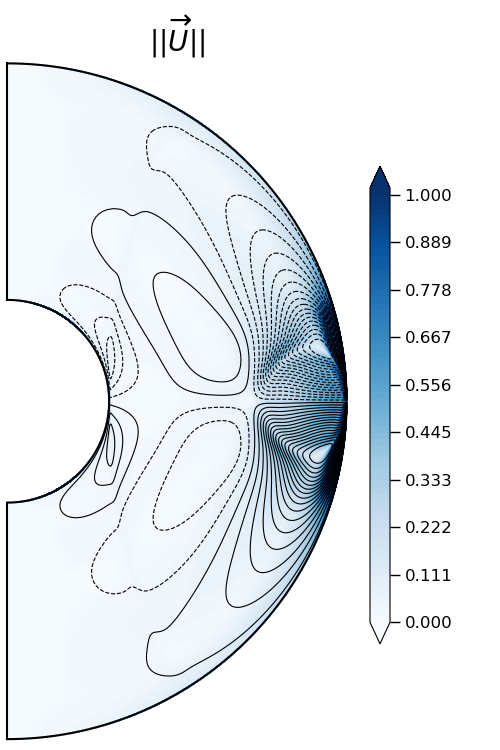}
\includegraphics[width=4.4cm]{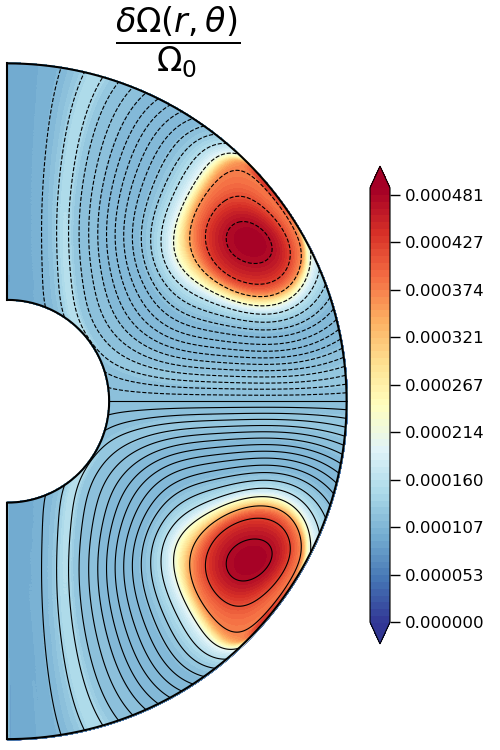}
\includegraphics[width=4.4cm]{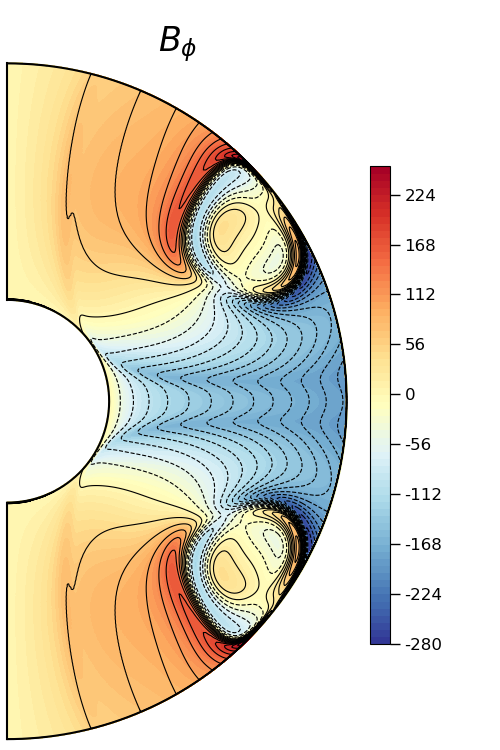}
\includegraphics[width=4.4cm]{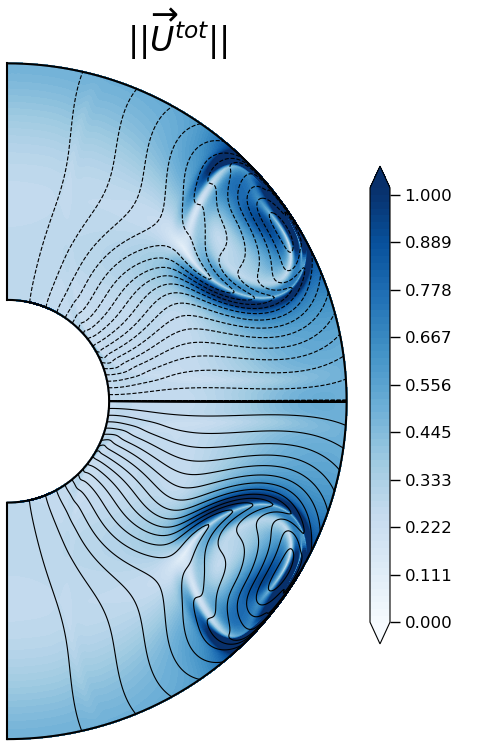}
\includegraphics[width=4.4cm]{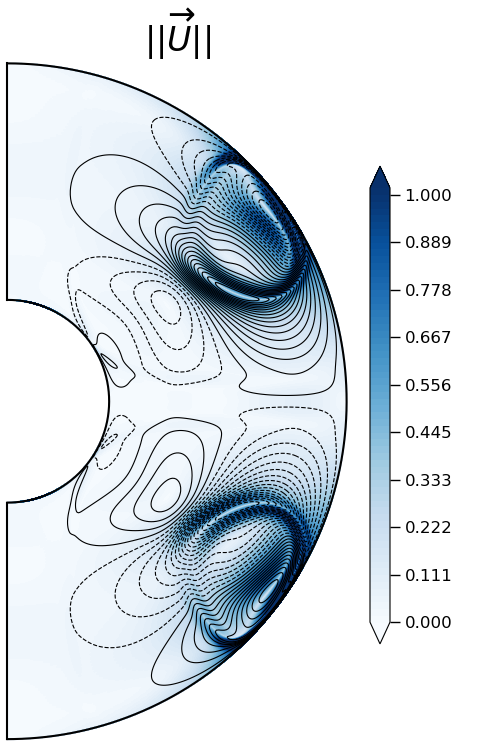}
\caption{Quasi-steady axisymmetric flow in the Eddington-Sweet regime when a dipolar (top row) or quadrupolar (bottom row) field is initially imposed. Panels of the first column: rotation rate normalised to the top value (colour) and poloidal field lines (black). Panels of the second column: toroidal field (colour) with the streamlines of the electrical-current function (black). Panels of the third column: norm (colour) and streamlines (black) of the total meridional circulation $\protect\vv{U}{_{\hspace*{-0.075cm} p}^{\text{tot}}} = \left(U_r + V_f\right) \protect\vv{e}_r + U_{\theta} \protect\vv{e}_{\theta}$. Panels of the fourth column: norm (colour) and streamlines (black) of the contraction-induced meridional circulation $\protect\vv{U}_{\hspace*{-0.05cm} p} = U_r  \protect\vv{e}_r + U_{\theta} \protect\vv{e}_{\theta}$. The dashed lines represent an anticlockwise electrical (panels of the second column) or fluid (panels of the third and fourth columns) circulation while the solid lines correspond to a clockwise direction. Parameters are $E= 10^{-5}$, $P_r \left( N_0/\Omega_0\right)^2 = 10^{-1}$, $P_m = 10^2$, $Re_c = 1$, $L_u = 10^5$ and $\rho_i/\rho_0 = 20.85$ (runs D$28$ and Q$13$ of Table \ref{parameters_viscous}).}
\label{typical_states_Edd}
\end{center}
\end{figure*}

Figure \ref{typical_states_Edd} displays the typical structure of the quasi-steady flows and fields obtained for a dipolar (top row) or a quadrupolar (bottom row) initial field. These simulations were performed at $Re_c = 1$, $L_u = 10^5$ and $P_r \left(N_0/\Omega_0\right)^2 = 10^{-1}$ (runs D$28$ and Q$13$ of Table \ref{parameters_viscous}) and thus satisfy $\tau_{\text{A}_{\text{p}}} / \tau_{\text{ED}} = 10^{-2}$ and $\tau_{\text{A}_{\text{p}}} /\tau_{\text{c}}=10^{-3}$. There are many similarities with the viscous case. From the panels of the first column, we again observe two regions that are magnetically decoupled. One occupies the major part of the spherical shell and is in quasi-solid rotation while the other, the DZ, exhibits a certain level of differential rotation. The amplitude of this differential rotation is similar for the dipolar and quadrupolar cases. We also note the presence of magnetic boundary layers: the Hartmann layer at the outer sphere, and the Shercliff layers wherever adjacent poloidal field lines are forced to rotate differently, namely along the tangent cylinder and around the DZ. As in the viscous case, the toroidal field is characterised by a strong amplitude at these locations, as indicated by the panels of the second column. 

We also see major differences with the viscous regime. First, although the contraction of the field lines is allowed, the DZ is confined near the outer sphere, whether a dipolar or quadrupolar field is initially imposed. This can be compared to the third panels of Figs. \ref{Numerical_Results_Viscous} and $\ref{Numerical_Results_Viscous_Quadrupole}$ in the viscous regime where the DZ was clearly advected towards the inner sphere. This difference is attributed to the effect of the contraction-induced meridional flow which now plays a significant role in the DZ advection. This flow is illustrated in the fourth column of Fig. \ref{typical_states_Edd} displaying the norm (colour) and the streamlines (black) of the poloidal velocity field $\vv{U}_{\hspace*{-0.05cm} p} = U_r \vv{e}_r + U_{\theta} \vv{e}_{\theta}$. For an initial dipolar field, this circulation is characterised by the presence of one cell of anticlockwise (clockwise) circulation in the northern (southern) hemisphere. This contraction-induced flow contributes to the total meridional circulation $\vv{U}_{\hspace*{-0.075cm} p}^{{\hspace*{0.05cm} \text{tot}}} = \left(U_r + V_f \right)\vv{e}_r + U_{\theta} \vv{e}_{\theta}$ displayed in the third column. As seen in the two top right panels, the induced flow inside the DZ tends to oppose contraction and the resulting total circulation becomes very weak, thus preventing the inward advection of the DZ. Outside the DZ, the total meridional flow is approximately parallel to the poloidal field lines close to the outer sphere, where the contraction velocity is maximum. In the deeper regions close to the inner sphere, the advection of poloidal field lines by the weaker contraction field is balanced by magnetic diffusion. 

For an initial quadrupolar field, we observe a strong circulation around the DZ while inside the DZ the flow is predominantly vertical, downwards (upwards) in the northern (southern) hemisphere. Again, away from the DZ, the contraction-induced meridional flow has only a negligible contribution to the total meridional circulation. Finally, in contrast to the viscous regime, for the parameters numerically reachable in this study, the quadrupolar configurations are stable with respect to MRI because the shear built in the DZs is not strong enough to counteract the stabilising effect of the poloidal field. By comparison, the contrast of differential rotation in run Q$5$ of the viscous regime is $\sim 130$ times larger and is not even unstable despite a weaker $L_u$ of $5\cdot10^4$.

\begin{figure}[!h]
\begin{center}
\hspace*{0.4cm}
\includegraphics[width=3.1cm]{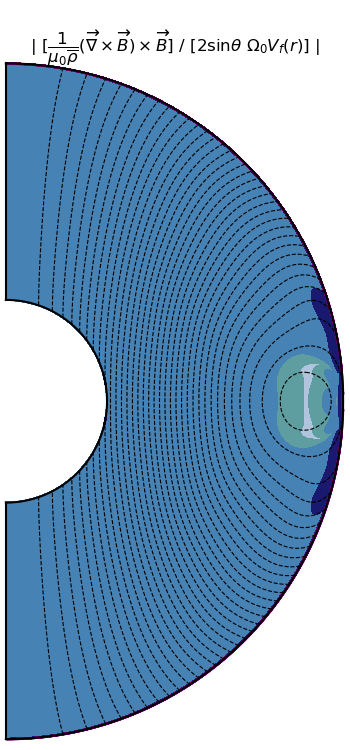}
\hspace*{0.8cm}
\includegraphics[width=4.4cm]{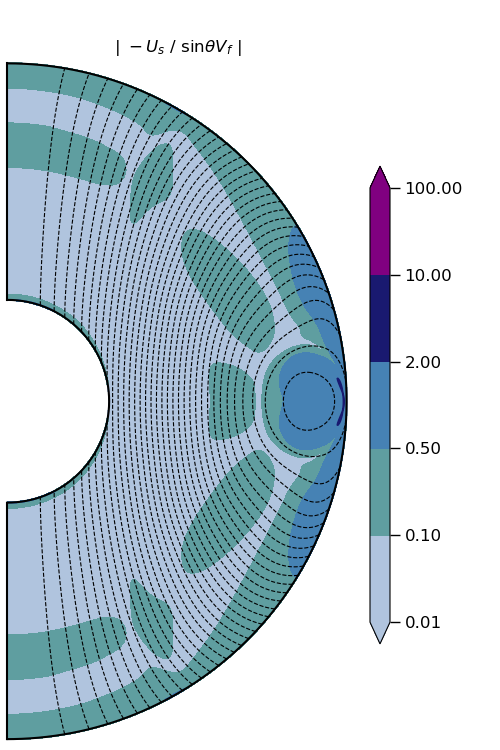}
\caption{$2$D maps comparing the relative importance of the Lorentz (left panel) and Coriolis (right panel) forces to the contraction in the presence of a dipolar field. In each panel the poloidal field lines are plotted in black. Parameters are the same as in Fig. \ref{typical_states_Edd}.}
\label{force_balances_outside_and_inside_DZ}
\end{center}
\end{figure}

In order to understand the flow dynamics inside and outside the DZ, we now examine the force balance in the AM equation Eq. \eqref{angular_momentum_evolution}, as was done in the viscous regime. The force amplitudes are analysed in the $2$D maps of Fig. \ref{force_balances_outside_and_inside_DZ} where we display the ratio of the Lorentz force (left panel) and of the Coriolis force (right panel) to the contraction. We can observe that inside the DZ, the contraction is now balanced by the Coriolis force because the toroidal component of the magnetic field tends to zero and the Lorentz force becomes negligible accordingly (see right panel of Fig. \ref{force_balances_outside_and_inside_DZ}). This implies that 

\begin{equation}
U_s = \sin{\theta} \displaystyle \frac{V_0 \rho_0 r_0^2}{\overline{\rho} r^2}
\label{coriolis_vs_contraction}.
\end{equation}

\noindent Here, contrary to the viscous case, the thermal diffusion weakens the stable stratification and enables a contraction-driven meridional circulation to exist. Inside the DZ, we also find that the thermal balance

\begin{equation}
U_r \displaystyle \frac{\text{d} \overline{S}}{\text{d} r} = \kappa \left \lbrack \left( \displaystyle \frac{\text{d} \ln{\overline{\rho}}}{\text{d} r} + \displaystyle \frac{\text{d} \ln{\overline{T}}}{\text{d} r} \right) \displaystyle \frac{\partial \delta S}{\partial r} + \vv{\nabla}^2 \delta S \right \rbrack
\label{anel_thermal_balance},
\end{equation}

\noindent and the thermal wind balance

\begin{equation}
2 r \Omega_0 \displaystyle \frac{\partial U_{\phi}}{\partial z} = \displaystyle \frac{g_0 r_0^2}{C_p r^2} \displaystyle \frac{\partial \delta S}{\partial \theta}
\label{anel_thermal_wind},
\end{equation}

\noindent are both satisfied. According to Eq. \eqref{coriolis_vs_contraction}, contraction then drives a meridional circulation $U_s \sim \mathcal{O} \left(V_f\right)$ which redistributes AM on an Eddington-Sweet timescale. Note that the circulation timescale $\tau_{\text{c}}$ can be quite different from the Eddington-Sweet timescale. Indeed, as stated in Sect. \ref{timescales_physical_processes}, the ratio $\tau_{\text{ED}} / \tau_{\text{c}}$ is measured by the dimensionless quantity $Re_c P_r \left(N_0/\Omega_0\right)^2$. In this numerical study of the Eddington-Sweet regime, $\tau_{\text{ED}} \ll \tau_{\text{c}}$ because $ P_r \left(N_0/\Omega_0\right)^2 \ll 1$ and because large contraction Reynolds numbers are too difficult to reach numerically.

Outside the DZ, the timescale of AM transport by the Alfvén waves is much shorter than the Eddington-Sweet timescale, and the Alfvén waves impose their dynamics. The left panel of Fig. \ref{force_balances_outside_and_inside_DZ} thus shows that the Lorentz force balances contraction and Eq. \eqref{static_balance_eq} holds, as in the viscous regime. In this case, a quasi-isorotation state along the field lines is obtained, verifying:

\begin{equation}
V_f(r) \left \lbrack \displaystyle \frac{\partial}{\partial r} \left( \displaystyle \frac{B_{\phi}}{r} \right) - \displaystyle \frac{B_{\phi}}{r} \displaystyle \frac{\text{d} \ln{\overline{\rho}}}{\text{d}r} \right \rbrack = \sin{\theta} \left( \vv{B}_p \cdot \vv{\nabla} \right) \delta \Omega
\label{quasi_ferraro_anel}.
\end{equation}

\noindent As a result, the estimate of the characteristic amplitude of the differential rotation along the field lines Eq. \eqref{estimate_diff_rot_ferraro} still holds, except that it must be weighted by $\left( \displaystyle \int_{r_i/r_0}^{1} \displaystyle \frac{\overline{\rho}}{\rho_0} \hspace*{0.05cm} \text{d} \left( r/r_0\right) \right)^{-1}$ 
accounting for the effect of the density stratification in the domain.

In the hydrodynamical case \citep{gouhier2020axisymmetric} we showed that the characteristic amplitude of the steady differential rotation resulting from the balance between the inward AM transport by the contraction and the AM redistribution by the Eddington-Sweet circulation should be $\mathcal{O} \left( P_r \left(N_0/\Omega_0\right)^2 Re_c \left( \int_{r_i/r_0}^{1} \left( \overline{\rho} / \rho_0 \right) \text{d} \tilde{r} \right)^{-1} \right) $. This global analysis does not apply directly in the present situation where the DZ is reduced to a small fraction of the spherical shell, confined near the outer sphere. As in Sect. \ref{DZ_description} and following \cite{oglethorpe2013spin}, to account for the DZ size and its effect on the differential rotation induced by the Eddington-Sweet circulation we introduce the lengthscale $L_{\hspace*{0.025cm} \text{DZ}} = 0.1 \hspace*{0.05cm} r_0$. Because of the density stratification, the contraction velocity is not very different between the outer and inner spheres. Thus, after using Eq. \eqref{coriolis_vs_contraction} and the continuity equation we have $U_r \approx V_0$. Then from Eq. \eqref{anel_thermal_balance} we get $\delta S \approx \left(V_0 L_{\hspace*{0.025cm} \text{DZ}}^2 / \kappa \right) \left( \text{d} \overline{S} / \text{d} r \right)$. Injecting this estimate in Eq. \eqref{anel_thermal_wind} yields finally to $\Delta \Omega_{\text{DZ}} / \Omega_0 \approx \left(g(r) / C_p\right) \left( \text{d} \overline{S} / \text{d} r \right) \left(V_0 r_0 / \kappa \Omega_0^2\right)  \left(L_{\hspace*{0.025cm} \text{DZ}} / r_0\right)^2 \approx P_r \left(N_0/\Omega_0\right)^2 Re_c \left(L_{\hspace*{0.025cm} \text{DZ}} / r_0\right)^2$, thus enabling us to recover the level of differential rotation inside the DZ up to a factor two. 

\begin{figure}[!h]
\begin{center}
\hspace*{-0.5cm}
\includegraphics[width=8.5cm]{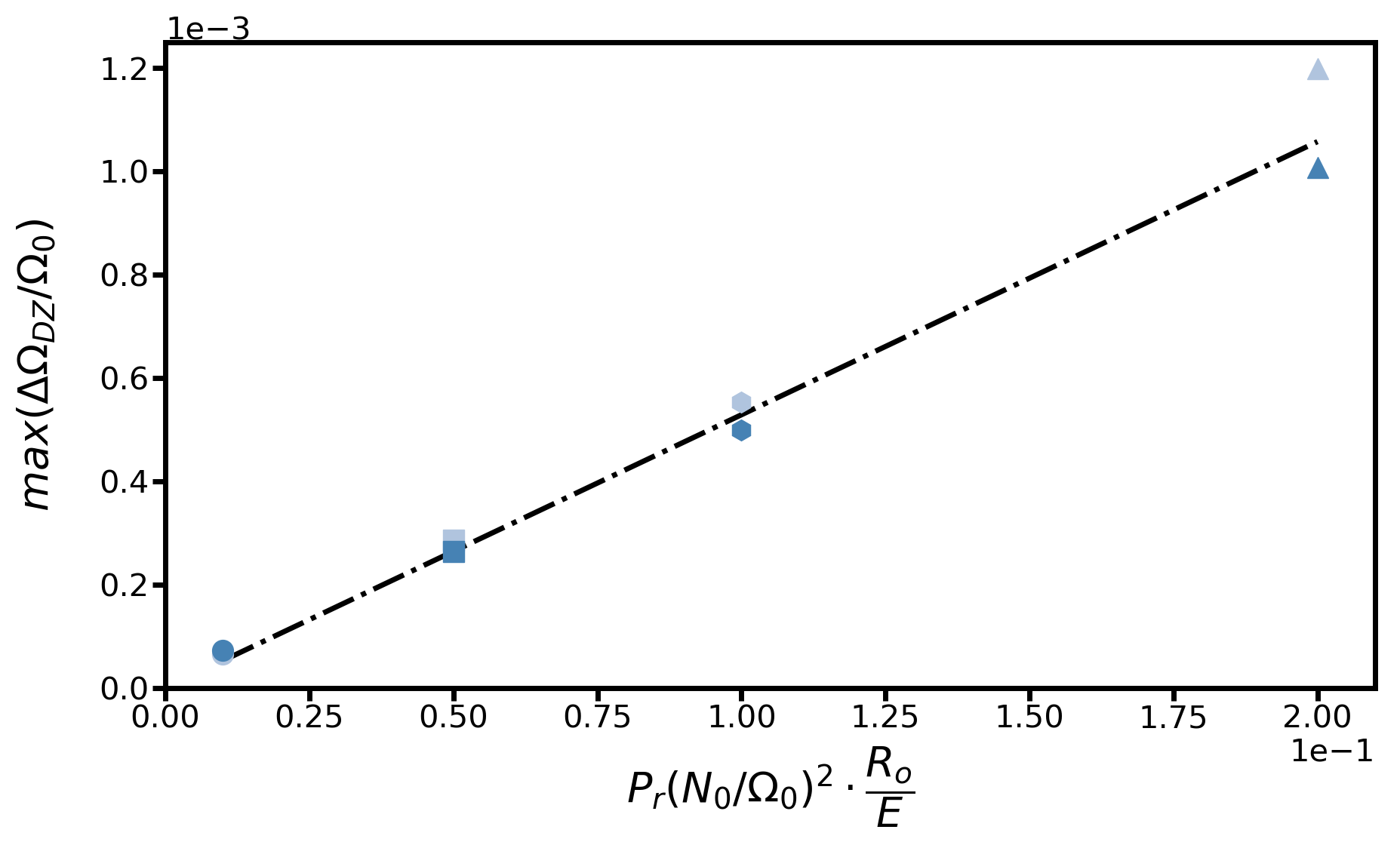}
\caption{Maximum contrast of differential rotation inside the DZ as a function of $P_r \left(N_0/\Omega_0\right)^2 Re_c$. The different symbols circle, square, hexagon and triangle respectively correspond to the simulations performed at $Re_c = 10^{-1}$, $5\cdot10^{-1}$, $1$ and $2$. The runs carried out at $L_u = 5\cdot10^4$ are presented in light blue and those at $L_u = 10^5$ in blue. The other parameters are $P_r \left(N_0/\Omega_0\right)=10^{-1}$, $E = 10^{-5}$, $P_m = 10^{2}$ and $\rho_i/\rho_0 = 20.85$ (runs D$21$-$24$ and D$27$-$30$ of Table \ref{parameters_viscous}).}
\label{rot_diff_scaling_dz_edd}
\end{center}
\end{figure}

In Fig. \ref{rot_diff_scaling_dz_edd} we plotted the maximum amplitude of the differential rotation inside the DZ as a function of $P_r \left( N_0/\Omega_0\right)^2 Re_c$ for the runs D$21$-$24$ and D$27$-$30$ of Table \ref{parameters_viscous} performed with $Re_c$ ranging from $10^{-1}$ to $2$ (identified with symbols) and $L_u$ from $5\cdot10^4$ (light blue) to $10^5$ (blue). The other parameters are fixed to $P_r \left(N_0/\Omega_0\right)^2 = 10^{-1}$, $E = 10^{-5}$, $P_m = 10^2$ and $\rho_i/\rho_0 = 20.85$. As expected, the maximum contrast of differential rotation follows a linear relation with $Re_c$. Moreover, we also observe that this level is almost independent of the Lundquist number, consistent with the balance Eq. \eqref{coriolis_vs_contraction} inside the DZ. However, for the highest contraction Reynolds number $Re_c=2$, we observe a clear discrepancy between the runs performed at $L_u = 5\cdot10^4$ and $L_u = 10^5$.

\begin{figure}[!h]
\begin{center}
\includegraphics[width=4.4cm]{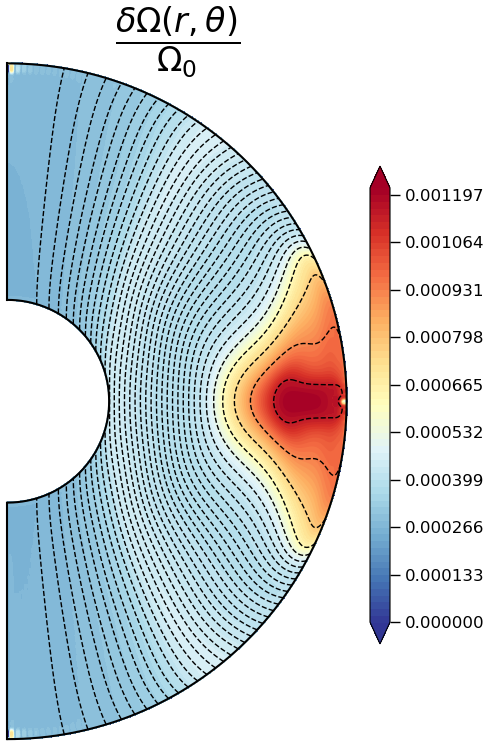}
\includegraphics[width=4.4cm]{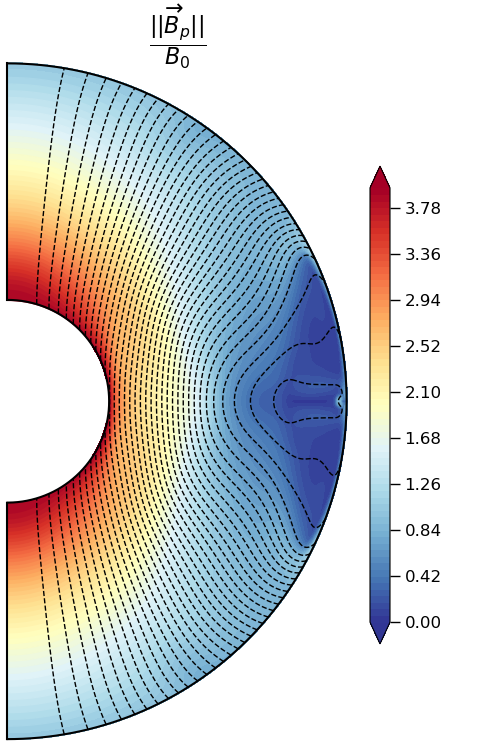}
\caption{Meridional cuts of the normalised differential rotation $\delta \Omega / \Omega_0$ (left panel) and of the normalised norm of the poloidal field $\left \| \protect\vv{B}_p \right \| / B_0$ (right panel). In black are also plotted the poloidal field lines to highlight the DZ. The parameters are $Re_c = 2$, $L_u = 5 \cdot 10^4$, $P_r \left(N_0/\Omega_0\right) = 10^{-1}$, $E=10^{-5}$, $P_m = 10^{2}$ and $\rho_i/\rho_0 = 20.85$ (run D$38$ of Table \ref{parameters_viscous}).}
\label{high_rec_structure}
\end{center}
\end{figure}

\noindent This deviation can be attributed to the fact that, in the lower $L_u$ case, the magnetic tension no longer prevents the advection of the magnetic field by the meridional flows. As shown in Fig. \ref{high_rec_structure}, this produces a significant deformation of the DZ geometry and a related expulsion of the magnetic flux outside the DZ. This phenomenon is discussed in more details below.

\subsubsection{Meridional circulation and differential rotation not confined to a dead zone}
\label{irrelevant_Edd_regime}

Simulations carried out at a smaller $P_r \left(N_0/\Omega_0\right)^2$ parameter (runs D$32$-D$39$ of Table \ref{parameters_viscous}) or at higher $Re_c$ (runs D$38$, Q$14$, Q$16$ and Q$17$ of Table \ref{parameters_viscous}), that is at higher values of $\tau_{\text{A}_{\text{p}}} / \tau_{\text{ED}}$ and $\tau_{\text{A}_{\text{p}}} / \tau_{\text{c}}$, exhibit different features. In the former case (smaller $P_r \left(N_0/\Omega_0\right)^2$), the differential rotation is no longer confined to the DZ as an AM redistribution by the Eddington-Sweet circulation occurs outside the DZ. In the second case (higher $Re_c$), the amplitude of the meridional circulation is strong enough to warp the DZ and expel the magnetic flux there. These two phenomena are now discussed.

\begin{figure}[!h]
\begin{center}
\includegraphics[width=4.4cm]{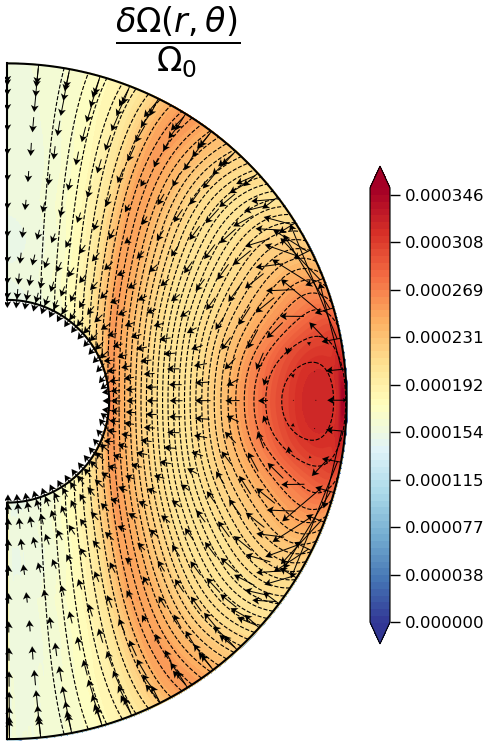}
\includegraphics[width=4.4cm]{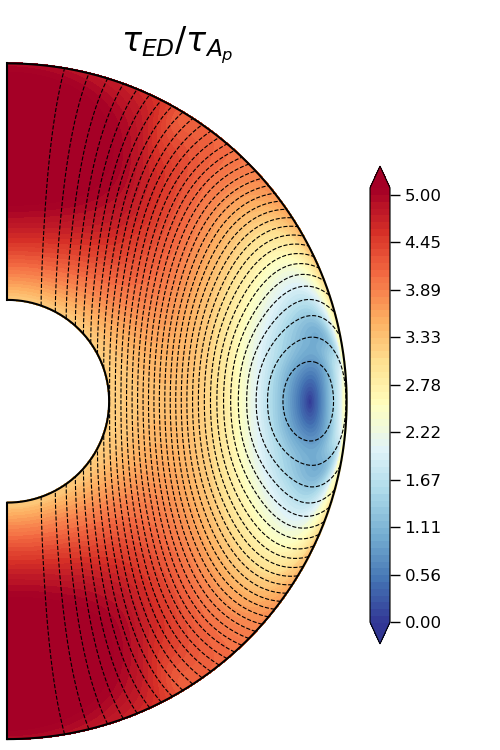}
\caption{Meridional cut of the normalised differential rotation $\delta \Omega / \Omega_0$ (left panel) and $2$D map comparing the Eddington-Sweet timescale to the Alfvén timescale (right panel). This one is locally estimated such as $\tau_{\text{A}_\text{p}} = r_0 \sqrt{\mu_0 \overline{\rho}} / \left \| \protect\vv{B}_p \right \|$. In these two panels, the poloidal field lines are also plotted in black. In addition, in the left panel, the vector lines of the total meridional velocity field $\protect\vv{U}{_{\hspace*{-0.075cm} p}^{\text{tot}}} = \left(U_r + V_f\right) \protect\vv{e}_r + U_{\theta} \protect\vv{e}_{\theta}$ are plotted as black arrows. The parameters are $Re_c = 1$, $L_u = 5 \cdot 10^4$, $P_r \left(N_0/\Omega_0\right) = 10^{-2}$, $E=10^{-5}$, $P_m = 10^{2}$ and $\rho_i/\rho_0 = 20.85$ (run D$36$ of Table \ref{parameters_viscous}).}
\label{edd_at_prn2o2_1d-2}
\end{center}
\end{figure}

The left panel of Fig. \ref{edd_at_prn2o2_1d-2} displays a meridional cut of the quasi-steady differential rotation (in colour) obtained for $Re_c = 1$, $L_u = 5 \cdot 10^4$ and $P_r \left(N_0/\Omega_0\right)=10^{-2}$ (run D$36$ of Table \ref{parameters_viscous}) on which are represented the vector lines of the total meridional velocity field as arrows and the poloidal field lines (in black). Compared to the simulation shown in Fig. \ref{typical_states_Edd}, the ratio  $\tau_{\text{A}_{\text{p}}} / \tau_{\text{ED}}$ has been increased by a factor $20$ (from $10^{-2}$ to $2 \cdot 10^{-1}$). Actually, if local values of this ratio are considered, a value of order $1$ can be reached, in particular in the vicinity of the DZ. This is shown in the right panel of Fig. \ref{edd_at_prn2o2_1d-2} that displays the distribution of the ratio of the Eddington-Sweet time to the local Alfvén time. In this regime, the differential rotation and the meridional circulation have spread out away from the DZ while the poloidal field lines, and thus the DZ, have not been affected by the circulation flow.

\begin{figure}[!h]
\begin{center}
\includegraphics[width=4.4cm]{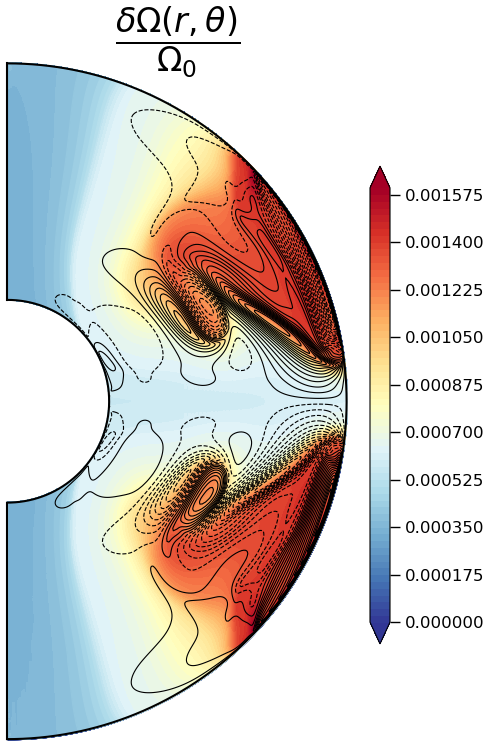}
\includegraphics[width=4.4cm]{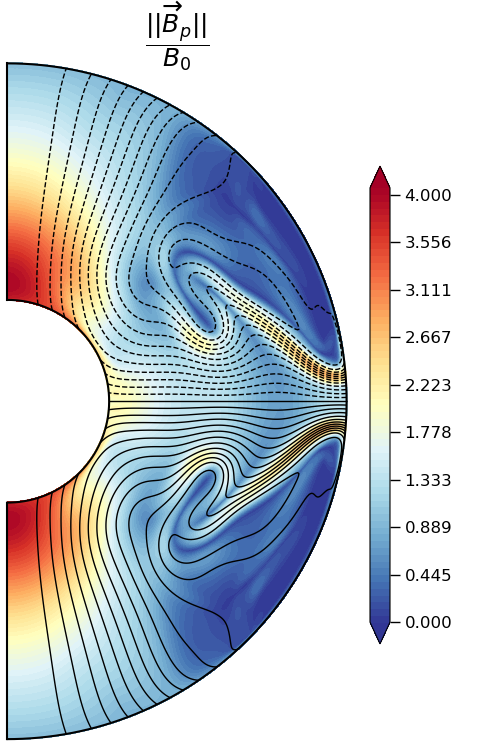}
\caption{Left panel: structure of the flow after $\sim 2 \hspace*{0.05cm} \tau_{\text{ED}}$ displayed through the coloured contours of the differential rotation normalised to the top value and the streamlines of the meridional flow in black. Right panel: norm of the poloidal field normalised to its initial value at the pole of the outer sphere in colour with the poloidal field lines in black. The parameters are $E=10^{-5}$, $P_r \left(N_0/\Omega_0\right)^2 = 10^{-1}$, $\rho_i/\rho_0 = 20.85$, $P_m = 10^2$, $Re_c = 3$ and $L_u = 5\cdot10^4$ (run Q$16$ of Table \ref{parameters_viscous}).}
\label{Magnetic_Flux_Expulsion_2_tauED}
\end{center}
\end{figure}

Another regime is encountered at sufficiently high $Re_c$, as presented in Fig. \ref{Magnetic_Flux_Expulsion_2_tauED}. The left panel of this figure displays the differential rotation in colour with the streamlines of the meridional circulation in black. The right panel shows, in colour, the norm of the poloidal field with the poloidal field lines in black. We observe that this meridional circulation significantly warps the DZ, thus leading to a reconfiguration of the magnetic field and the differential rotation. Compared to the simulation shown in the top panel of Fig. \ref{typical_states_Edd}, the ratio $\tau_{\text{A}_{\text{p}}} / \tau_{\text{c}}$ has been increased by a factor of $6$ (from $10^{-3}$ to $6 \cdot 10^{-3}$). This is apparently sufficient for the Lorentz force not to be able to confine the circulation in the DZ. The magnetic field is then advected and partly dissipated in the vicinity of the original DZ. The dissipation process is reminiscent of the  phenomenon of magnetic flux expulsion studied by \cite{weiss1966expulsion}, whereby an eddy advects the magnetic field to such small scales that magnetic diffusion is efficient. We indeed observe that the magnetic flux ends up being expelled from the regions where the meridional circulation exists and the poloidal field is found concentrated in free layers separating the quasi-solid rotation region from the one in differential rotation.

\section{Summary and conclusions}
\label{conclusion}

In this work we investigated the dynamics of a contracting radiative spherical layer embedded in a large-scale magnetic field. The aim was to determine the differential rotation that results from the combined effects of contraction and magnetic fields. The contraction is modelled through an imposed radial velocity field $\vv{V}_f$ and the gas dynamics is modelled using either the Boussinesq or the anelastic approximations. The parametric study has been guided by the results obtained without magnetic field \citep{gouhier2020axisymmetric} highlighting two hydrodynamical regimes, namely the viscous regime in the strongly stratified cases and the Eddington-Sweet regime in the weakly stratified cases.

We find that the contracting layer first evolves towards a quasi-steady configuration characterised by two magnetically decoupled regions. In the first region all the poloidal field lines connect to the outer sphere. The rotation is quasi-uniform in this region because the contraction only allows very small deviations from Ferraro's isorotation law along the field lines and the outer sphere rotates uniformly. The second region, called the DZ, is decoupled from the first one as the poloidal field lines loop-back on themselves or connect to the inner sphere. In addition, the poloidal field amplitude vanishes at some point within the DZ. A significant level of differential rotation can be produced in these DZs, the inward AM transport by the contraction being balanced either by a viscous transport or by an Eddington-Sweet circulation. The exact amplitude of the differential rotation also depends on the size, the shape and the location of the DZ.

In a second step, after a time of the order of the contraction time, the shear built in the DZ can trigger a powerful axisymmetric instability that profoundly modifies the subsequent evolution of the flow. Indeed, for an initial quadrupolar field in the viscous regime, we observe that if the field strength is low enough an MRI-type instability grows and produces a multi-cellular meridional circulation organised at small scales in the radial direction. This flow advects and eventually enables to efficiently dissipate the magnetic energy. The new field configuration is strongly modified, and the differential rotation which is no longer constrained to the DZ spreads to most of the spherical layer while its amplitude increases. This instability has not been observed for the quadrupolar field in the Eddington-Sweet regime because numerical limitations did not allow us to reach significant levels of differential rotation. However we anticipate that for realistic contraction Reynolds numbers and Lundquist numbers, the differential rotation in the DZ of the quadrupolar field will also trigger an instability in the Eddington-Sweet regime. By contrast, the dipolar field configuration does not lead to an instability. Indeed, in this case, the DZ is symmetric with respect to the equator and the contraction produces maximum rotation rates along the equator. The latitudinal differential rotation thus increases away from the rotation axis which implies stability with respect to the MRI. We note that the same configuration in an expanding layer would lead to minimum rotation rates along the equator and thus to differential rotations potentially unstable to MRI. 

If we intent to extrapolate to a more complex geometry of the initial poloidal field, the dipolar topology with a single equatorially symmetric DZ appears exceptional. Thus, we expect that generically negative latitudinal gradients of the rotation rate, potentially unstable to the MRI, are present in DZs. Rather than the topology of the poloidal field, what can prevent the MRI to develop is its intensity. Indeed, according to \cite{balbus1998instability}, the magnetic tension stabilises the flow if the perturbation length scales $\lambda_r$ are smaller than $\left(B_r / \Omega\right) \cdot \left(\sqrt{2} \pi  / \sqrt{ \mu_0 \hspace*{0.05cm} \overline{\rho} \left|q\right|}\right)$. Applying this criteria to the degenerate core of a typical subgiant of $1.1$  \(\textup{M}_\odot\) and $2$ \(\textup{R}_\odot\), we find that, assuming a $\mathcal{O}(1)$ shear $\left|q\right|$, a rotation rate $\Omega = 3.1 \cdot 10^{-6}$ rad $\cdot$ s$^{-1}$ and a mean core density $\overline{\rho}_{\text{c}} = 2.1 \cdot 10^5$ kg$\cdot$m$^{-3}$, 
fields higher than $3 \cdot 10^5$ G will stabilise all the perturbations smaller than the degenerate core size of  $0.06$ \(\textup{R}_\odot\). In practice the radial wavelength of the unstable modes is constrained by the stable stratification rather than by the core size and thus even lower field intensities will be stabilising. For example, in our simulations $\lambda_r$ is $\sim 44$ times smaller than the outer radius of the spherical layer. The critical field in our numerical simulations is reached for a Lorentz number $L_o = B_0 / \sqrt{\mu_0 \rho_0} r_0 \Omega_0$ equal to $\sim 10^{-2}$. For the sub-giant core rotation and density given above, this corresponds to a $\sim 10^4$ G 
critical field strength. As $P_r \left(N_0/\Omega_0\right)^2$ and $P_m$ of the simulations are not too far from realistic values in subgiant cores, and the shear should remain limited to $\mathcal{O}(1)$ even for more realistic contraction Reynolds numbers, this critical field extrapolated from the simulations might be of the right order of magnitude. To be more precise, a closer look at the MRI driven by a negative rotation latitudinal gradient in a radiative zone will be necessary.

Our numerical study thus points towards the following scenario: during a first period of the order of the contraction timescale, a contracting radiative layer embedded in a large scale poloidal field tends to rotate uniformly except in localised DZs where the contraction induces a significant differential rotation. If the field is weak enough and not purely dipolar, the development of a powerful axisymmetric MRI reconfigures the field and diminishes its intensity. The magnetic coupling then becomes  inefficient in the major part of the radiative layer and the contraction can force the differential rotation there. 

Such a scenario could potentially explain the evolution of the rotation of the subgiants between the end of the MS and the tip of the RGB. As mentioned in the introduction, the asteroseismic data can be reproduced by assuming a uniform rotation during a first period after the end of the MS followed by a second period where the contraction is left free to enforce differential rotation \citep{spada2016angular}. This is consistent with the two young subgiants in near solid-body found by \cite{deheuvels2020seismic}. At their age, the post-MS contraction should have increased their core rotation by a factor of four which means that the period of uniform rotation should last at least a contraction time scale. This time-scale is compatible with our scenario. 

Our simulations are nevertheless far to describe the full complexity of a magnetic and contracting subgiant. In particular, an expanding layer and boundary conditions mimicking the effect of a convective envelope should be added. The role of non-axisymmetric instabilities should also be considered, especially in the magnetic configuration that results from the axisymmetric instability. Non-axisymmetric MRI or Tayler instability might indeed be present and take part to the AM transport particularly along the giant branch as already invoked \citep{cantiello2014angular,fuller2019slowing}.

As far as intermediate-mass stars are concerned, the occurrence of a powerful contraction-driven instability could help explain the dichotomy between Ap/Bp and Vega-like magnetisms. Indeed, strong Ap/Bp-like magnetic fields are  expected to survive the instability during the PMS while below a certain field intensity the axisymmetric MRI would change the pre-existing large scale field into a small-scale field of smaller amplitude leading to Vega-like magnetism. This is in line with the scenario proposed by \cite{auriere2007weak} except that the instability invoked in this paper was a non-axisymmetric instability produced during the initial winding-up of the poloidal field by the differential rotation. Numerical investigations of this process confirmed the presence of such instabilities but not their ability to profoundly modify the pre-existing poloidal field \cite{jouve2015three, jouve2020interplay}. By contrast, the axisymmetric MRI found in the present paper has a very strong impact on the initial poloidal field destroying its large scale structure and even diminishing its amplitude. To test this scenario, the threshold field strength that separates MRI stable and from MRI unstable configurations is crucial because it should be compatible with the observed $~ 300$ G lower bound of Ap/Bp surface magnetic fields. Again, this calls for further numerical and theoretical investigations of the critical field of the MRI driven by rotation latitudinal gradients in radiative zones.

Part of the above discussion is based on extrapolations of our numerical results to 
%
stellar conditions. Our simulations are indeed a simplified version of a contracting star. Among the simplifications, the ratio between the contraction time and the rotation time is larger in stars than in our simulations ($\tau_{\text{c}} / \tau_{\Omega} \sim 3.4 \cdot 10^8 - 1.1 \cdot 10^{11}$ in stars while this ratio is comprised between $10^3$ and $5 \cdot 10^5$ in our simulations). However the physical model derived from our simulation does not depend critically on this ratio. Indeed, by running our simulations for $5-6$ contraction times, we observed that, after a contraction time, a powerful axisymmetric MHD instability develops. This leads to a complete reconfiguration of the initial magnetic field and to the subsequent development of differential rotation in most part of the spherical shell. This process should not be affected by increasing the ratio $\tau_{\text{c}} / \tau_{\Omega}$ to stellar values. In the same spirit, the ratio $\tau_{\Omega} / \tau_{\nu}$, the Ekman number, is much lower in stars than in numerical simulations. But as shown in \cite{gouhier2020axisymmetric}, the hydrodynamical AM transport is not affected when this ratio is decreased by various orders of magnitude. Thus, despite the simplifications inherent to numerical simulations, the physical model derived from these simulations seems robust enough to apply to stars, especially in the viscous regime where the MRI has been observed. A question that remains to be addressed in future works concerns the occurrence of the
MRI in more realistic Eddington-Sweet regimes which will require to explore the strongly non-linear regime corresponding to very large ratio $\tau_{\text{c}} / \tau_{\nu}$.

\begin{acknowledgements}

The authors acknowledge the developers of MagIC (\url{https://github.com/magic-sph/magic}) for the open-source code thanks to which the simulations were performed. This work was granted access to the HPC resources of CALMIP supercomputing center under the allocation P$1118$. LJ acknowledges funding by the Institut Universitaire de France. The authors wish to thank Sébastien Deheuvels for very fruitful discussions.

\end{acknowledgements}

\bibliographystyle{aa}
\bibliography{biblio.bib}

\begin{thebibliography}{75}
\expandafter\ifx\csname natexlab\endcsname\relax\def\natexlab#1{#1}\fi

\bibitem[{Acevedo-Arreguin {et~al.}(2013)Acevedo-Arreguin, Garaud, \&
  Wood}]{acevedo2013dynamics}
Acevedo-Arreguin, L., Garaud, P., \& Wood, T.~S. 2013, Monthly Notices of the
  Royal Astronomical Society, 434, 720

\bibitem[{Acheson \& Hide(1973)}]{acheson1973hydromagnetics}
Acheson, D. \& Hide, R. 1973, Reports on Progress in Physics, 36, 159

\bibitem[{Alecian {et~al.}(2013)Alecian, Wade, Catala, Grunhut, Landstreet,
  Bagnulo, B{\"o}hm, Folsom, Marsden, \& Waite}]{alecian2013high}
Alecian, E., Wade, G., Catala, C., {et~al.} 2013, Monthly Notices of the Royal
  Astronomical Society, 429, 1001

\bibitem[{Augustson {et~al.}(2016)Augustson, Brun, \&
  Toomre}]{augustson2016magnetic}
Augustson, K.~C., Brun, A.~S., \& Toomre, J. 2016, The Astrophysical Journal,
  829, 92

\bibitem[{Auri{\`e}re {et~al.}(2007)Auri{\`e}re, Wade, Silvester,
  Ligni{\`e}res, Bagnulo, Bale, Dintrans, Donati, Folsom, Gruberbauer,
  {et~al.}}]{auriere2007weak}
Auri{\`e}re, M., Wade, G., Silvester, J., {et~al.} 2007, Astronomy \&
  Astrophysics, 475, 1053

\bibitem[{Balbus \& Hawley(1991)}]{balbus1991powerful}
Balbus, S.~A. \& Hawley, J.~F. 1991, The Astrophysical Journal, 376, 214

\bibitem[{Balbus \& Hawley(1994)}]{balbus1994stability}
Balbus, S.~A. \& Hawley, J.~F. 1994, Monthly Notices of the Royal Astronomical
  Society, 266, 769

\bibitem[{Balbus \& Hawley(1998)}]{balbus1998instability}
Balbus, S.~A. \& Hawley, J.~F. 1998, Reviews of Modern Physics, 70, 1

\bibitem[{Blaz{\`e}re {et~al.}(2016)Blaz{\`e}re, Petit, Ligni{\`e}res,
  Auri{\`e}re, Ballot, B{\"o}hm, Folsom, Gaurat, Jouve, Ariste,
  {et~al.}}]{blazere2016detection}
Blaz{\`e}re, A., Petit, P., Ligni{\`e}res, F., {et~al.} 2016, Astronomy \&
  Astrophysics, 586, A97

\bibitem[{Braginsky \& Roberts(1995)}]{braginsky1995equations}
Braginsky, S.~I. \& Roberts, P.~H. 1995, Geophysical \& Astrophysical Fluid
  Dynamics, 79, 1

\bibitem[{Braithwaite \& Spruit(2004)}]{braithwaite2004fossil}
Braithwaite, J. \& Spruit, H.~C. 2004, Nature, 431, 819

\bibitem[{Braithwaite \& Spruit(2017)}]{braithwaite2017magnetic}
Braithwaite, J. \& Spruit, H.~C. 2017, Royal Society Open Science, 4, 160271

\bibitem[{Brun {et~al.}(2005)Brun, Browning, \& Toomre}]{brun2005simulations}
Brun, A.~S., Browning, M.~K., \& Toomre, J. 2005, The Astrophysical Journal,
  629, 461

\bibitem[{Bugnet {et~al.}(2021)Bugnet, Prat, Mathis, Astoul, Augustson,
  Garc{\'\i}a, Mathur, Amard, \& Neiner}]{bugnet2021magnetic}
Bugnet, L., Prat, V., Mathis, S., {et~al.} 2021, Astronomy \& Astrophysics,
  650, A53

\bibitem[{Cally(1991)}]{cally1991phase}
Cally, P. 1991, Journal of Plasma Physics, 45, 453

\bibitem[{Cantiello \& Braithwaite(2011)}]{cantiello2011magnetic}
Cantiello, M. \& Braithwaite, J. 2011, Astronomy \& Astrophysics, 534, A140

\bibitem[{Cantiello {et~al.}(2016)Cantiello, Fuller, \&
  Bildsten}]{cantiello2016asteroseismic}
Cantiello, M., Fuller, J., \& Bildsten, L. 2016, The Astrophysical Journal,
  824, 14

\bibitem[{Cantiello {et~al.}(2014)Cantiello, Mankovich, Bildsten,
  Christensen-Dalsgaard, \& Paxton}]{cantiello2014angular}
Cantiello, M., Mankovich, C., Bildsten, L., Christensen-Dalsgaard, J., \&
  Paxton, B. 2014, The Astrophysical Journal, 788, 93

\bibitem[{Ceillier {et~al.}(2013)Ceillier, Eggenberger, Garc{\'\i}a, \&
  Mathis}]{ceillier2013understanding}
Ceillier, T., Eggenberger, P., Garc{\'\i}a, R., \& Mathis, S. 2013, Astronomy
  \& Astrophysics, 555, A54

\bibitem[{Charbonneau \& MacGregor(1993)}]{charbonneau1993angular}
Charbonneau, P. \& MacGregor, K. 1993, The Astrophysical Journal, 417, 762

\bibitem[{Clune {et~al.}(1999)Clune, Elliott, Miesch, Toomre, \&
  Glatzmaier}]{clune1999computational}
Clune, T.~C., Elliott, J., Miesch, M., Toomre, J., \& Glatzmaier, G.~A. 1999,
  Parallel Computing, 25, 361

\bibitem[{Deheuvels {et~al.}(2020)Deheuvels, Ballot, Eggenberger, Spada, Noll,
  \& den Hartogh}]{deheuvels2020seismic}
Deheuvels, S., Ballot, J., Eggenberger, P., {et~al.} 2020, Astronomy \&
  Astrophysics, 641, A117

\bibitem[{Deheuvels {et~al.}(2014)Deheuvels, Do{\u{g}}an, Goupil, Appourchaux,
  Benomar, Bruntt, Campante, Casagrande, Ceillier, Davies,
  {et~al.}}]{deheuvels2014seismic}
Deheuvels, S., Do{\u{g}}an, G., Goupil, M., {et~al.} 2014, Astronomy \&
  Astrophysics, 564, A27

\bibitem[{Donati \& Landstreet(2009)}]{donati2009magnetic}
Donati, J.-F. \& Landstreet, J. 2009, Annual Review of Astronomy and
  Astrophysics, 47, 333

\bibitem[{Dormy {et~al.}(1998)Dormy, Cardin, \& Jault}]{dormy1998mhd}
Dormy, E., Cardin, P., \& Jault, D. 1998, Earth and Planetary Science Letters,
  160, 15

\bibitem[{Eggenberger {et~al.}(2019)Eggenberger, Deheuvels, Miglio,
  Ekstr{\"o}m, Georgy, Meynet, Lagarde, Salmon, Buldgen, Montalb{\'a}n,
  {et~al.}}]{eggenberger2019asteroseismology}
Eggenberger, P., Deheuvels, S., Miglio, A., {et~al.} 2019, Astronomy \&
  Astrophysics, 621, A66

\bibitem[{Eggenberger {et~al.}(2012)Eggenberger, Montalb{\'a}n, \&
  Miglio}]{eggenberger2012angular}
Eggenberger, P., Montalb{\'a}n, J., \& Miglio, A. 2012, Astronomy \&
  Astrophysics, 544, L4

\bibitem[{Ferraro(1937)}]{ferraro1937non}
Ferraro, V.~C. 1937, Monthly Notices of the Royal Astronomical Society, 97, 458

\bibitem[{Fuller {et~al.}(2015)Fuller, Cantiello, Stello, Garc{\'\i}a, \&
  Bildsten}]{fuller2015asteroseismology}
Fuller, J., Cantiello, M., Stello, D., Garc{\'\i}a, R.~A., \& Bildsten, L.
  2015, Science, 350, 423

\bibitem[{Fuller {et~al.}(2019)Fuller, Piro, \& Jermyn}]{fuller2019slowing}
Fuller, J., Piro, A.~L., \& Jermyn, A.~S. 2019, Monthly Notices of the Royal
  Astronomical Society, 485, 3661

\bibitem[{Gallet \& Bouvier(2013)}]{gallet2013improved}
Gallet, F. \& Bouvier, J. 2013, Astronomy \& Astrophysics, 556, A36

\bibitem[{Garaud(2002)}]{garaud2002rotationally}
Garaud, P. 2002, Monthly Notices of the Royal Astronomical Society, 335, 707

\bibitem[{Garaud \& Acevedo-Arreguin(2009)}]{garaud2009penetration2}
Garaud, P. \& Acevedo-Arreguin, L. 2009, The Astrophysical Journal, 704, 1

\bibitem[{Garaud \& Brummell(2008)}]{garaud2008penetration1}
Garaud, P. \& Brummell, N. 2008, The Astrophysical Journal, 674, 498

\bibitem[{Garaud \& Garaud(2008)}]{garaud2008dynamics}
Garaud, P. \& Garaud, J.-D. 2008, Monthly Notices of the Royal Astronomical
  Society, 391, 1239

\bibitem[{Garaud {et~al.}(2015)Garaud, Medrano, Brown, Mankovich, \&
  Moore}]{garaud2015excitation}
Garaud, P., Medrano, M., Brown, J., Mankovich, C., \& Moore, K. 2015, The
  Astrophysical Journal, 808, 89

\bibitem[{Gastine \& Wicht(2012)}]{gastine2012effects}
Gastine, T. \& Wicht, J. 2012, Icarus, 219, 428

\bibitem[{Gouhier {et~al.}(2021)Gouhier, Ligni{\`e}res, \&
  Jouve}]{gouhier2020axisymmetric}
Gouhier, B., Ligni{\`e}res, F., \& Jouve, L. 2021, Astronomy \& Astrophysics,
  648, A109

\bibitem[{Heyvaerts \& Priest(1983)}]{heyvaerts1983coronal}
Heyvaerts, J. \& Priest, E. 1983, Astronomy and Astrophysics, 117, 220

\bibitem[{Hollerbach \& R{\"u}diger(2005)}]{hollerbach2005new}
Hollerbach, R. \& R{\"u}diger, G. 2005, Physical Review Letters, 95, 124501

\bibitem[{Jones {et~al.}(2011)Jones, Boronski, Brun, Glatzmaier, Gastine,
  Miesch, \& Wicht}]{jones2011anelastic}
Jones, C., Boronski, P., Brun, A., {et~al.} 2011, Icarus, 216, 120

\bibitem[{Jouve {et~al.}(2015)Jouve, Gastine, \&
  Ligni{\`e}res}]{jouve2015three}
Jouve, L., Gastine, T., \& Ligni{\`e}res, F. 2015, Astronomy \& Astrophysics,
  575, A106

\bibitem[{Jouve {et~al.}(2020)Jouve, Ligni{\`e}res, \&
  Gaurat}]{jouve2020interplay}
Jouve, L., Ligni{\`e}res, F., \& Gaurat, M. 2020, Astronomy \& Astrophysics,
  641, A13

\bibitem[{Kirillov {et~al.}(2014)Kirillov, Stefani, \&
  Fukumoto}]{kirillov2014local}
Kirillov, O.~N., Stefani, F., \& Fukumoto, Y. 2014, Journal of Fluid Mechanics,
  760, 591

\bibitem[{Knobloch(1992)}]{knobloch1992stability}
Knobloch, E. 1992, Monthly Notices of the Royal Astronomical Society, 255, 25P

\bibitem[{{Lantz}(1992)}]{lantz1992dynamical}
{Lantz}, S.~R. 1992, PhD thesis, Cornell University, United States

\bibitem[{Ligni{\`e}res {et~al.}(1999)Ligni{\`e}res, Califano, \&
  Mangeney}]{lignieres1999shear}
Ligni{\`e}res, F., Califano, F., \& Mangeney, A. 1999, Astronomy \&
  Astrophysics, 349, 1027

\bibitem[{Ligni{\`e}res {et~al.}(2014)Ligni{\`e}res, Petit, Auri{\`e}re, Wade,
  \& B{\"o}hm}]{lignieres2013dichotomy}
Ligni{\`e}res, F., Petit, P., Auri{\`e}re, M., Wade, G.~A., \& B{\"o}hm, T.
  2014, Proceedings of the International Astronomical Union, 9, 338

\bibitem[{Ligni{\`e}res {et~al.}(2009)Ligni{\`e}res, Petit, B{\"o}hm, \&
  Auri{\`e}re}]{lignieres2009first}
Ligni{\`e}res, F., Petit, P., B{\"o}hm, T., \& Auri{\`e}re, M. 2009, Astronomy
  \& Astrophysics, 500, L41

\bibitem[{Maeder(2008)}]{maeder2008physics}
Maeder, A. 2008, Physics, Formation and Evolution of Rotating Stars (Springer
  Science \& Business Media)

\bibitem[{Marques {et~al.}(2013)Marques, Goupil, Lebreton, Talon, Palacios,
  Belkacem, Ouazzani, Mosser, Moya, Morel, {et~al.}}]{marques2013seismic}
Marques, J., Goupil, M., Lebreton, Y., {et~al.} 2013, Astronomy \&
  Astrophysics, 549, A74

\bibitem[{Menou {et~al.}(2004)Menou, Balbus, \& Spruit}]{menou2004local}
Menou, K., Balbus, S.~A., \& Spruit, H.~C. 2004, The Astrophysical Journal,
  607, 564

\bibitem[{Mestel \& Weiss(1987)}]{mestel1987magnetic}
Mestel, L. \& Weiss, N. 1987, Monthly Notices of the Royal Astronomical
  Society, 226, 123

\bibitem[{Mosser {et~al.}(2017)Mosser, Belkacem, Pin{\c{c}}on, Takata, Vrard,
  Barban, Goupil, Kallinger, \& Samadi}]{mosser2017dipole}
Mosser, B., Belkacem, K., Pin{\c{c}}on, C., {et~al.} 2017, Astronomy \&
  Astrophysics, 598, A62

\bibitem[{Mosser {et~al.}(2012)Mosser, Elsworth, Hekker, Huber, Kallinger,
  Mathur, Belkacem, Goupil, Samadi, Barban,
  {et~al.}}]{mosser2012characterization}
Mosser, B., Elsworth, Y., Hekker, S., {et~al.} 2012, Astronomy \& Astrophysics,
  537, A30

\bibitem[{Oglethorpe \& Garaud(2013)}]{oglethorpe2013spin}
Oglethorpe, R. \& Garaud, P. 2013, The Astrophysical Journal, 778, 166

\bibitem[{Petit {et~al.}(2011)Petit, Ligni{\`e}res, Auri{\`e}re, Wade, Alina,
  Ballot, B{\"o}hm, Jouve, Oza, Paletou, {et~al.}}]{petit2011detection}
Petit, P., Ligni{\`e}res, F., Auri{\`e}re, M., {et~al.} 2011, Astronomy \&
  Astrophysics, 532, L13

\bibitem[{Petit {et~al.}(2010)Petit, Ligni{\`e}res, Wade, Auri{\`e}re,
  B{\"o}hm, Bagnulo, Dintrans, Fumel, Grunhut, Lanoux,
  {et~al.}}]{petit2010rapid}
Petit, P., Ligni{\`e}res, F., Wade, G., {et~al.} 2010, Astronomy \&
  Astrophysics, 523, A41

\bibitem[{Petitdemange {et~al.}(2013)Petitdemange, Dormy, \&
  Balbus}]{petitdemange2013axisymmetric}
Petitdemange, L., Dormy, E., \& Balbus, S. 2013, Physics of the Earth and
  Planetary Interiors, 223, 21

\bibitem[{Roberts(1967)}]{roberts1967singularities}
Roberts, P. 1967, Proceedings of the Royal Society of London. Series A.
  Mathematical and Physical Sciences, 300, 94

\bibitem[{R{\"u}diger {et~al.}(2018)R{\"u}diger, Gellert, Hollerbach, Schultz,
  \& Stefani}]{rudiger2018stability}
R{\"u}diger, G., Gellert, M., Hollerbach, R., Schultz, M., \& Stefani, F. 2018,
  Physics Reports, 741, 1

\bibitem[{R{\"u}diger {et~al.}(2015)R{\"u}diger, Gellert, Spada, \&
  Tereshin}]{rudiger2015angular}
R{\"u}diger, G., Gellert, M., Spada, F., \& Tereshin, I. 2015, Astronomy \&
  Astrophysics, 573, A80

\bibitem[{R{\"u}diger {et~al.}(2016)R{\"u}diger, Schultz, \&
  Kitchatinov}]{rudiger2016instability}
R{\"u}diger, G., Schultz, M., \& Kitchatinov, L. 2016, Monthly Notices of the
  Royal Astronomical Society, 456, 3004

\bibitem[{Shercliff(1956)}]{shercliff1956flow}
Shercliff, J. 1956, Journal of Fluid Mechanics, 1, 644

\bibitem[{Spada {et~al.}(2016)Spada, Gellert, Arlt, \&
  Deheuvels}]{spada2016angular}
Spada, F., Gellert, M., Arlt, R., \& Deheuvels, S. 2016, Astronomy \&
  Astrophysics, 589, A23

\bibitem[{Spruit(1999)}]{spruit1999differential}
Spruit, H. 1999, Astronomy \& Astrophysics, 349, 189

\bibitem[{Spruit(2002)}]{spruit2002dynamo}
Spruit, H. 2002, Astronomy \& Astrophysics, 381, 923

\bibitem[{Stefani {et~al.}(2006)Stefani, Gundrum, Gerbeth, R{\"u}diger,
  Schultz, Szklarski, \& Hollerbach}]{stefani2006experimental}
Stefani, F., Gundrum, T., Gerbeth, G., {et~al.} 2006, Physical Review Letters,
  97, 184502

\bibitem[{Stello {et~al.}(2016{\natexlab{a}})Stello, Cantiello, Fuller,
  Garc{\'\i}a, \& Huber}]{stello2016suppression}
Stello, D., Cantiello, M., Fuller, J., Garc{\'\i}a, R.~A., \& Huber, D.
  2016{\natexlab{a}}, Publications of the Astronomical Society of Australia, 33

\bibitem[{Stello {et~al.}(2016{\natexlab{b}})Stello, Cantiello, Fuller, Huber,
  Garc{\'\i}a, Bedding, Bildsten, \& Aguirre}]{stello2016prevalence}
Stello, D., Cantiello, M., Fuller, J., {et~al.} 2016{\natexlab{b}}, Nature,
  529, 364

\bibitem[{Wade {et~al.}(2005)Wade, Drouin, Bagnulo, Landstreet, Mason,
  Silvester, Alecian, B{\"o}hm, Bouret, Catala, {et~al.}}]{wade2005discovery}
Wade, G.~A., Drouin, D., Bagnulo, S., {et~al.} 2005, Astronomy \& Astrophysics,
  442, L31

\bibitem[{Weiss(1966)}]{weiss1966expulsion}
Weiss, N.~O. 1966, Proceedings of the Royal Society of London. Series A.
  Mathematical and Physical Sciences, 293, 310

\bibitem[{Wicht(2002)}]{wicht2002inner}
Wicht, J. 2002, Physics of the Earth and Planetary Interiors, 132, 281

\bibitem[{Wood \& Brummell(2012)}]{wood2012transport}
Wood, T.~S. \& Brummell, N.~H. 2012, The Astrophysical Journal, 755, 99

\bibitem[{Zahn(1992)}]{zahn1992circulation}
Zahn, J.-P. 1992, Astronomy and Astrophysics, 265, 115

\end{thebibliography}

\appendix

\onecolumn

\section{Electrical-current function}
\label{elec_current_fonc}

In this appendix we determine the stream function that is constant along the streamlines of the poloidal component of the current density $\vv{j}_p$. The toroidal component of the magnetic field is related to $\vv{j}_p$ through

\begin{equation}
\vv{j}_p = \vv{\nabla} \times \left( B_{\phi} \vv{e}_{\phi}\right) = \displaystyle \frac{1}{r \sin{\theta}} \displaystyle \frac{\partial}{\partial \theta} \left( \sin{\theta} B_{\phi} \right) \vv{e}_r - \displaystyle \frac{1}{r} \displaystyle \frac{\partial}{\partial r} \left( r B_{\phi} \right) \vv{e}_{\theta}.
\end{equation}

\noindent As the divergence of the curl is zero and the problem is axisymmetric we can define a vector potential $\chi$ as

\begin{equation}
j_r = \displaystyle \frac{1}{r \sin{\theta}} \displaystyle \frac{\partial}{\partial \theta} \left( \sin{\theta} \chi \right) ~ \text{;} \quad j_{\theta} = \displaystyle \frac{-1}{r} \displaystyle \frac{\partial}{\partial r} \left( r \chi \right),
\end{equation}

\noindent from which we define the electrical-current stream function $\mathcal{J}_p = r \sin{\theta} \chi$:

\begin{equation}
j_r = \displaystyle \frac{1}{r^2 \sin{\theta}} \displaystyle \frac{\partial \mathcal{J}_p}{\partial \theta} ~ \text{;} \quad j_{\theta} = \displaystyle \frac{-1}{r \sin{\theta}} \displaystyle \frac{\partial \mathcal{J}_p}{\partial r}.
\end{equation}

\noindent We thus obtain a relationship between the toroidal field and the electrical-current stream function

\begin{equation}
\displaystyle \frac{1}{r \sin{\theta}} \displaystyle \frac{\partial}{\partial \theta} \left( \sin{\theta} B_{\phi} \right) = \displaystyle \frac{1}{r^2 \sin{\theta}} \displaystyle \frac{\partial \mathcal{J}_p}{\partial \theta} ~ \text{;} \quad  - \displaystyle \frac{1}{r} \displaystyle \frac{\partial}{\partial r} \left( r B_{\phi} \right) = \displaystyle \frac{-1}{r \sin{\theta}} \displaystyle \frac{\partial \mathcal{J}_p}{\partial r}.
\end{equation}

\noindent By integrating the first equation we have 

\begin{equation}
\mathcal{J}_p = r \displaystyle \int \displaystyle \frac{\partial}{\partial \theta} \left( \sin{\theta} B_{\phi} \right) \text{d} \theta = r \sin{\theta} B_{\phi} + f(r),
\end{equation}

\noindent and substituting in the second equation we deduce

\begin{equation}
\displaystyle \frac{f^{'}(r)}{- r \sin{\theta}} = 0 \quad \Longrightarrow \quad f(r) = \text{cte}
.
\end{equation}

\noindent As the electrical-current stream function is zero at the poles, we conclude that $\text{cte} = 0$ hence

\begin{equation}
\mathcal{J}_p = r \sin{\theta} B_{\phi}
.
\end{equation}

\noindent 

\section{Hartmann boundary layer equations}
\label{hartmann_equations}

\begin{figure}[!h]
\begin{center}
\includegraphics[width=8.7cm]{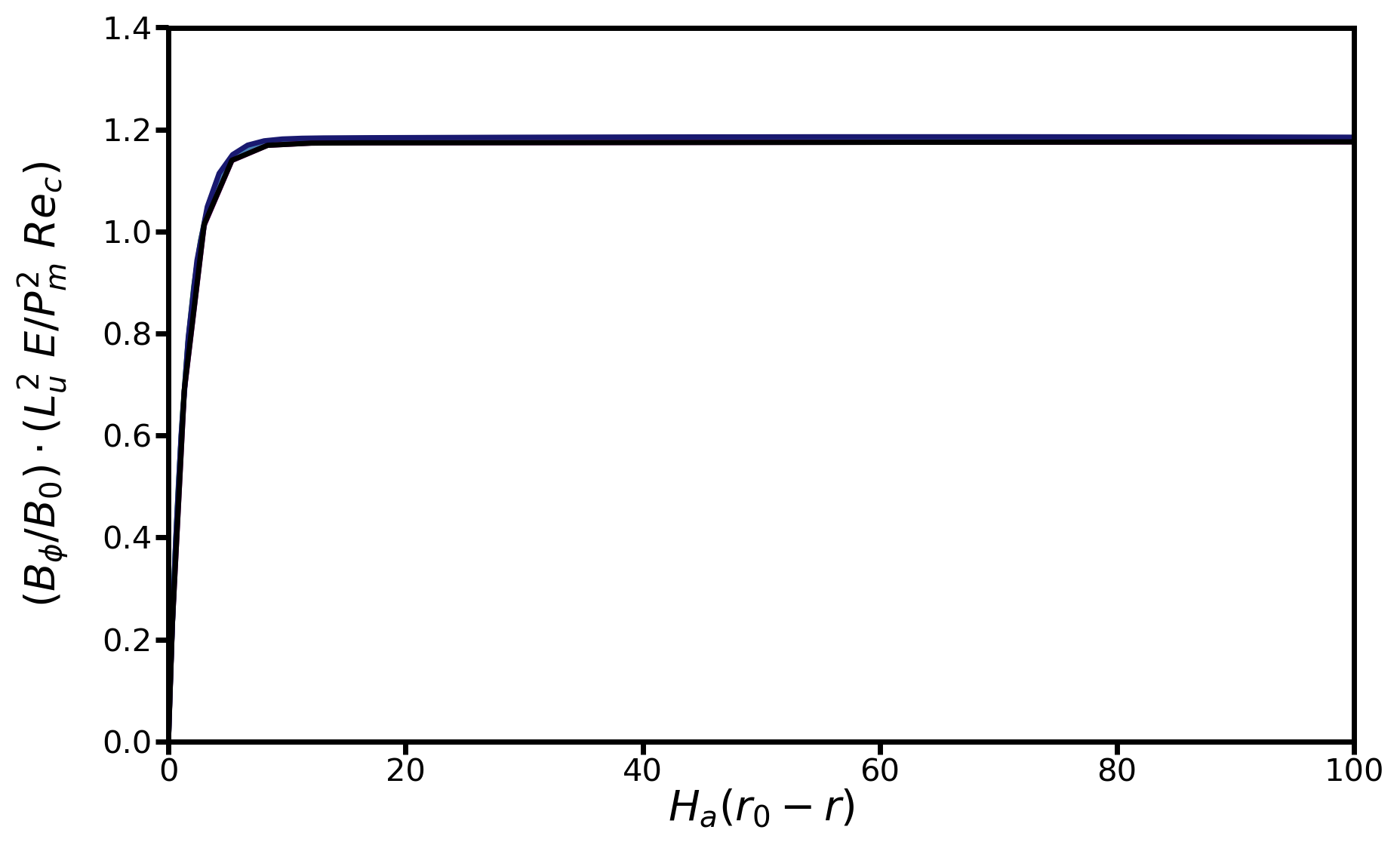}
\includegraphics[width=8.7cm]{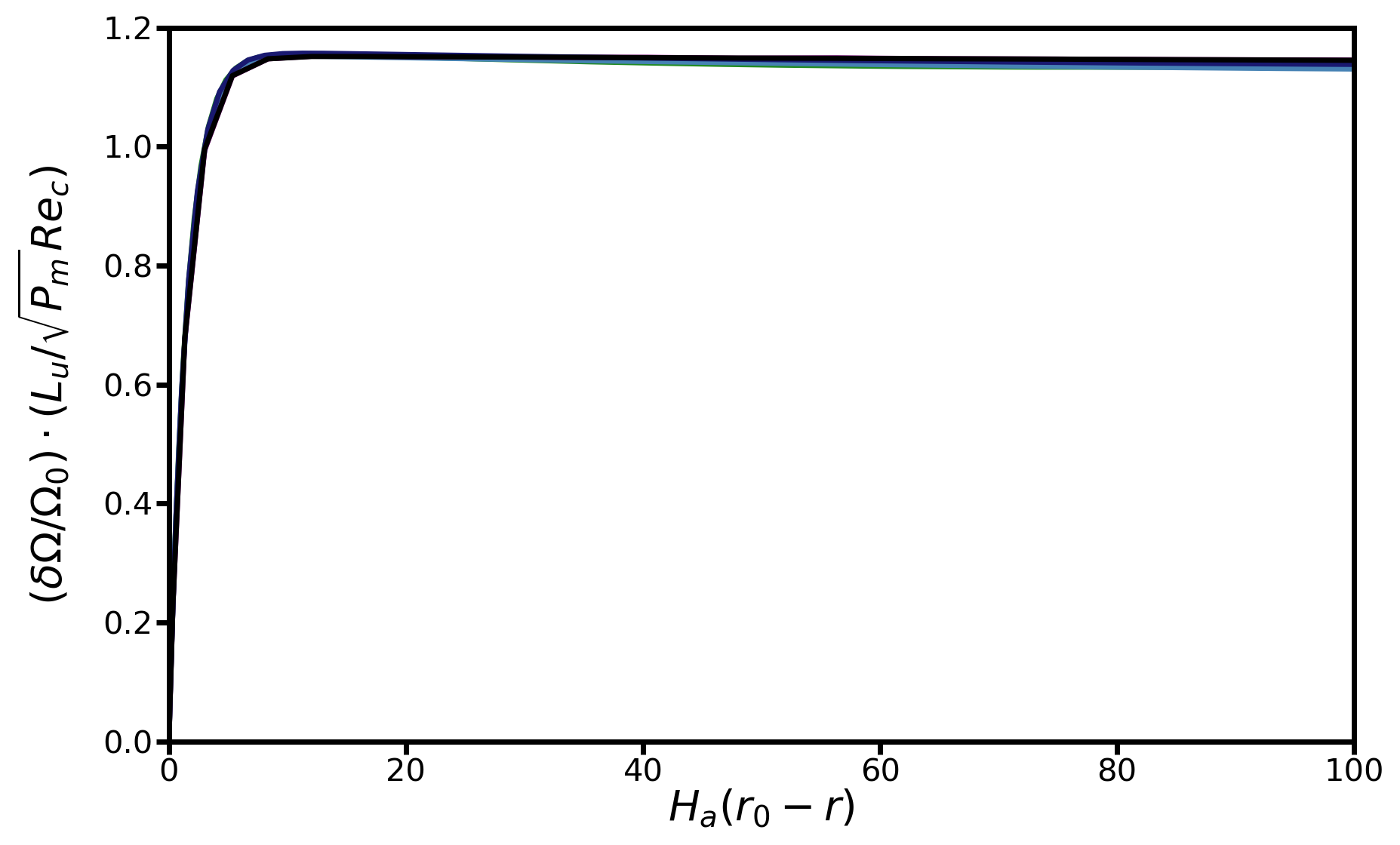}
\caption{Toroidal field normalised to $B_0$ (left panel) and differential rotation normalised to $\Omega_0$ (right panel) respectively rescaled by $L_u^2 E / P_m^2 Re_c$ and $L_u / \sqrt{P_m} Re_c$, as a function of the stretched coordinate $\xi = H_a \left(r_0 - r\right)$. The curves are plotted at $\theta = \pi/4$ for various $Re_c$ and $L_u$ with no contraction term in the induction equation. We thus have $Re_c = 1$, $L_u = 5\cdot10^3$ in green, $Re_c = 1$, $L_u = 10^4$ in light blue, $Re_c = 1$, $L_u = 5 \cdot 10^4$ in blue, $Re_c = 5$, $L_u = 10^4$ in purple and $Re_c = 5$, $L_u = 5\cdot 10^4$ in black. The other parameters are $E = 10^{-4}$, $P_r \left(N_0/\Omega_0\right)^2 = 10^{4}$ and $P_m = 10^2$ (runs D$3$ to D$7$ of Table \ref{parameters_viscous}).}  
\label{Bphi_Rot_Diff_Hartmann}
\end{center}
\end{figure}

In the present appendix the Hartmman boundary layer equations are derived with the aim of relating the jump of the toroidal field across the layer with the differential rotation jump. The Hartmann layer occurs at the electrically insulating outer boundary where the azimuthal velocity is fixed and a magnetic field perpendicular to the flow is present. When the contraction term is removed from the induction equation, the governing equations for the azimuthal flow and the toroidal field component in the steady linear limit read:

\begin{equation}
2 \Omega_0 \left(U_s - \sin{\theta} \displaystyle \frac{V_0  r_0^2}{r^2}\right) = \displaystyle \frac{1}{\mu_0 \rho_0} \left \lbrack \displaystyle \frac{B_{\theta}}{r} \displaystyle \frac{\partial B_{\phi}}{\partial \theta} + \displaystyle \frac{\cot{\theta}}{r} B_{\theta} B_{\phi} + B_r \displaystyle \frac{\partial B_{\phi}}{\partial r} + \displaystyle \frac{B_r B_{\phi}}{r} \right \rbrack + \nu \left \lbrack \displaystyle \frac{\partial^2 U_{\phi}}{\partial r^2} + \displaystyle \frac{2}{r} \displaystyle \frac{\partial U_{\phi}}{\partial r} + \displaystyle \frac{1}{r^2} \displaystyle \frac{\partial^2 U_{\phi}}{\partial \theta^2} + \displaystyle \frac{\cot{\theta}}{r^2} \displaystyle \frac{\partial U_{\phi}}{\partial \theta} - \displaystyle \frac{U_{\phi}}{r^2 \sin^2 \theta} \right \rbrack 
\label{balance_viscous_lorentz_coriolis},
\end{equation}

\begin{equation}
\begin{array}{lll}
U_r \displaystyle \frac{\partial B_{\phi}}{\partial r} + \displaystyle \frac{U_{\theta}}{r} \displaystyle \frac{\partial B_{\phi}}{\partial \theta} - B_r \displaystyle \frac{\partial U_{\phi}}{\partial r} - \displaystyle \frac{B_{\theta}}{r} \displaystyle \frac{\partial U_{\phi}}{\partial \theta} + U_{\phi} \left(\displaystyle \frac{B_r}{r} + \displaystyle \frac{\cot{\theta} B_{\theta}}{r} \right) - B_{\phi} \left(\displaystyle \frac{U_r}{r} + \displaystyle \frac{\cot{\theta} U_{\theta}}{r} \right) = \\\\ \eta \left \lbrack \displaystyle \frac{\partial^2 B_{\phi}}{\partial r^2} + \displaystyle \frac{2}{r} \displaystyle \frac{\partial B_{\phi}}{\partial r} + \displaystyle \frac{1}{r^2} \displaystyle \frac{\partial^2 B_{\phi}}{\partial \theta^2} + \displaystyle \frac{\cot{\theta}}{r^2} \displaystyle \frac{\partial B_{\phi}}{\partial \theta} - \displaystyle \frac{B_{\phi}}{r^2 \sin^2 \theta} \right \rbrack 
\label{balance_diffusion_om_effect}.
\end{array}
\end{equation}

\noindent Since in our simulations the Elsasser number $\Lambda = B_0^2 / \mu_0 \rho_0 \Omega_0 \eta$ is $\gg 1$, the Coriolis term becomes negligible with respect to the Lorentz force \citep{acheson1973hydromagnetics}, then, assuming $U_r \ll U_{\phi}$ and conserving the highest radial derivatives in each term, we get the Hartmann boundary layer equations (see also reviews by \cite{roberts1967singularities, acheson1973hydromagnetics, dormy1998mhd}): 

\begin{equation}
\displaystyle \frac{B_r}{\mu_0 \rho_0} \displaystyle \frac{\partial B_{\phi}^{\text{H}}}{\partial r} = - \nu \displaystyle \frac{\partial^2 U_{\phi}^{\text{H}}}{\partial r^2} ~ \text{,} \quad
B_r \displaystyle \frac{\partial U_{\phi}^{\text{H}}}{\partial r} = - \eta \displaystyle \frac{\partial^2 B_{\phi}^{\text{H}}}{\partial r^2}
\label{hart_equations}.
\end{equation}

\noindent where the index $\text{H}$ denotes a boundary layer flow. As $B_r(r_0,\theta) = - B_0 \cos{\theta}$ (see Eq. \eqref{dipole}), then after introducing the stretched coordinate $\xi =  \left(B_0 r_0 / \sqrt{\mu_0 \rho_0 \eta \nu} \right) \left( r_0 - r \right)$, Sys. \eqref{hart_equations} is rewritten as:

\begin{equation}
\displaystyle \frac{\cos{\theta}}{\sqrt{\mu_0 \rho_0}} \displaystyle \frac{\partial B_{\phi}^{\text{H}}}{\partial \xi} = - r_0 \sqrt{\displaystyle \frac{\nu}{\eta}} \hspace*{0.05cm} \displaystyle  \frac{\partial^2 U_{\phi}^{\text{H}}}{\partial \xi^2} ~ \text{,} \quad
\sqrt{\mu_0 \rho_0} \cos{\theta} \displaystyle \frac{\partial U_{\phi}^{\text{H}}}{\partial \xi} = - r_0 \sqrt{\displaystyle \frac{\eta}{\nu}} \hspace*{0.05cm} \displaystyle \frac{\partial^2 B_{\phi}^{\text{H}}}{\partial \xi^2}
\label{dim_hart_equations_strechted_coordinate}.
\end{equation}

\noindent By deriving the second of these two equations, $U_{\phi}^{\text{H}}$ can be eliminated to yield:

\begin{equation}
\displaystyle \frac{\partial^3 B_{\phi}^{\text{H}}}{\partial \xi^3} - \displaystyle \frac{\cos^2 \theta}{r_0^2} \displaystyle \frac{\partial B_{\phi}^{\text{H}}}{\partial \xi} = 0,
\end{equation}

\noindent whose the solution is

\begin{equation}
B_{\phi}^{\text{H}}(\xi,\theta) = C_1 + \displaystyle \frac{C_2 r_0}{\cos{\theta}} \exp{\left(- \cos{\theta} \displaystyle \frac{\xi}{r_0}\right)} + \displaystyle \frac{C_3 r_0}{\cos{\theta}} \exp{\left(\cos{\theta} \displaystyle \frac{\xi}{r_0}\right)}
.
\end{equation}

\noindent From the evanescent condition when $\xi \to \infty$, $C_1 = C_3 = 0$, and from the vacuum condition at the outer sphere $B_{\phi}^{\text{H}}(0,\theta) + B_{\phi}^{\text{I}}(r_0,\theta) = 0$, $C_2 = - \left(\cos{\theta} / r_0 \right) B_{\phi}^{\text{I}}(r_0,\theta) $, hence: 

\begin{equation}
B_{\phi}^{\text{H}}(\xi,\theta) = - B_{\phi}^{\text{I}}(r_0,\theta) \exp{\left(- \cos{\theta} \displaystyle \frac{\xi}{r_0}\right)}.
\end{equation}

\noindent This time, the index $I$ stands for interior flow. We now integrate the second equation of Sys. \eqref{dim_hart_equations_strechted_coordinate}:

\begin{equation}
U_{\phi}^{\text{H}}(\xi,\theta) = - \displaystyle \frac{1}{\sqrt{\mu_0 \rho_0}} \sqrt{\displaystyle \frac{\eta}{\nu}} B_{\phi}^{\text{I}}(r_0,\theta) \exp{\left(- \cos{\theta}\displaystyle \frac{\xi}{r_0}\right)} + C_4
.
\end{equation}

\noindent Again, from the evanescent equation when $\xi \to \infty$ we readily get $C_4 = 0$. Then, as $U_{\phi}^{\text{H}}(\xi,\theta) + U_{\phi}^{\text{I}}(r_0,\theta) = 0$ at the outer sphere we obtain the following relationship between the interior flows:

\begin{equation}
U_{\phi}^{\text{I}}(r_0,\theta) = \displaystyle \frac{1}{\sqrt{\mu_0 \rho_0}} \sqrt{\displaystyle \frac{\eta}{\nu}} B_{\phi}^{\text{I}}(r_0,\theta),
\label{relationship_between_interior_flows_hartmann}
\end{equation}

\noindent and,

\begin{equation}
U_{\phi}^{\text{H}}(\xi,\theta) = - U_{\phi}^{\text{I}}(r_0,\theta) \exp{\left(- \cos{\theta}\displaystyle \frac{\xi}{r_0}\right)}
.
\end{equation}

\noindent Thus, using the estimate Eq. \eqref{estimate_bphi}, we deduce from Eq. \eqref{relationship_between_interior_flows_hartmann} the order of magnitude of the jump on the differential rotation across the Hartmann layer

\begin{equation}
U_{\phi}^{\text{I}}(r_0,\theta) \approx \mathcal{O} \left( \displaystyle \frac{r_0 \mu_0 \rho_0 V_0 \Omega_0}{\sqrt{\mu_0 \rho_0} B_0} \sqrt{\displaystyle \frac{\eta}{\nu}} \right) \quad \Rightarrow \quad \displaystyle \frac{\Delta \Omega^{\text{I}}(r_0,\theta)}{\Omega_0} \approx \mathcal{O} \left( \displaystyle \frac{\sqrt{\mu_0 \rho_0} V_0}{B_0} \sqrt{\displaystyle \frac{\eta}{\nu}} \right) = \mathcal{O} \left( \displaystyle \frac{\sqrt{P_m} Re_c}{L_u} \right)
\label{final_estimate_jump_rot_diff_hart}.
\end{equation}

\noindent In Fig. \ref{Bphi_Rot_Diff_Hartmann} we plotted the normalised toroidal field $B_{\phi}/B_0$ (left panel) and the normalised differential rotation $\delta \Omega/\Omega_0$ (right panel) as a function of the stretched coordinate $\xi$ at the particular location $\theta = \pi/4$. This was done for runs obtained at various $L_u$ ranging from $5\cdot10^3$ to $5\cdot10^4$, both for $Re_c = 1$ and $5$ (runs D$3$ to D$7$ of Table \ref{parameters_viscous}). These quantities have been rescaled with their characteristic amplitude as given by Eqs. \eqref{estimate_bphi} and \eqref{final_estimate_jump_rot_diff_hart}. We can observe that the different curves overlap and are of $\mathcal{O}(1)$ after rescaling. This enable us to conclude that the $\mathcal{O} \left(r_0 \mu_0 \rho_0 V_0 \Omega_0 / B_0 \right) $ jump on the toroidal magnetic field at the outer sphere induces a $\mathcal{O} \left(r_0 \sqrt{\mu_0 \rho_0} V_0 \Omega_0 \sqrt{\eta} / B_0 \sqrt{\nu} \right) $ jump on the azimuthal velocity field across the Hartmann layer or equivalently, a $\mathcal{O} \left( \sqrt{\mu_0 \rho_0} V_0 \sqrt{\eta} / B_0 \sqrt{\nu} \right) $ jump on the normalised differential rotation. As a result, the quasi-solid rotation region is in differential rotation with the outer sphere and the characteristic amplitude of this differential rotation is given by Eq. \eqref{final_estimate_jump_rot_diff_hart}.

\section{Approximate solutions of differential rotation in the dead zone}
\label{second_hydrostatic_state}
\subsection{Case $1$ - No effect of the contraction on the field lines}
\label{resolution_no_induct}

\begin{figure}[!h]
\begin{center}
\includegraphics[width=12cm]{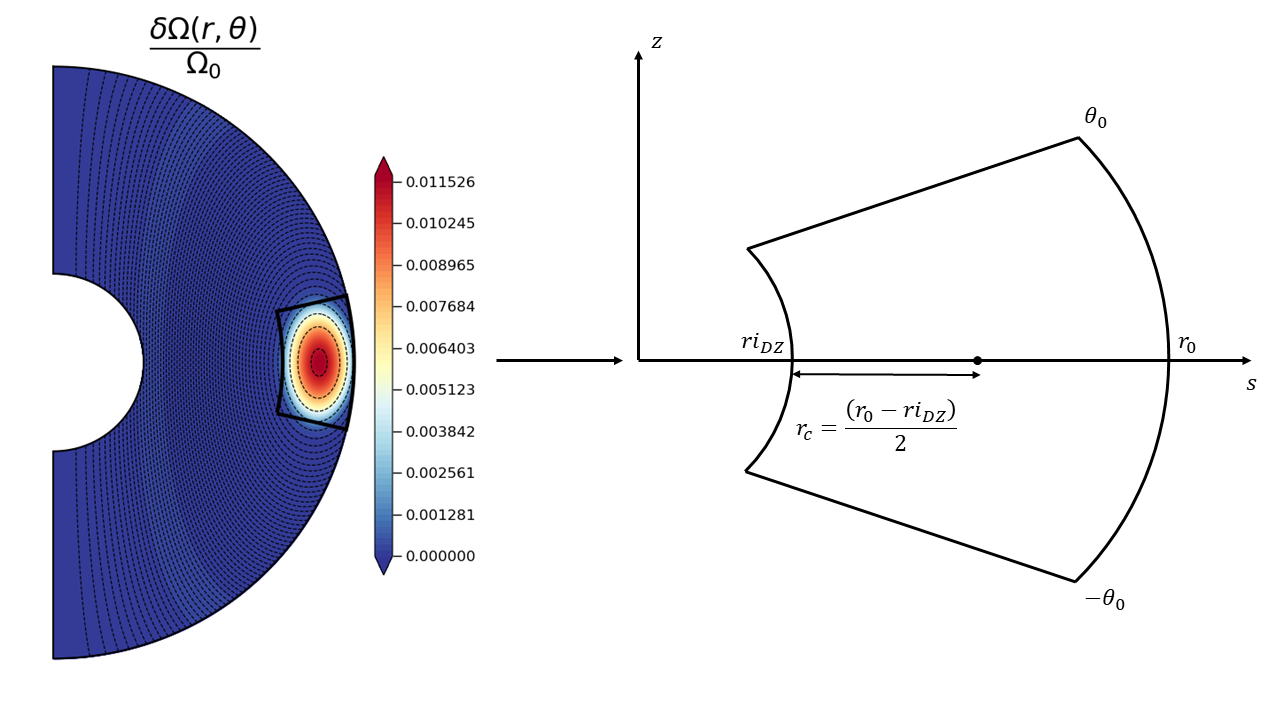}
\caption{Sketch of the conical domain chosen to represent the DZ when the field lines are not advected by the contraction. The displayed meridional cut of the normalised rotation rate is the same as in the first panel of Fig. \ref{Numerical_Results_Viscous}. The conical domain, delimited by thick black lines on the meridional cut, is defined by $r \in \left \lbrack r_{i_{\hspace*{0.025cm} \text{DZ}}} ; r_0 \right \rbrack $ and $\theta \in \left \lbrack - \theta_0 ; \theta_0 \right \rbrack$ with $r_{i_{\hspace*{0.025cm} \text{DZ}}} = 0.77 \hspace*{0.025cm} r_0$, $\theta_0 \approx \pi/14$ and $-\theta_0 \approx - \pi/14$ or equivalently, in terms of colatitude, $\theta \in \left \lbrack 3 \pi / 7 ; 4 \pi / 7 \right \rbrack$.}  
\label{conical_domain}
\end{center}
\end{figure}

This appendix is intended to solve in the DZ the viscous balance Eq. \eqref{differential_equation}, whose dimensionless form reads 

\begin{equation}
\tilde{r} \displaystyle \frac{\partial^2 \delta \tilde{\Omega}}{\partial r^2} + 4 \displaystyle \frac{\partial \delta \tilde{\Omega}}{\partial r} + \displaystyle \frac{1}{\tilde{r}} \displaystyle \frac{\partial^2 \delta \tilde{\Omega}}{\partial \theta^2} + \displaystyle \frac{3 \cot{\theta}}{\tilde{r}} \displaystyle \frac{\partial \delta \tilde{\Omega}}{\partial \theta} = \displaystyle \frac{-2}{E \tilde{r}^2}
\label{viscous_balance},
\end{equation}

\noindent To do so, the DZ is assimilated to a conical domain as sketched in Fig. \ref{conical_domain}. This domain, denoted $\mathcal{D}$, is such that:

\begin{equation}
\mathcal{D} = \left \lbrace r_{i_{\hspace*{0.025cm} \text{DZ}}} \leq r \leq r_0 \hspace*{0.1cm} ; \hspace*{0.1cm} - \theta_0 \leq \theta \leq \theta_0 \hspace*{0.05cm} \right \rbrace
\label{domain_def},
\end{equation}

\noindent where $r_{i_{\hspace*{0.025cm} \text{DZ}}} = 0.77 r_0$ and the latitude $\theta_0$ is $\approx \pi/14$. In addition, we assume that the term $\left(3 \cos{\theta} /r \right) \partial \delta \Omega / \partial \theta$ is negligible as compared to the others. This assumption has been verified a posteriori. Thus, Eq. \eqref{viscous_balance} is rewritten as follows:

\begin{equation}
\tilde{r}^3 \displaystyle \frac{\partial^2 \delta \tilde{\Omega}}{\partial r^2} + 4 \tilde{r}^2 \displaystyle \frac{\partial \delta \tilde{\Omega}}{\partial r} + \tilde{r} \displaystyle \frac{\partial^2 \delta \tilde{\Omega}}{\partial \theta^2} = \displaystyle \frac{-2}{E}
\label{viscous_balance_simplified}.
\end{equation}

\noindent Outside the domain $\mathcal{D}$ the flow is assumed to be in solid rotation and symmetrical with respect to the equatorial plane so that we adopt the following homogenous boundary conditions:

\begin{equation}
\delta \Omega(r_0,\theta) = \delta \Omega(r_{i_{\hspace*{0.025cm} \text{DZ}}},\theta) = \delta \Omega(r,\theta_0) = \left. \displaystyle \frac{\partial \delta \Omega(r,\theta)}{\partial \theta} \right |_{\theta = 0} = 0
\label{boundary_conditions}.
\end{equation}

\noindent The method for solving Eq. \eqref{viscous_balance_simplified} was excerpted from a lecture of the Ira A. Fulton College of Engineering and Technology at Brigham Young University (\url{https://www.et.byu.edu/~vps/ME505/IEM/08 02.pdf}). The broad lines are now given.

\begin{itemize}
    \item We first build a set of basis functions that will be used to express the solution.
    \item They are obtained by looking for separable solutions of the eigenvalue problem 
    \begin{equation}
    \tilde{r}^3 \displaystyle \frac{\partial^2 \delta \tilde{\Omega}_{\text{h}}}{\partial r^2} + 4 \tilde{r}^2 \displaystyle \frac{\partial \delta \tilde{\Omega}_{\text{h}}}{\partial r} + \tilde{r} \displaystyle \frac{\partial^2 \delta \tilde{\Omega}_{\text{h}}}{\partial \theta^2} = \tilde{\lambda} \tilde{r} \delta \tilde{\Omega}_{\text{h}}
    \label{determination_eigenvalue_problems},
    \end{equation}
    satisfying the boundary conditions Eq. \eqref{boundary_conditions}.
    \item We finally determine a general solution using the orthogonality properties of the basis functions.
\end{itemize}

\noindent Separable solutions, $\delta \tilde{\Omega}_{\text{h}} = \tilde{g}(\theta) \tilde{f}(r)$, of Eq. \eqref{determination_eigenvalue_problems} must verify: 

\begin{equation}
\tilde{r}^3 \tilde{f}^{''}(r) + 4 \tilde{r}^2 \tilde{f}^{'}(r) + \tilde{r} \tilde{g}^{''}(\theta) = \tilde{\lambda} \hspace*{0.05cm} \tilde{r} \tilde{g}(\theta) \tilde{f}(r) \quad \Longrightarrow \quad \displaystyle \frac{\tilde{r}^2 \tilde{f}^{''}(r) + 4 \tilde{r} \tilde{f}^{'}(\tilde{r})}{\tilde{f}(r)} + \displaystyle \frac{\tilde{g}^{''}(\theta)}{\tilde{g}(\theta)} = \tilde{\lambda}
\label{eq_sep_var},
\end{equation}

\noindent with the boundary conditions

\begin{equation}
g(\theta_0) = g^{'}(0) = 0 \quad \text{and,} \quad f(r_0) = f(r_{i_{\hspace*{0.025cm} \text{DZ}}}) = 0
\label{bc_on_eigenfunctions}.
\end{equation}

\noindent The problem thus reduces to the solving the two following sub-eigenvalue problems:

\begin{equation}
\tilde{g}^{''}(\theta) - \tilde{\nu} \tilde{g}(\theta) = 0 \quad \text{and,} \quad \tilde{r}^2 \tilde{f}^{''}(r) + 4 \tilde{r} \tilde{f}^{'}(r) - \tilde{\mu} \tilde{f}(r) = 0 \quad \text{with} \quad \tilde{\nu} + \tilde{\mu} = \tilde{\lambda}
\label{the_two_eigenvalue_problems}.
\end{equation}

\noindent The differential equation in $\theta$ is a classical Sturm-Liouville problem while the differential equation in $r$ is known as the Euler's problem. We first deal with the Sturm-Liouville problem.

In order to avoid unphysical solutions, the eigenvalues must be strictly negative. The solution then reads 

\begin{equation}
\tilde{g}_k(\theta) = \tilde{A}_k \cos{\left(\sqrt{\left|\tilde{\nu}_k\right|} \hspace*{0.05cm} \theta\right)} + \tilde{B}_k \sin{\left(\sqrt{\left|\tilde{\nu}_k\right|} \hspace*{0.05cm} \theta\right)}
\label{sol1}.
\end{equation}

\noindent Its latitudinal derivative is readily

\begin{equation}
\tilde{g}_k^{'}(\theta) = - \tilde{A}_k \sqrt{\left|\tilde{\nu_k}\right|} \sin{\left(\sqrt{\left|\tilde{\nu_k}\right|} \hspace*{0.05cm} \theta\right)} + \tilde{B}_k \sqrt{\left|\tilde{\nu_k}\right|} \cos{\left(\sqrt{\left|\tilde{\nu_k}\right|} \hspace*{0.05cm} \theta\right)}
\label{derivative_sol1}.
\end{equation}

\noindent From the condition of symmetry, $\tilde{B}_k=0$ and $\tilde{g}_k(\theta) = \tilde{A}_k \cos{\left(\sqrt{\left|\tilde{\nu}_k\right|} \hspace*{0.05cm} \theta\right)}$. Besides, $\tilde{g}_k(\theta_0) = \tilde{A}_k \cos{\left(\sqrt{\left|\tilde{\nu}_k\right|} \hspace*{0.05cm} \theta_0 \right)} = 0$. Since we are looking for a non-trivial solution (i.e., $\tilde{A}_k\neq0$) we obtain

\begin{equation}
\sqrt{\left|\tilde{\nu}_k\right|} \theta_0 = k \pi - \displaystyle \frac{\pi}{2} \quad \text{with} \quad k \in \mathbb{Z},
\end{equation}

\noindent thus concluding that

\begin{equation}
\tilde{\nu}_{k} = - \left(\displaystyle \frac{\left(2 k - 1 \right) \pi}{2 \theta_0} \right)^2  \quad \text{with} \quad k \in \mathbb{Z}
\label{eigenvalues_SL},
\end{equation}

\noindent are the sought eigenvalues and are associated with the eigenfunctions

\begin{equation}
\tilde{g}_k(\theta) = \tilde{A}_k \cos{\left(\left( \displaystyle \frac{\left(2k - 1\right) \pi}{2 \theta_0} \right) \theta\right)} \quad \text{with} \quad k \in \mathbb{Z}
\label{eigenfunctions_SL}.
\end{equation}

We are now going to solve the Euler's problem. Here the eigenvalues must be $< - 9 /4$ to avoid unphysical solutions. In that case, the solution reads:

\begin{equation}
\tilde{f}_n(r) = \tilde{C}_n \hspace*{0.05cm} \tilde{r}^{ \displaystyle \frac{-1}{2}\left( 3 + i \sqrt{\left|9 + 4 \hspace*{0.05cm} \tilde{\mu}_n\right|} \right)} + 
\tilde{D}_n \hspace*{0.05cm} \tilde{r}^{ \displaystyle \frac{-1}{2}\left( 3 - i \sqrt{\left|9 + 4 \hspace*{0.05cm} \tilde{\mu}_n\right|} \right)}
\label{sol_euler},
\end{equation}

\noindent which can be rewritten as

\begin{equation}
\tilde{f}_n(r) = \displaystyle \frac{\tilde{C}_n}{\tilde{r}^{3/2}} \hspace*{0.05cm} \cos{\left( \displaystyle \frac{1}{2} \sqrt{\left|9 + 4 \hspace*{0.05cm} \tilde{\mu}_n\right|} \ln{\left(\tilde{r}\right)} \right)} + 
\displaystyle \frac{\tilde{D}_n}{\tilde{r}^{3/2}} \hspace*{0.05cm} \sin{\left( \displaystyle \frac{1}{2} \sqrt{\left|9 + 4 \hspace*{0.05cm} \tilde{\mu}_n\right|} \ln{\left(\tilde{r}\right)} \right)}
\label{sol_euler_2}.
\end{equation}

\noindent From the condition $\tilde{f}_n(1) = 0$ we have $C_n = 0$ and

\begin{equation}
\tilde{f}_n(r) = \displaystyle \frac{\tilde{D}_n}{\tilde{r}^{3/2}} \hspace*{0.05cm} \sin{\left( \displaystyle \frac{1}{2} \sqrt{\left|9 + 4 \hspace*{0.05cm} \tilde{\mu}_n\right|} \ln{(\tilde{r})} \right)}.
\end{equation}

\noindent The second boundary condition leads to

\begin{equation}
\displaystyle \frac{\tilde{D}_n}{\tilde{r}_{i_{\hspace*{0.05cm} \text{DZ}}}^{3/2}} \hspace*{0.05cm} \sin{\left( \displaystyle \frac{1}{2} \sqrt{ \left|9 + 4 \hspace*{0.05cm} \tilde{\mu}_n \right|} \ln{\left(\tilde{r}_{i_{\hspace*{0.05cm} \text{DZ}}}\right)} \right)} = 0.
\end{equation}

\noindent Excluding the non-trivial solution $\tilde{D}_n=0$ we obtain

\begin{equation}
\displaystyle \frac{1}{2} \sqrt{\left|9 + 4 \hspace*{0.05cm} \tilde{\mu}_n \right|} \ln{\left(\tilde{r}_{i_{\hspace*{0.05cm} \text{DZ}}}\right)} = n \pi  \quad \text{with} \quad n \in \mathbb{Z}.
\end{equation}

\noindent The sought eigenvalues are thus

\begin{equation}
\tilde{\mu}_{n} = - \left(\displaystyle \frac{3}{2}\right)^2 -  \left(\displaystyle \frac{n\pi}{\ln{\left( \tilde{r}_{i_{\hspace*{0.05cm} \text{DZ}}}\right)}}\right)^2  \quad \text{with} \quad n \in \mathbb{Z},
\end{equation}

\noindent and are associated with the eigenfunctions

\begin{equation}
\tilde{f}_n(r) = \displaystyle \frac{\tilde{D}_n}{\tilde{r}^{3/2}} \sin{\left(\displaystyle \frac{n \pi \ln{\left(\tilde{r}\right)}}{\ln{\left(\tilde{r}_{i_{\hspace*{0.05cm} \text{DZ}}}\right)}}\right)} \quad \text{with} \quad n \in \mathbb{Z}.
\end{equation}

\noindent The general solution $\delta \Omega(r,\theta)$ is now expanded over the basis of the eigenfunctions satisfying the boundary conditions Eq. \eqref{bc_on_eigenfunctions}:

\begin{equation}
\delta \tilde{\Omega}(r,\theta) = \sum_{n=1}^{\infty} \sum_{k=1}^{\infty} \displaystyle \frac{\tilde{A}_{nk}}{\tilde{r}^{3/2}} 
\sin{\left(\left(\displaystyle \frac{n \pi}{\ln{\left(\tilde{r}_{i_{\hspace*{0.05cm} \text{DZ}}}\right)}}\right) \ln{\left(\tilde{r}\right)} \right)}  \cos{\left(\left( \displaystyle \frac{\left(2k - 1\right) \pi}{2 \theta_0} \right) \theta\right)} \quad \text{with} \quad n, k \in \mathbb{Z}
\label{general_sol}.
\end{equation}

\noindent Thus, this solution verifies the boundary conditions Eq. \eqref{boundary_conditions}. We now need to find the $A_{nk}$ coefficients. These are determined using the orthogonality properties of the eigenfunctions. After determining the different partial derivatives of the function $\delta \Omega(r,\theta)$ and after substituting their expressions in Eq. \eqref{viscous_balance_simplified} we obtain the following relationship:

\begin{equation}
\sum_{n=1}^{\infty} \sum_{k=1}^{\infty} \displaystyle \frac{\tilde{A}_{nk}}{\sqrt{\tilde{r}}} \left \lbrack  - \left(\displaystyle \frac{3}{2}\right)^2 - \left(\displaystyle \frac{n\pi}{\ln \left(\tilde{r}_{i_{\hspace*{0.05cm} \text{DZ}}}\right)}\right)^2 - \left( \displaystyle \frac{\left(2k - 1\right) \pi}{2 \theta_0} \right)^2 \right \rbrack 
\sin{\left(\left(\displaystyle \frac{n \pi}{\ln{\left(\tilde{r}_{i_{\hspace*{0.05cm} \text{DZ}}}\right)}}\right)\ln{\left(\tilde{r}\right)} \right)} 
\cos{\left(\left( \displaystyle \frac{\left(2k - 1\right) \pi}{2 \theta_0} \right) \theta\right)} = \displaystyle \frac{-2}{E} 
\label{relation}.
\end{equation}

\noindent As,

\begin{equation}
\tilde{\lambda}_{nk} =  - \left \lbrack \underbrace{\left(\displaystyle \frac{3}{2}\right)^2 + \left(\displaystyle \frac{n\pi}{\ln \left(\tilde{r}_{i_{\hspace*{0.05cm} \text{DZ}}}\right)}\right)^2}_{\left|\tilde{\mu}_n\right|} + 
\underbrace{\left( \displaystyle \frac{\left(2k - 1\right) \pi}{2 \theta_0} \right)^2}_{\left|\tilde{\nu}_k \right|} \right \rbrack
\quad \text{with} \quad n, k \in \mathbb{Z},
\end{equation}

\noindent Eq. \eqref{relation} is rewritten as follows

\begin{equation}
\sum_{n=1}^{\infty} \sum_{k=1}^{\infty} \displaystyle \frac{\tilde{A}_{nk} \tilde{\lambda}_{nk}}{\tilde{r}^{3/2}}
\sin{\left(\left(\displaystyle \frac{n \pi}{\ln{\left(\tilde{r}_{i_{\hspace*{0.05cm} \text{DZ}}}\right)}}\right)\ln{\left(\tilde{r}\right)} \right)} 
\cos{\left(\left( \displaystyle \frac{\left(2k - 1\right) \pi}{2 \theta_0} \right) \theta\right)} = \displaystyle \frac{-2}{\tilde{r}E}.
\label{relation2}
\end{equation}

\noindent Multiplying each members of this relationship by $\tilde{r}^{-1/2} \sin{\left(\left(\displaystyle \frac{n \pi}{\ln{\left(\tilde{r}_{i_{\hspace*{0.05cm} \text{DZ}}}\right)}}\right)\ln{\left(\tilde{r}\right)} \right)} 
\cos{\left(\left( \displaystyle \frac{\left(2k - 1\right) \pi}{2 \theta_0} \right) \theta\right)}$ then integrating over the domain $\mathcal{D}$ the resulting relationship, we obtain the coefficients $A_{nk}$:

\begin{equation}
\tilde{A}_{nk} = \left( \displaystyle \frac{4 \left(n\pi\right)^2 + \left(\ln{\left(\tilde{r}_{i_{\hspace*{0.05cm} \text{DZ}}}\right)} \right)^2}{E \tilde{\lambda}_{nk} \left(\tilde{r}_{i_{\hspace*{0.05cm} \text{DZ}}}-1\right) \left(n\pi\right)^2 \theta_0 } \right)
\left( \displaystyle \frac{-4 \theta_0}{\left(2k-1\right)\pi} \left(-1 \right)^{k} \right)
\left( \displaystyle \frac{4 n \pi \ln{\left(\tilde{r}_{i_{\hspace*{0.05cm} \text{DZ}}}\right)}}{4 \left(n\pi\right)^2 + 9 \left( \ln{\left(\tilde{r}_{i_{\hspace*{0.05cm} \text{DZ}}}\right)} \right)^2} \left \lbrack \left(-1\right)^n \tilde{r}_{i_{\hspace*{0.05cm} \text{DZ}}}^{3/2} - 1 \right \rbrack \right) \quad \text{with} \quad n, k \in \mathbb{Z},
\end{equation}

\noindent and the final solution, which reads under dimensional form

\begin{equation}
\begin{array}{lll}
\displaystyle \frac{\delta \Omega(r,\theta)}{\Omega_0} = Re_c \sum_{n=1}^{\infty} \sum_{k=1}^{\infty} \left \lbrack 
\left( \displaystyle \frac{\left(2n\pi\right)^2 + \left(\ln{\left(\displaystyle \frac{r_{i_{\hspace*{0.025cm} \text{DZ}}}}{r_0}\right)} \right)^2}{\lambda_{nk} \left(\displaystyle \frac{r_{i_{\hspace*{0.025cm} \text{DZ}}}}{r_0}-1\right) \left(n\pi\right)^2 \theta_0 } \right)
\left( \displaystyle \frac{-4 \theta_0}{\left(2k-1\right)\pi} \left(-1 \right)^{k} \right)
\left( \displaystyle \frac{4 n \pi \ln{\left(\displaystyle \frac{r_{i_{\hspace*{0.025cm} \text{DZ}}}}{r_0}\right)}}{\left(2n\pi\right)^2 + \left( 3\ln{\left(\displaystyle \frac{r_{i_{\hspace*{0.025cm} \text{DZ}}}}{r_0}\right)} \right)^2} 
\left ( \left(-1\right)^n \left(\displaystyle \frac{r_{i_{\hspace*{0.025cm} \text{DZ}}}}{r_0}\right)^{3/2} - 1 \right ) \right) \right.
\\\\ \hspace*{3.5cm} 
\left( \displaystyle \frac{r_0}{r} \right)^{3/2} \sin{\left(\left(\displaystyle \frac{n \pi}{\ln{\left(\displaystyle \frac{r_{i_{\hspace*{0.025cm} \text{DZ}}}}{r_0}\right)}}\right) \ln{\left(\displaystyle \frac{r}{r_0}\right)} \right)}  
\cos{\left(\left( \displaystyle \frac{\left(2k - 1\right) \pi}{2 \theta_0} \right) \theta\right)} \huge \Bigg{]} \normalsize \quad \text{with} \quad n, k \in \mathbb{Z}
\end{array}
\label{analytical_solution}.
\end{equation}

\subsection{Case $2$ - Advection of the field lines by the contraction}
\label{resolution_induct_mod}

\begin{figure}[!h]
\begin{center}
\includegraphics[width=12cm]{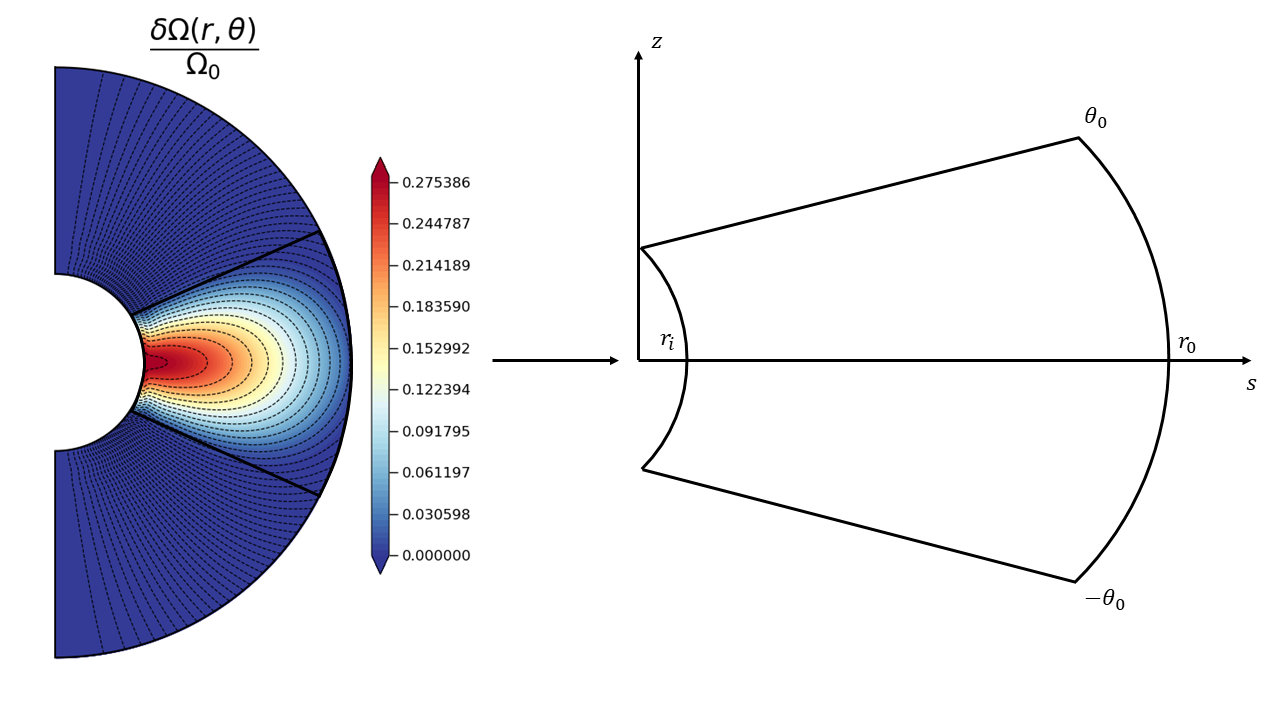}
\caption{Sketch of the conical domain chosen to represent the DZ when the field lines are advected by the contraction. The displayed meridional cut of the normalised rotation rate is the same as in the third panel of Fig. \ref{Numerical_Results_Viscous}. The conical domain, delimited by thick black lines on the meridional cut, is defined by $r \in \left \lbrack r_i ; r_0 \right \rbrack $ and $\theta \in \left \lbrack - \theta_0 ; \theta_0 \right \rbrack$ where $\theta_0 \approx 4\pi/23$ and $-\theta_0 \approx - 4\pi/23$ or equivalently, in terms of colatitude, $\theta \in \left \lbrack 15 \pi / 46 ; 31 \pi / 46 \right \rbrack$.}  
\label{conical_domain2}
\end{center}
\end{figure}

When the contraction term is introduced in the induction equation, the poloidal field lines are advected and the DZ connects to the inner sphere (see Fig. \ref{conical_domain2}). This modifies the boundary conditions Eq. \eqref{boundary_conditions} since at the inner sphere, the azimuthal velocity field now satisfies a stress-free condition hence:

\begin{equation}
\delta \Omega(r_0,\theta) = \left. \displaystyle \frac{\partial \delta \Omega(r,\theta)}{\partial \theta} \right |_{r = r_i} = \delta \Omega(r,\theta_0) = \left. \displaystyle \frac{\partial \delta \Omega(r,\theta)}{\partial \theta} \right |_{\theta = 0} = 0
\label{new_boundary_conditions}.
\end{equation}

\noindent Thus, the solution of the Sturm-Liouville problem is unchanged and the sought eigenvalues are again Eq. \eqref{eigenvalues_SL} and are associated with the eigenfunctions Eq. \eqref{eigenfunctions_SL}. For the Euler's problem the eigenvalues must still be $< 9/4$ to avoid unphysical solutions. By taking Eq. \eqref{sol_euler_2} and after using the condition $\tilde{f}_n(1) = 0$ we have

\begin{equation}
\tilde{f}_n(r) = \displaystyle \frac{\tilde{D}_n}{\tilde{r}^{3/2}} \hspace*{0.05cm} \sin{\left( \displaystyle \frac{1}{2} \sqrt{\left|9 + 4 \hspace*{0.05cm} \tilde{\mu}_n\right|} \ln{\left(\tilde{r}\right)} \right)}
\label{new_sol_euler}.
\end{equation}

\noindent By deriving this expression to apply the stress-free condition to it, we get the transcendental equation

\begin{equation}
3 \tan{\left( \displaystyle \frac{1}{2} \sqrt{|9+4 \hspace*{0.05cm} \tilde{\mu}_n|} \ln(\tilde{r}_i) \right)} - \sqrt{|9+4 \hspace*{0.05cm} \tilde{\mu}_n|} = 0
\label{transcendental_equation}.
\end{equation}

\noindent After numerically solving this equation we obtain the eigenvalues $\mu_n$. The general solution $\delta \Omega(r,\theta)$ of Eq. \eqref{viscous_balance_simplified} is then expanded on the basis of the eigenfunctions satisfying the boundary conditions Eq. \eqref{new_boundary_conditions} on the domain $\mathcal{D}$

\begin{equation}
\delta \tilde{\Omega}(r,\theta) = \sum_{n=1}^{\infty} \sum_{k=1}^{\infty} \displaystyle \frac{\tilde{A}_{nk}}{\tilde{r}^{3/2}} 
\sin{\left( \displaystyle \frac{1}{2} \sqrt{\left|9 + 4 \hspace*{0.05cm} \tilde{\mu}_n \right|} \ln{\left(\tilde{r}\right)} \right)} \cos{\left(\left( \displaystyle \frac{\left(2k - 1\right) \pi}{2 \theta_0} \right) \theta\right)} \quad \text{with} \quad n, k \in \mathbb{Z}
\label{sol_generale_new}.
\end{equation}

\noindent As previously, after determining the different partial derivatives of the function $\delta \Omega(r,\theta)$, and after substituting them in Eq. \eqref{viscous_balance_simplified} we get:

\begin{equation}
\sum_{n=1}^{\infty} \sum_{k=1}^{\infty} \displaystyle \frac{\tilde{A}_{nk}}{\sqrt{\tilde{r}}} \left \lbrack  \underbrace{\tilde{\mu}_n - \left( \displaystyle \frac{\left(2k - 1\right) \pi}{2 \theta_0} \right)^2}_{\tilde{\lambda}_{nk}} \right \rbrack 
\sin{\left( \displaystyle \frac{1}{2} \sqrt{\left|9 + 4 \hspace*{0.05cm} \tilde{\mu}_n \right|} \ln{\left(\tilde{r}\right)} \right)} 
\cos{\left(\left( \displaystyle \frac{\left(2k - 1\right) \pi}{2 \theta_0} \right) \theta\right)} = \displaystyle \frac{-2}{E} 
\label{relation_new}.
\end{equation}

\noindent Again, using the orthogonality properties enables us to determine the $A_{nk}$ coefficients

\begin{equation}
A_{nk} = \displaystyle \frac{16 \left(1 + \left| 9 + 4 \hspace*{0.025cm} \tilde{\mu}_n \right| \right) \left(-1\right)^k}{E \hspace*{0.025cm} \lambda_{nk} \left(2k-1\right) \left(9 + \left| 9 + 4 \hspace*{0.025cm} \tilde{\mu}_n \right|\right) \pi} \cdot
\displaystyle \frac{\left \lbrack 2 \sqrt{\left| 9 + 4 \hspace*{0.025cm} \tilde{\mu}_n \right|} \left(\tilde{r}_i^{3/2} \cos{\left(\displaystyle \frac{1}{2} \sqrt{\left| 9 + 4 \hspace*{0.025cm} \tilde{\mu}_n \right|} \ln{(\tilde{r}_i)} \right)} -1 \right) - 6 \hspace*{0.025cm} \tilde{r}_i^{3/2} \sin{\left(\displaystyle \frac{1}{2} \sqrt{\left| 9 + 4 \hspace*{0.025cm} \tilde{\mu}_n \right|} \ln{(\tilde{r}_i)}\right)} \right \rbrack}{\left \lbrack \left| 9 + 4 \hspace*{0.025cm} \tilde{\mu}_n \right| \left(1 - \tilde{r}_i\right) + \tilde{r}_i \left( \cos{\left(\sqrt{\left| 9 + 4 \hspace*{0.025cm} \tilde{\mu}_n \right|}\ln{(\tilde{r}_i)}\right)}  + \sqrt{\left| 9 + 4 \hspace*{0.025cm} \tilde{\mu}_n \right|} \sin{\left(\sqrt{\left| 9 + 4 \hspace*{0.025cm} \tilde{\mu}_n \right|}\ln{(\tilde{r}_i)}\right)} -1 \right) \right \rbrack}
\label{ank_coeff_advection},
\end{equation}

\noindent and so the final solution which, under dimensional form, reads:

\begin{equation}
\begin{array}{lll}
\displaystyle \frac{\delta \Omega(r,\theta)}{\Omega_0} = Re_c \displaystyle \sum_{n=1}^{\infty} \displaystyle \sum_{k=1}^{\infty}
\displaystyle \frac{\left \lbrack 2 \sqrt{\left| 9 + 4 \hspace*{0.025cm} \mu_n \right|} \left(\left(\displaystyle \frac{r_i}{r_0}\right)^{3/2} \cos{\left(\displaystyle \frac{1}{2} \sqrt{\left| 9 + 4 \hspace*{0.025cm} \mu_n \right|} \ln{\left(\displaystyle \frac{r_i}{r_0}\right)} \right)} -1 \right) - 6 \left(\displaystyle \frac{r_i}{r_0}\right)^{3/2} \sin{\left(\displaystyle \frac{1}{2} \sqrt{\left| 9 + 4 \hspace*{0.025cm} \mu_n \right|} \ln{\left(\displaystyle \frac{r_i}{r_0}\right)}\right)} \right \rbrack}{\left \lbrack \left| 9 + 4 \hspace*{0.025cm} \mu_n \right| \left(1 - \displaystyle \frac{r_i}{r_0}\right) + \displaystyle \frac{r_i}{r_0} \left( \cos{\left(\sqrt{\left| 9 + 4 \hspace*{0.025cm} \mu_n \right|}\ln{\left(\displaystyle \frac{r_i}{r_0}\right)}\right)}  + \sqrt{\left| 9 + 4 \hspace*{0.025cm} \mu_n \right|} \sin{\left(\sqrt{\left| 9 + 4 \hspace*{0.025cm} \mu_n \right|}\ln{\left(\displaystyle \frac{r_i}{r_0}\right)}\right)} - 1 \right) \right \rbrack} \\\\ \hspace*{3.3cm}
\displaystyle \frac{16 \left(1 + \left| 9 + 4 \hspace*{0.025cm} \mu_n \right|\right) \left(-1\right)^k}{\lambda_{nk} \left(2k-1\right) \left(9 + \left| 9 + 4 \hspace*{0.025cm} \mu_n \right|\right) \pi} \displaystyle \left(\frac{r_0}{r}\right)^{3/2} 
\sin{\left( \displaystyle \frac{1}{2} \sqrt{\left|9 + 4 \hspace*{0.025cm} \mu_n \right|} \ln{\left(\displaystyle \frac{r}{r_0}\right)} \right)} \cos{\left(\left( \displaystyle \frac{\left(2k - 1\right) \pi}{2 \theta_0} \right) \theta\right)}
,\end{array}
\label{final_sol_advection}
\end{equation}

\noindent where $n, k \in \mathbb{Z}$.

\end{document}